\RequirePackage[l2tabu, orthodox]{nag}
\documentclass[prd,aps,notitlepage,nofootinbib,floatfix,reprint,groupedaddress]{revtex4-1}
\usepackage[T1]{fontenc}
\usepackage[utf8]{inputenc}
\usepackage{color}
\usepackage{array}
\usepackage{float}
\usepackage{calc}
\usepackage{units}
\usepackage{url}
\usepackage{multirow}
\usepackage{amsmath}
\usepackage{amssymb}
\usepackage{physics}
\usepackage{graphicx}
\usepackage[svgnames]{xcolor} 
\usepackage[makeroom]{cancel}
\usepackage{booktabs}
\usepackage{pdfsync}
\DeclareMathOperator{\sinc}{sinc}
\DeclareMathOperator{\asinc}{asinc}
\usepackage{siunitx}
\numberwithin{equation}{section}
\numberwithin{figure}{section}
\numberwithin{table}{section}
\usepackage{listings}
\usepackage{subcaption}
\usepackage{slashed}
\usepackage{bm}
\usepackage[bordercolor=white,backgroundcolor=gray!30,linecolor=black,colorinlistoftodos]{todonotes}
\DeclareFontFamily{U}{wncy}{}
\DeclareFontShape{U}{wncy}{m}{n}{<->wncyr10}{}
\DeclareSymbolFont{mcy}{U}{wncy}{m}{n}
\DeclareMathSymbol{\Sh}{\mathord}{mcy}{"58}

\usepackage{hhline}

\usepackage{enumitem}		
\usepackage{footmisc}
\usepackage{natbib}

\usepackage[unicode=true, bookmarks=true,bookmarksnumbered=true,bookmarksopen=false, 
breaklinks=false,pdfborder={0 0 1},backref=false,colorlinks=true]{hyperref}
\usepackage{breakurl}
\usepackage{bookmark}
\hypersetup{pdftitle={essay},
  pdfauthor={},
  linkcolor=magenta, urlcolor=magenta, citecolor=blue}
\usepackage{cleveref}
\usepackage{subdepth}

\usepackage{pdfcomment}

\newcommand{\MODALLSS}{\texttt{MODAL-LSS}}
\newcommand{\LPICOLA}{\texttt{L-PICOLA}}
\newcommand{\GADGET}{\texttt{GADGET-3}}

\begin{document}

\title{Advancing the matter bispectrum estimation of large-scale structure: a comparison of dark matter codes}
\author{Johnathan Hung}
\email{jmch2@damtp.cam.ac.uk}
\affiliation{Centre for Theoretical Cosmology, DAMTP, University of Cambridge, CB3 0WA, United Kingdom}
\author{James Fergusson}
\email{J.Fergusson@damtp.cam.ac.uk}
\affiliation{Centre for Theoretical Cosmology, DAMTP, University of Cambridge, CB3 0WA, United Kingdom}
\author{E.P.S. Shellard}
\email{E.P.S.Shellard@damtp.cam.ac.uk}
\affiliation{Centre for Theoretical Cosmology, DAMTP, University of Cambridge, CB3 0WA, United Kingdom}

\date{\today}

\begin{abstract}
Cosmological information from forthcoming galaxy surveys, such as LSST and Euclid, will soon exceed that available from the CMB.  Higher order correlation functions, like the bispectrum, will be indispensable for realising this potential.   The interpretation of this data faces many challenges because gravitational collapse of matter is a complex non-linear process, typically modelled by computationally expensive N-body simulations.  Proposed alternatives using fast dark matter codes (e.g. 2LPT or particle-mesh) are primarily evaluated on their ability to reproduce clustering statistics linked to the matter power spectrum.  The accuracy of these codes can be tested in more detail by looking at higher-order
 statistics, and in this paper we will present an efficient and optimal methodology (\MODALLSS{}) to reconstruct the full bispectrum of any 3D density field.  We make quantitative comparisons between a number of fast dark matter codes and \GADGET{} at redshift $z=0.5$. This will serve as an important diagnostic tool for dark matter/halo mock catalogues and lays the foundation for realistic high precision
analysis with the galaxy bispectrum. In particular, we show that the lack of small-scale power in the bispectrum of fast codes can be ameliorated by a simple `boosting' technique for the power spectrum. We also investigate the covariance of the \MODALLSS{} bispectrum estimator, demonstrating the plateauing of non-Gaussian errors in contrast to simple Gaussian extrapolations.  This has important consequences for the extraction of information from the bispectrum and hence parameter estimation. Finally we make quantitative comparisons of simulation bispectra with theoretical models, discussing the initial parameters required to create mock catalogues with accurate bispectra.
\end{abstract}




\pagebreak

\maketitle

\section{Introduction}

In the standard description of Cosmology the early Universe went through a phase
of accelerated expansion known as inflation. Through this inflationary period 
quantum fluctuations of the primordial fields became classical perturbations 
which are in turn the seeds for late-time observables such as the anisotropies 
of the Cosmic Microwave Background (CMB) and the distribution of large-scale 
structure (LSS) of the Universe such as dark matter halos and galaxies. 
Extensive work has been done with CMB anisotropies, culminating in the 
tight constraints on parameters such as $f_{nl}$ given by the latest Planck
results \citep{planck2015}. However, the constraining power of the CMB has
nearly reached its limits and will ultimately be superseded by observations of
the large-scale structure of the Universe; this is simply because the
three-dimensional galaxy distribution can provide more information than the
two-dimensional map of the CMB. This goal is facilitated by upcoming large
data sets offered by galaxy surveys such as the Dark Energy Survey (DES)
\citep{DES1, DES2}, the Large Synoptic Survey Telescope (LSST) \citep{LSST},
the ESA Euclid Satellite \citep{Euclid} and the Dark Energy Spectroscopic
Instrument (DESI) \citep{DESI}. One of the most active areas of cosmological
research today is therefore to understand the collapse of matter and evolution
of large scale structure in the Universe. Extra value can be obtained from the
addition of LSS observational data as it can be cross-correlated and combined
with CMB data, e.g. through weak lensing \citep{lensing}, for a wealth of new
information.

Standard single field slow-roll inflation generates only small primordial
non-Gaussianities (PNG) that slow roll supressed \citep{maldacena}, which
is consistent with the null detection presented in latest Planck results \citep{Planck}. Due to the
linearity of CMB physics and the approximately Gaussian initial conditions most CMB information is encoded in the power
spectrum $C_l$. This is not the case for LSS as non-linear gravitational
interaction trandfers information from the power spectrum to higher order correlators. For example, at mildly
non-linear scales the bispectrum is the primary diagnostic as it exceeds the
power spectrum in terms of cosmological information. A recent comprehensive
forecasting of constraints from the galaxy power spectrum and bispectrum \citep{forecast}
has shown that the galaxy bispectrum leads to 5 times better bounds than
the power spectrum alone, giving much tighter constraints for local-type PNG
than current limits from Planck. This work is more complete and realistic than
previous forecasts, e.g. \citep{forecast1,forecast2,forecast3,forecast4}, as
they combined in their analysis different factors that were previously considered
independently. The bispectrum has a stronger dependence on cosmologica
parameters so can provide tighter constraints than the power spectrum for the same signal to noise and can help break degeneracies in parameter space , notably those between $\sigma_8$
and bias \citep{bias}. Many inflationary scenarios, such as those inspired by 
fundamental theories like superstring theory, or alternatives to inflation 
typically yield small, but measurable, PNGs that would be tell-tale signatures of
new physics. In addition to constraining and testing early universe theories,
the bispectrum can be used to test alternative scenarios such as those that
modify standard Einstein gravity. Measurements of the galaxy bispectrum has
been done for existing galaxy survey data from the Baryon Oscillation
Spectroscopic Survey (BOSS) \citep{BOSS1, BOSS2, sdssi, sdssii, sdssiii}. 

There are many complications when extracting information from LSS compared to 
the CMB. At the time when recombination took place and CMB photons were released 
(i.e. redshift $z=1100$), inhomogeneities in the universe were small, therefore 
CMB physics is linear and can be well modelled by perturbation theories. By 
contrast, we still do not have a solid theoretical understanding of the
non-linear gravitational evolution of matter and galaxy formation. A
combination of perturbation theory, e.g. an effective field theory (EFT)
approach \citep{EFT}, and nonlinear halo models has been shown to characterise
the dark matter power spectrum and bispectrum very well at small and large scales,
but the bispectrum at mildly non-linear regimes remain poorly understood
\citep{Andrei}. 

This paper is outlined as follows: in \Cref{previous} we will give an overview
on non-Gaussianity and the three-point correlator of LSS, including in particular
a summary of the \MODALLSS{} method for reconstructing any theoretical
bispectrum or the full bispectrum of an observational or simulated data set.
The main results of this paper, including quantitative bi-spectral comparisons
between different dark matter codes, non-Gaussian covariances of the \MODALLSS{}
estimator, and comparisons between simulations and theory, will be presented
in \Cref{work}, where we also address the difficulties in the latter. We
conclude our paper in \Cref{sec:conclusions}.

\section{Previous work}
\label{previous}

\subsection{Basics of non-Gaussianity}
\label{sec:basics-non-gauss}

At early times before matter collapsed to form structures, the matter distribution
in the Universe was highly uniform. In the absence of any primordial
non-Gaussianity, $\delta$ is Gaussian distributed and can be fully described by its
two-point correlation function, or in Fourier space its power spectrum:
\begin{align}
  \expval{\delta(\mathbf{k}) \delta(\mathbf{k}')}=(2\pi)^3 
  \delta_D(\mathbf{k}+\mathbf{k}') P(k),
  \label{PS}
\end{align}
where $\delta_D$ is the Dirac delta function. At late times this is no longer the
case as gravitational collapse induces non-Gaussianities. For mildly non-linear scales the primary diagnostic
is the three point correlation function or bispectrum
$B_\delta(k_1, k_2, k_3)$:
\begin{align}
  &\expval{\delta(\mathbf{k}_1) \delta(\mathbf{k}_2) \delta(\mathbf{k}_3)}
    \nonumber \\
  &\qquad=(2\pi)^3 \delta_D (\mathbf{k}_1+\mathbf{k}_2+\mathbf{k}_3)
    B_\delta(k_1,k_2,k_3).
    \label{bispectrum}
\end{align}
Due to statistical isotropy and homogeneity the bispectrum only depends on
the wavenumbers $k_i$. Additionally the delta function, arising from momentum conservation, imposes the triangle
condition on the wavevectors so the three $k_i$ when taken as lengths must be able to form a triangle.

\subsection{Bispectrum shapes
  \label{sec:shap-non-gauss}}

Bispectra are naturaly 3D objects unlike power spectra which are only 1D. The particular dependence of a bispectra on the three $k_i$ is known as its shape. The shapes of popular interest
in CMB analysis are inspired by various inflationary scenarios, but we are more
interested in a few phenomenological shapes that will help us capture
the behaviour of the matter bispectrum at late times. Here we present
a few of these templates popular in the literature, i.e. the tree-level
bispectrum and its extensions, the nine-parameter model and the 3-shape
model. This enables us to investigate any primordial non-Gaussianities
through observational data by subtracting off the dominant contributions
from gravitational collapse.

\subsubsection{Tree-level bispectrum}

By solving the dark matter equations of motion perturbatively,
at lowest order we can derive the tree-level bispectrum
\citep{Bernardeau}:
\begin{align}
  &B^{\text{tree}}(k_1,k_2,k_3)=
    \nonumber \\
  &\quad 2P_{\text{lin}}(k_1) P_{\text{lin}}(k_2)F^{(s)}_2
    (\mathbf{k}_1,\mathbf{k}_2)+2\,\text{perms.},
    \label{eq:tree_shape}
\end{align}
where the $F^{(s)}_2$ kernel takes the form
\begin{align}
  F^{(s)}_2(\mathbf{k}_1,\mathbf{k}_2)=\frac{5}{7}+
  \frac{1}{2}\frac{\mathbf{k}_1\cdot\mathbf{k}_2}{k_1 k_2}
  \left(\frac{k_1}{k_2}+\frac{k_2}{k_1}\right)+\frac{2}{7}
  \frac{(\mathbf{k}_1\cdot\mathbf{k}_2)^2}{k_1^2 k_2^2}.
  \label{F2_EdS}
\end{align}
and $P_{\text{lin}}$ is the linear power spectrum. This technically
only applies in an Einstein-de Sitter universe for which $\Omega_m=1$
and $\Omega_\Lambda=0$, and hence the linear growth factor $D_1=a$. 
We are interested instead in the late time universe where 
$\Omega_\Lambda>0$ so we modify $F^{(s)}_2$ to become
\begin{align}
  F^{(s),\Lambda}_2(\mathbf{k}_1,\mathbf{k}_2)
  &=\frac{1}{2}(1+\epsilon)+
    \frac{1}{2}\frac{\mathbf{k}_1\cdot\mathbf{k}_2}{k_1 k_2}
    \left(\frac{k_1}{k_2}+\frac{k_2}{k_1}\right) \nonumber \\
  &\qquad +\frac{1}{2}(1-\epsilon)
    \frac{(\mathbf{k}_1\cdot\mathbf{k}_2)^2}{k_1^2 k_2^2},
    \label{F2}
\end{align}
where $\epsilon\approx-(3/7)\Omega_m^{-1/143}$ (\citep{bouchet}, and 
correcting for a mistake in \citep{Bernardeau}). The tree-level 
bispectrum is a very useful shape for characterising the matter 
bispectrum at large scales where density perturbations are small. 
It fails at smaller scales when perturbation theory breaks down so 
we need additional shapes for a good fit to the bispectrum in those 
regimes. The authors of \citep{Andrei,transients} have extended the 
tree-level shape by replacing $P_{\text{lin}}$ by the non-linear 
power spectrum $P_{\text{NL}}$ and we shall follow their
example here.

\subsubsection{Nine-parameter model}

The tree-level bispectrum fails to describe the matter bispectrum 
accurately even at mildly non-linear regimes. A way of extending
perturbation theories without resorting to loop corrections is with
phenomenological corrections to the kernel $F^{(s)}_2$ by fitting
to simulations. One such example was introduced in \citep{9param}
which proposed
\begin{align}
  &F^{\text{eff}}_2(\mathbf{k}_1,\mathbf{k}_2)=
    \frac{5}{7}a(n_1,k_1)a(n_2,k_2) \nonumber \\
  &\quad+\frac{1}{2}\frac{\mathbf{k}_1\cdot
    \mathbf{k}_2}{k_1 k_2}\left(\frac{k_1}{k_2}
    +\frac{k_2}{k_1}\right)
    b(n_1,k_1)b(n_2,k_2) \nonumber \\
  &\quad+\frac{2}{7}\frac{(\mathbf{k}_1\cdot
    \mathbf{k}_2)^2}{k_1^2 k_2^2}
    c(n_1,k_1)c(n_2,k_2),
    \label{9param}
\end{align}
where
\begin{align}
  a(n,k)&=\frac{1+\sigma_8^{a_6}(z) [0.7Q_3(n)]^{1/2}
          (qa_1)^{n(k)+a_2}}{1+(qa_1)^{n(k)+a_2}} \\
  b(n,k)&=\frac{1+0.2a_3(n(k)+3)(qa_7)^{n(k)+3+a_8}}
          {1+(qa_7)^{n(k)+3.5+a_8}} \\
  c(n,k)&=\frac{1+4.5a_4/[1.5+(n(k)+3)^4]
          (qa_5)^{n(k)+3+a_9}}{1+(qa_5)^{n(k)+3.5+a_9}}.
\end{align}
Here $q=k/k_{\text{NL}}$, where $k_{\text{NL}}$ which is
the scale at which perturbation theory breaks down
and is found by solving the equation
$k_{\text{NL}}^3 P_{\text{lin}}(k_{\text{NL}},z)=2\pi^2$.
The functions $n(k)$ and $Q_3(n)$ are defined as:
\begin{align}
  n(k)&=\frac{d\log P_{\text{lin}}(k)}{d\log k} \\
  Q_3(n)&=\frac{4-2^n}{1+2^{n+1}}.  \\
\end{align}
The 9 parameters $a_i$ were fitted to simulations with an error
threshold of 10\% in the $k$-range of $0.03\,h^{-1}\text{Mpc}\leq
k\leq 0.4\,h^{-1}\text{Mpc}$ and redshift range of
$0\leq z \leq 1.5$, and take the values of 
\begin{align}
  &a_1=0.484 \quad &&a_2=3.740 \quad &&a_3=-0.849 \nonumber \\
  &a_4=0.392 \quad &&a_5=1.013 \quad &&a_6=-0.575 \nonumber \\
  &a_7=0.128 \quad &&a_8=-0.722 \quad &&a_9=-0.926. \nonumber \\
\end{align}

\subsubsection{Local shape}

The local, or squeezed, bispectrum shape is another popular example. Its
name derives from the local type non-Gaussianity which is generated simply
by adding a term proportional to the square of the Gaussian field $\phi_{G}$: to itself
\begin{align}
  \label{eq:local_NG}
  \phi_{NG}=\phi_{G}+f_{nl}(\phi_{G}^2-\langle\phi_{G}^2\rangle),
\end{align}
where $f_{nl}$ is the non-linearity parameter that gives the degree of
non-Gaussianity, and the term in angle brackets is added to ensure
$\phi_{NG}$ has zero mean. It can be shown that the bispectrum of
$\phi_{NG}$ takes the form
\begin{align}
  \label{eq:local_shape}
  &B^{\text{local}}(k_1,k_2,k_3) = 
    \frac{1}{3} [P_\phi(k_1)P_\phi(k_2)
    \nonumber \\
  &\qquad\qquad+P_\phi(k_2)P_\phi(k_3)+
    P_\phi(k_3)P_\phi(k_1)],
\end{align}
where $P_\phi(k)\propto k^{n_s}$ is the power spectrum of 
$\phi_G$ and $n_s$ is the scalar spectral index. There are two 
ways of promoting this into late times. The easy, and incorrect, 
way is to replace $P_\phi$ with the linear power spectrum:
\begin{align}
  \label{eq:squeez_shape}
  &B^{\text{squeez}}(k_1,k_2,k_3) = 
    \frac{1}{3} [P_{\text{lin}}(k_1)P_{\text{lin}}(k_2)
    \nonumber \\
  &\qquad\qquad+P_{\text{lin}}(k_2)P_{\text{lin}}(k_3)+
    P_{\text{lin}}(k_3)P_{\text{lin}}(k_1)].
\end{align}
Since the linear power spectrum $P_{\text{lin}}(k)\propto k^{n_s-4}$ 
for large $k$, $B^{\text{squeez}}$ peaks for squeezed triangle
configurations where one of the $k$'s is much smaller than the other
two, e.g. $k_1\ll k_2,k_3$. This shape is, however, not the correct
extension since at large scales $B\propto D^3_1$ where $D_1$ is the 
linear growth factor, whereas $P_{\text{lin}}$ grows as $D^2_1$.
Using $\delta(\mathbf{k},z)=M(k,z)\phi_{NG}(\mathbf{k})$ and\footnote{
  $T(k)$ denotes the transfer function, $\Omega_M$ is the present-day
  matter density parameter, and $H_0$ is the Hubble parameter.}
$M(k,z)=\frac{2 D_1(z)T(k) k^2}{3\Omega_M H_0^2}$ we obtain
\begin{align}
  \label{eq:local_late_shape}
  &B^{\text{local,late}}(k_1,k_2,k_3) 
    \nonumber \\
  &= M(k_1)M(k_2)M(k_3) B^{\text{local}}(k_1,k_2,k_3) 
    \nonumber \\
  &\propto
    \sqrt{\frac{P_{\text{lin}}(k_1)P_{\text{lin}}(k_2)
    P_{\text{lin}}(k_3)}{(k_1k_2k_3)^{n_s}}}(
    k_1^{n_s-2}k_2^{n_s-2}k_3^{2} \nonumber \\
  &\qquad+k_1^{2}k_2^{n_s-2}k_3^{n_s-2}+k_1^{n_s-2}k_2^{2}k_3^{n_s-2}).
\end{align}

\subsubsection{Constant shape}

Another useful shape is the constant shape produced by equilateral
triangles $k_1=k_2=k_3$:
\begin{align}
  \label{eq:const_shape}
  B^{\text{const}}(k_1,k_2,k_3) &= B,
\end{align}
where $B$ is, expectedly, a constant. This is the bispectrum shape 
obtained by a set of Poisson-distributed point sources, for instance 
the late time matter distribution at small scales which consists of 
point-like dark matter halo particles. The constant shape is therefore 
ideal for describing the late time matter bispectrum at small scales.

\subsubsection{3-shape model}

The authors of \citep{Andrei} have proposed a benchmark model that
utilises 3 basic bispectrum shapes to build a phenomenological model
for the matter bispectrum calibrated to simulations, very much akin
to the HALOFIT model \citep{halofit} which was introduced
to capture the behaviour of the matter power spectrum. For greater
flexibility of the model they allowed the shapes to have
scale-dependent amplitudes $f_i(K)$ with $K=k_1+k_2+k_3$ for a
better fit to the data. The 3-shape bispectrum is the following
linear combination of the constant, squeezed and tree-level shapes:
\begin{align}
  &B_{\text{3-shape}}(k_1,k_2,k_3) \nonumber \\
  &= f_{1h}(K)B^{\text{const}}(k_1,k_2,k_3)
    +f_{2h}(K)B^{\text{squeez}}(k_1,k_2,k_3)
    \nonumber \\
  &\quad
    +f_{3h}(K)B^{\text{treeNL}}(k_1,k_2,k_3),
    \label{eqn:3-shape}
\end{align}
where $B^{\text{const}}$ and $B^{\text{squeez}}$ are given by
\Cref{eq:const_shape,eq:squeez_shape}. The tree-level shape is based
on \Cref{eq:tree_shape} except we have replaced the linear power
spectrum with the non-linear power spectrum obtained from simulations:
\begin{align}
  &B^{\text{treeNL}}(k_1,k_2,k_3)=
    \nonumber \\
  &\quad 2P_{\text{NL}}(k_1) P_{\text{NL}}(k_2)F^{(s),\Lambda}_2
    (\mathbf{k}_1,\mathbf{k}_2)+2\,\text{perms.},
    \label{treeNL}
\end{align}
The amplitudes $f_i(K)$ are found by fitting each of these shapes to
the three halo model components. For a comprehensive review on the
halo model bispectrum please see \citep{Andrei}. The one-halo bispectrum
has been shown to correlate very well with the constant shape with the
following choice of Lorentzian fitting function:
\begin{equation}
  f_{1h}(K)=\frac{A}{(1+bK^2)^2},
\end{equation}
where $A$ and $b$ are redshift-dependent functions through
the linear growth factor $D(z)$:
\begin{align}
  A&=\frac{2.45\times10^6D(z)^8}{0.8+0.2\,D(z)^{-3}}\\
  b&=0.054\,D(z)^{2.2}.
\end{align}
The two-halo bispectrum has a strong correlation with the squeezed
shape but has several notable shortcomings \citep{halo1,halo2,halo3}.
To resolve these deficiencies Valageas and Nishimichi developed a
halo-PT model \citep{haloPT1,haloPT2} that combines the halo model
with perturbation theory. The fitting function
\begin{align}
  f_{2h}(K)=\frac{C}{(1+DK^{-1})^3}.
\end{align}
with this choice of coefficients $C$ and $D$ 
\begin{align}
  C&=140\,D(z)^{-5/4}\\
  D&=1.9\,D(z)^{-3/2}
\end{align}
gives a good fit to simulations. Finally, the three-halo bispectrum 
is simply non-linear tree-level shape predicted for large scales so an exponential fitting 
function is introduced to suppress it at small scales:
\begin{equation}
  f_{3h}(K)=\exp(-K/E).
\end{equation}
An approximate fit for $E$ to simulations is
\begin{align}
  E=7.5\,k_{\text{NL}}(z).
\end{align}

\subsection{Estimating Non-Gaussianity}
\label{estimate}

Generally bispectra can be parameterised by $f_{nl}B^{th}$, where the non-linearity parameter
$f_{nl}$ can be thought of as the amplitude of this particular
bispectrum and $B^{th}$ described the shape. Our goal is to find an optimal estimator for
$f_{nl}$ for a given shape.

It can be shown that the optimal estimator for $f_{nl}$ in the limit of weak non-Gaussianity and under the assumptions of statistical
isotropy and homogeneity
takes the form:
\begin{align}
  &\hat{f}_{nl}=
    \frac{(2\pi)^6}{N_{th}} \int_{\mathbf{k}_1,\mathbf{k}_2,\mathbf{k}_3}
    \frac{\delta_D(\mathbf{k}_1+\mathbf{k}_2+\mathbf{k}_3)
    B^{th}(k_1,k_2,k_3)}{P(k_1)P(k_2)P(k_3)} \nonumber \\
  &\qquad\times\left(\delta_{\mathbf{k}_1}\delta_{\mathbf{k}_2}
    \delta_{\mathbf{k}_3}-3\expval{\delta_{\mathbf{k}_1}
    \delta_{\mathbf{k}_2}}\delta_{\mathbf{k}_3}\right).
    \label{fnl}
\end{align}  
where $\int_{\mathbf{k}_1,\mathbf{k}_2,\mathbf{k}_3}=\int
\frac{d^3k_1}{(2\pi)^3}\frac{d^3k_2}{(2\pi)^3}\frac{d^3k_3}{(2\pi)^3}$.

The purpose of the linear term used above ( $\expval{\delta_{\mathbf{k}_1}
  \delta_{\mathbf{k}_2}}\delta_{\mathbf{k}_3}$ ), analogous to that used in CMB analysis, is that it suppresses mode couplings 
due to anisotropic effects e.g. incomplete survey coverage. Clearly this is not an issue 
for the work on simulations in this paper so we will neglect it, noting that it could be important for observational analysis. To work out the normalisation 
factor $N_{th}$ we impose the condition that $\langle\hat{f}_{nl}\rangle=1$ 
if the theoretical model is indeed the correct underlying bispectrum, i.e. 
if $B^{th}=B^{\text{correct}}_\delta$ where $\expval{\delta(\mathbf{k}_1) 
  \delta(\mathbf{k}_2)\delta(\mathbf{k}_3)}\equiv(2\pi)^3 
\delta_D(\mathbf{k}_1+\mathbf{k}_2+\mathbf{k}_3) 
f_{nl} B^{\text{correct}}_\delta(k_1,k_2,k_3)$. After taking the statistical 
average of $\hat{f}_{nl}$ over different realisations of $\delta$ we get
\begin{align}
  &\expval{\hat{f}_{nl}}= \nonumber \\
  &\quad\frac{1}{N_{th}}\frac{V}{\pi} 
    \int_{\mathcal{V}_B}dV_k\,
    k_1k_2k_3 \frac{B^{th}(k_1,k_2,k_3)B_\delta(k_1,k_2,k_3)}{P(k_1)P(k_2)P(k_3)}, 
    \label{expectation}
\end{align}
where $dV_k\equiv dk_1dk_2dk_3$, and the superscript 
`$\text{correct}$' has been dropped for brevity.
$\mathcal{V}_B$ is the bispectrum domain defined by the 
triangle condition imposed on the wavenumbers $k_i$ such that
$\mathbf{k}_1+\mathbf{k}_2+\mathbf{k}_3=0$, together with a
chosen resolution limit $k_1,k_2,k_3<k_{max}$.
Setting $B^{th}=B_\delta$ and demanding $\expval{\hat{f}_{nl}}=1$
gives the normalisation factor as 
\begin{align}
  N_{th} &= \frac{V}{\pi} \int_{\mathcal{V}_B}dV_k\,
           k_1k_2k_3 \frac{[B(k_1,k_2,k_3)]^2}{P(k_1)P(k_2)P(k_3)}. 
           \label{nth}
\end{align}

The form of \Cref{expectation} suggests we should define inner products
between bispectra as\footnote{We use square brackets $[\,\,]$ for inner
  products to avoid confusion with expectation values, which are labelled
  with angle brackets $\expval{}$.}
\begin{align}
  \left[B_i,B_j\right] \equiv \frac{V}{\pi} \int_{\mathcal{V}_B}dV_k\,
  k_1k_2k_3 \frac{B_i(k_1,k_2,k_3)B_j(k_1,k_2,k_3)}{P(k_1)P(k_2)P(k_3)}.
  \label{inner_product}
\end{align}
This naturally motivates the definition of the signal-to-noise (SN)
weighted bispectrum,
\begin{align}
  B_i^{SN}(k_1,k_2,k_3) \equiv \sqrt{\frac{k_1k_2k_3}{P(k_1)P(k_2)P(k_3)}}
  B_i(k_1,k_2,k_3).
  \label{SN}
\end{align}
This SN-weighted bispectrum is relevant for observations of the
matter bispectrum and is useful for providing forecasts for 
future surveys. 

\begin{figure*}[!htb]
  \begin{subfigure}[b]{0.42\textwidth}
    \includegraphics[width=\textwidth]{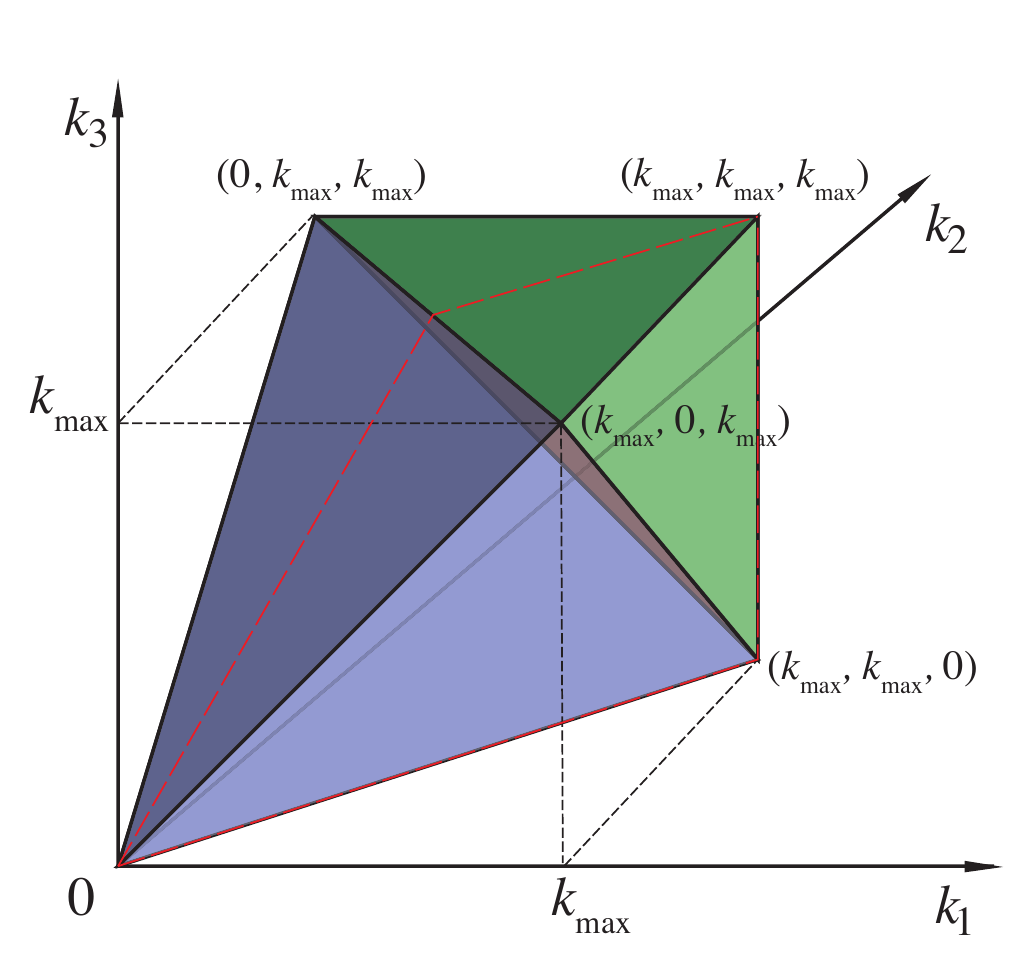} 
    \caption{The full tetrapyd bispectrum domain consists of a tetrahedral region (blue)
      defined by the wavevector triangle condition $\mathbf{k}_1+\mathbf{k}_2+\mathbf{k}_3=0$, 
      together with a pyramidal region (green) bounded by the resolution limit $k_{\text{max}}$. 
      To show the internal structure of the tetrapyd we split it along the red dashed line 
      to obtain \Cref{tetrapyd_split}. \citep{Andrei}}
    \label{tetrapyd}
  \end{subfigure}
  ~
  \begin{subfigure}[b]{0.42\textwidth}
    \includegraphics[width=\textwidth]{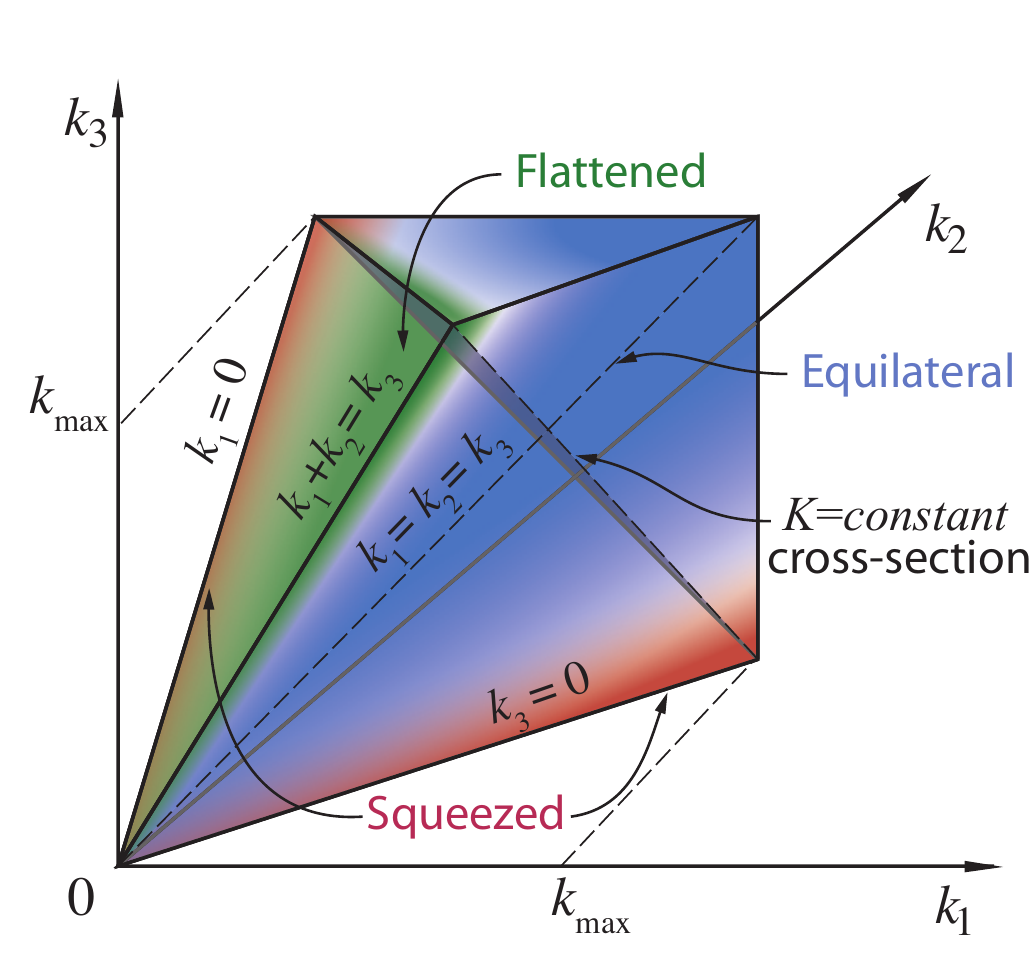}
    \caption{The split 3D tetrapyd region showing only the back half with 
      $k_1<k_2$. Colour-coded regions show the location of the `squeezed' (red), 
      `flattened' (green) and `equilateral' or `constant' (blue) shape signals.
      The scale dependence of the bispectrum is reflected by
      the $K\equiv k_1+k_2+k_3=\text{const.}$ cross sectional planes.
      \citep{Andrei}}
    \label{tetrapyd_split}
  \end{subfigure}
  \caption{Tetrahedral geometry of the allowed bispectrum combination $B(k_1,k_2,k_3)$.}
\end{figure*}

\begin{figure*}
  \begin{subfigure}[b]{0.32\textwidth}
    \includegraphics[width=\linewidth]{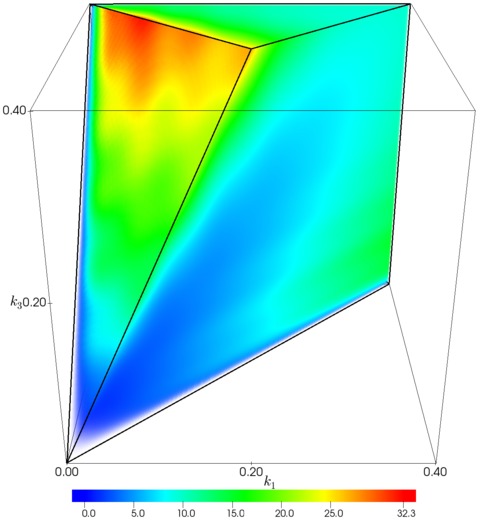}
    \caption{Tree-level shape, $k_{max}=0.4$}
  \end{subfigure}
  ~
  \begin{subfigure}[b]{0.32\textwidth}
    \includegraphics[width=\linewidth]{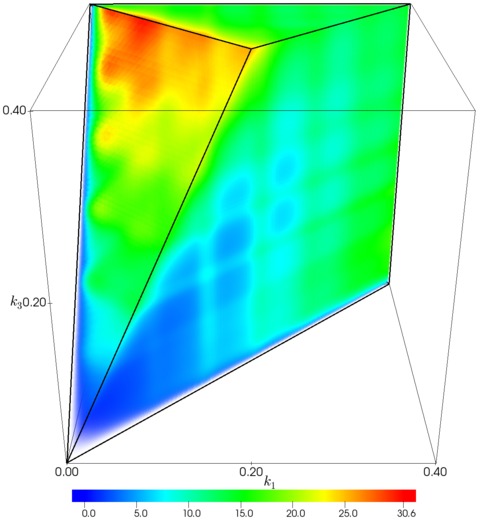}
    \caption{Nine-parameter model, $k_{max}=0.4$}
  \end{subfigure}
  ~
  \begin{subfigure}[b]{0.32\textwidth}
    \includegraphics[width=\linewidth]{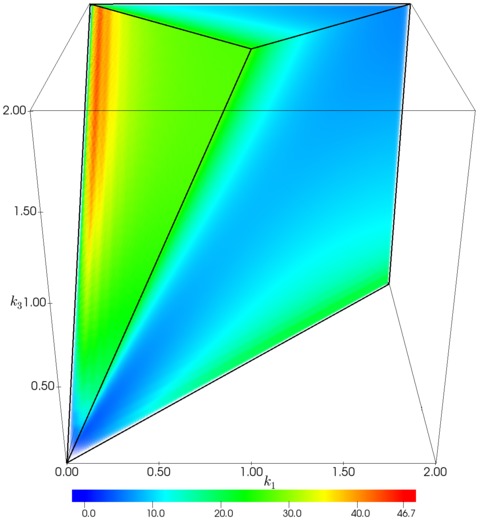}
    \caption{Scaled Tree-level shape, $k_{max}=2$}
  \end{subfigure}
  
  \begin{subfigure}[b]{0.32\textwidth}
    \includegraphics[width=\linewidth]{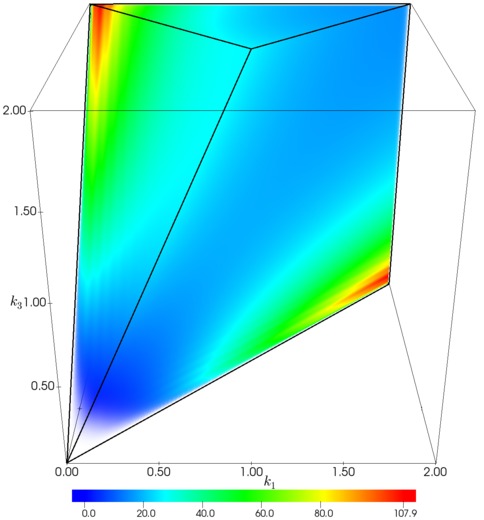}
    \caption{Scaled squeezed shape, $k_{max}=2$}
  \end{subfigure}
  ~
  \begin{subfigure}[b]{0.32\textwidth}
    \includegraphics[width=\linewidth]{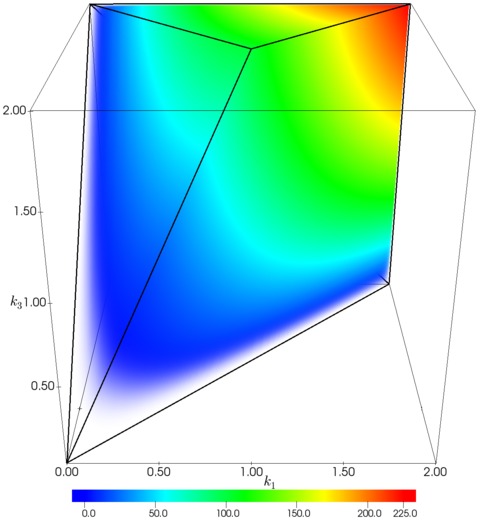}
    \caption{Scaled constant shape, $k_{max}=2$}
  \end{subfigure}
  ~
  \begin{subfigure}[b]{0.32\textwidth}
    \includegraphics[width=\linewidth]{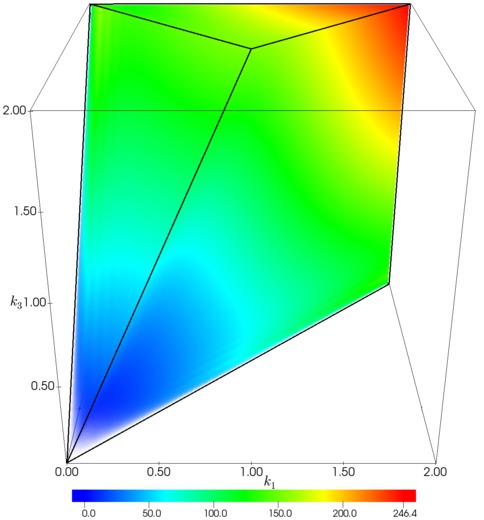}
    \caption{3-shape model, $k_{max}=2$}
  \end{subfigure}
  \caption{The bispectrum shapes introduced in \Cref{sec:shap-non-gauss} plotted 
    at redshift $z=0.5$ up to various $k_{max}$.}
  \label{fig:theory_bis}
\end{figure*}

The bispectrum domain $\mathcal{V}_B$ takes the form of a tetrapyd in
$k$-space as shown in \Cref{tetrapyd}. It is the union of a tetrahedral
region and a triangular pyramid on top. Plotting the full tetrapyd 
obscures it inner structure, and we have found it useful to split it 
in half to make apparent its internal morphology. As illustrated in 
\Cref{tetrapyd_split}, different bispectrum shapes can be distinguished 
through the regions in the tetrapyd where they give the strongest signal.
In \Cref{fig:theory_bis} we show the bispectra shapes introduced in
\Cref{sec:shap-non-gauss}. The bispectra plots are in this paper generated 
with \texttt{ParaView} \citep{paraview}, an open source scientific
visualisation tool.

\subsubsection{Correlators Between Bispectra
  \label{corr}}

Using \Cref{expectation} we can further 
define 4 correlators between bispectra.
The shape correlator, $\mathcal{S}$, is defined by
\begin{align}
  \mathcal{S}(B_i,B_j) \equiv \frac{\left[B_i,B_j\right]}
  {\sqrt{\left[B_i,B_i\right]\left[B_j,B_j\right]}},
  \label{eqn:shape}
\end{align}
and is restricted to $-1\leq\mathcal{S}\leq1$. It can be thought 
of as the cosine between $B_i$ and $B_j$. To quantify how well
the magnitudes of $B_i$ and $B_j$ match each other we define the
amplitude correlator $\mathcal{A}$ as
\begin{align}
  \mathcal{A}(B_i,B_j) \equiv \sqrt{\frac{\big[B_i,B_i\big]}
  {\big[B_j,B_j\big]}}.
\end{align}
We can combine the information given by the shape and amplitude
correlators into a single quantity known as the total
correlator $\mathcal{T}$:
\begin{align}
  &\mathcal{T}(B_i,B_j)
    \equiv 1- \sqrt{\frac{
    \left[B_j-B_i,B_j-B_i\right]}{\left[B_j,B_j\right]}} \nonumber \\
  &\quad= 1-\sqrt{1-2\mathcal{S}(B_i,B_j)\mathcal{A}(B_i,B_j)+
    \mathcal{A}^2(B_i,B_j)}.
    \label{total}
\end{align}
The total correlator is a stringent test of correlation between
bispectra, as both misalignment ($\mathcal{S}<1$) or a 
difference in amplitude ($\mathcal{A}\neq1$) lead to a 
decrease in $\mathcal{T}$. Later on we will use $\mathcal{T}$ 
to test the ability of \MODALLSS{} to reconstruct theoretical
bispectra (see \Cref{sec:modal-methodology}).

We can interpret $\mathcal{T}$ physically as follows.
Let $B_T$ be the true bispectrum and $B_A$ be an approximation to $B_T$. Now suppose we constrain each of these templates with \Cref{fnl} to obtain $f^T_{NL}$ and $f^A_{NL}$.  The variance of each estimate is given by 
\begin{align}
  \sigma^2_i = \expval{{f^i_{NL}}^2} = N_i^{-1} = \left[B_i,B_i\right]^{-1}
\end{align}
and the variance of the difference between the two estimates is given by
\begin{align}
  \sigma^2_{diff} &= \expval{\left(f^T_{NL} - f^A_{NL} \right)^2} \nonumber\\
                  &= \frac{1}{(N_T N_A)^2} \left[N_A B_T - N_T B_A , N_A B_T - N_T B_A  \right] \nonumber\\
                  &= \frac{N_A  - 2 \left[ B_A, B_T\right] +  N_T}{N_A N_T}
\end{align}
If we take the ratio of $\sigma_{diff}$ and $\sigma_A$ then we get
\begin{align}
  \frac{\sigma^2_{diff}}{\sigma^2_A} &=1 - 2 \frac{1}{N_T}\left[ B_A, B_T\right] +  \frac{N_A}{N_T} \nonumber\\
                                     & \left( 1 -  \mathcal{T}(B_T,B_A) \right)^2
\end{align}
This allows us to identify $1-\mathcal{T}$ as the coefficient of variation $c_v$
\citep{everitt}. Therefore if $B_A$ is used as a proxy for
$B_T$, $1-\mathcal{T}$ gives us the standard deviation between our estimate of $f_{NL}$ and the true value as a fraction of our error bar, ie:
\begin{align}
  \sigma_{diff} = \left( 1 -  \mathcal{T} \right) \sigma
\end{align}

$\mathcal{T}$ is appropriate for comparing theoretical bispectra,
but its performance is easily degraded by cosmic variance and hence 
another correlator is needed when simulation/observational data is 
involved. The $f_{nl}$ correlator, named as such due to its similarity 
to the $\langle\hat{f}_{nl}\rangle$ parameter in \Cref{expectation} 
above, again combines the shape and amplitude correlators:
\begin{align}
  f_{nl}(B_i,B_j)
  &\equiv \frac{\big[B_i,B_j\big]}{\big[B_j,B_j\big]} \nonumber \\
  &= \mathcal{S}(B_i,B_j)\mathcal{A}(B_i,B_j).
    \label{fnl_corr}
\end{align}
This can be interpreted as simply correlation between our estimate of $f_{NL}$ with the true value, normalised by the true value.
\begin{align}
  \frac{\expval{ f^T_{NL} f^A_{NL}}}{\expval{ {f^T_{NL}}^2 }} &= \frac{1}{N_A} \left[B_T,B_A\right] \nonumber\\
                                                              &=  f_{nl}(B_T,B_A)
\end{align}

\subsection{\MODALLSS{} Methodology}
\label{sec:modal-methodology}

For general bispectra the 9-dimensional integral in
the $\hat{f}_{NL}$ estimator (\Cref{fnl}) is computationally intractable.
This computation barrier has been solved by a separable method introduced
in \citep{MODAL}. This \texttt{MODALl} method
has been applied to Planck CMB analysis with great success \citep{Planck}. This approach was adapted analyse the bispectrum of the large scale structure of the universe in \citep{MODAL-LSS}, which
iwas aptly named \MODALLSS{}. Here we outline the \MODALLSS{} methodology.

\subsubsection{\MODALLSS{} Basis}

\begin{figure*}
  \begin{align}
    \vcenter{\hbox{\includegraphics[width=0.2\linewidth]{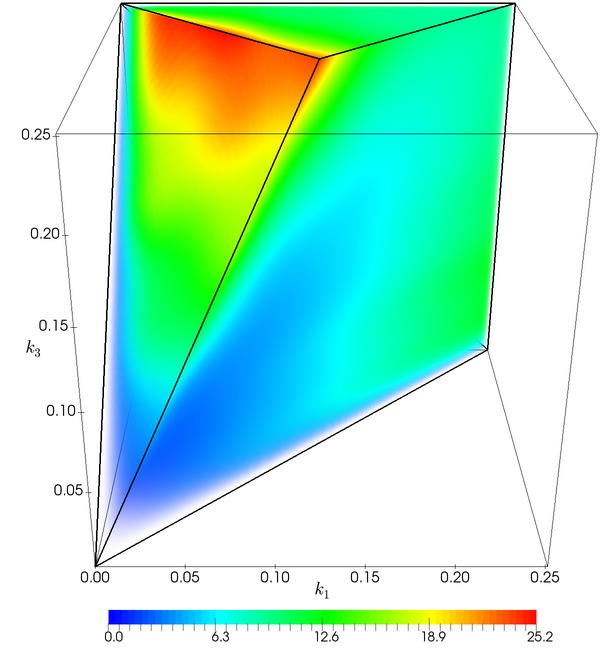}}}
    =\,\alpha_1
    \vcenter{\hbox{\includegraphics[width=0.2\linewidth]{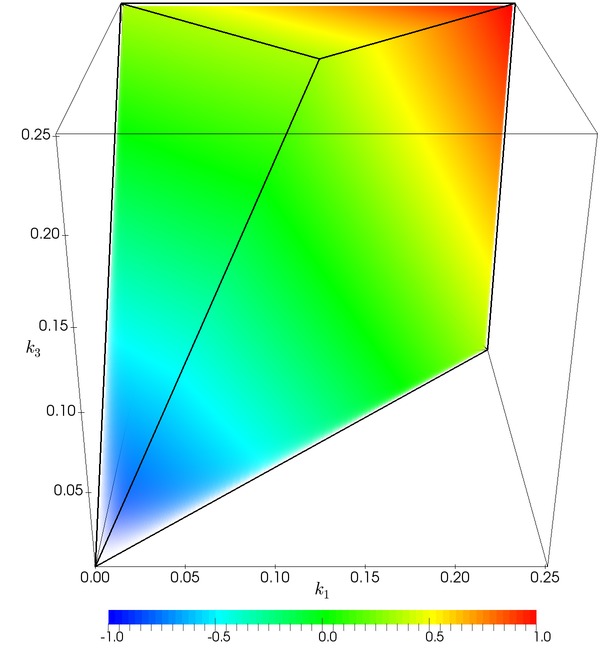}}}
    +\,\alpha_2
    \vcenter{\hbox{\includegraphics[width=0.2\linewidth]{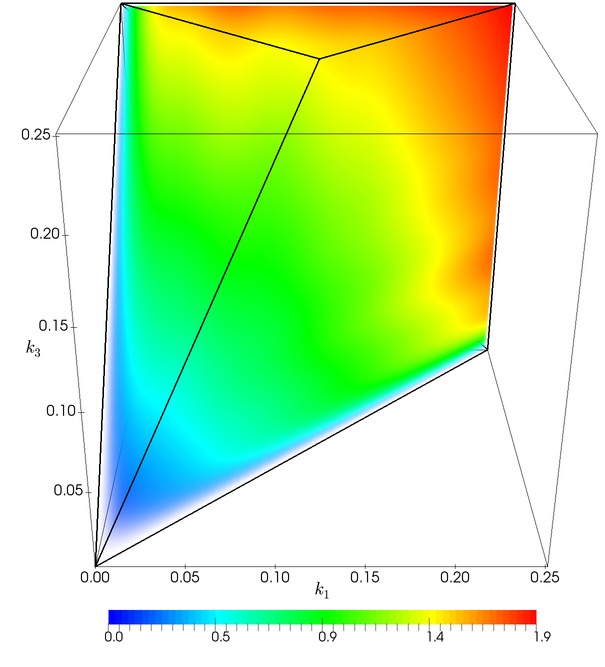}}}
    +\,\alpha_3
    \vcenter{\hbox{\includegraphics[width=0.2\linewidth]{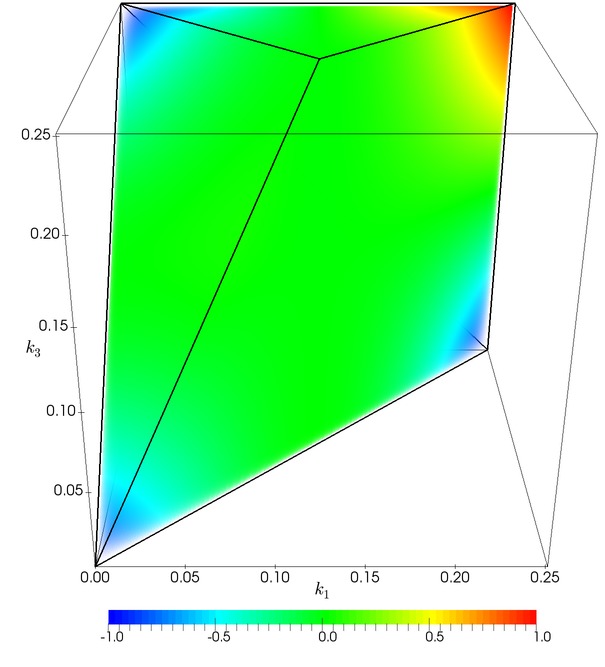}}}
    +\cdots
  \end{align}
  \caption{A cartoon demonstrating the \MODALLSS{} expansion.
    Here we are expanding the tree level bispectrum (\Cref{eq:tree_shape})
    as a linear combination of the $Q^{\MODALLSS{}}_n$ basis functions
    (represented by the tetrapyds), each of which is weighted by
    an $\alpha_n$ coefficient.} 
  \label{fig:modal_expansion_cartoon}
\end{figure*}

We first approximate the SN-weighted theoretical bispectrum in \Cref{SN} 
by expanding it in a general seperable basis (see also
\Cref{fig:modal_expansion_cartoon}):
\begin{align}
  &\sqrt{\frac{k_1k_2k_3}{P(k_1)P(k_2)P(k_3)}}B^{th}(k_1,k_2,k_3) \nonumber \\
  &\quad\approx\sum_n^{n_{max}} \alpha^Q_n
    Q^{\MODALLSS{}}_n(k_1/k_{max},k_2/k_{max},k_3/k_{max}).
    \label{expand}
\end{align}
The basis functions ${Q^{\MODALLSS{}}_n}$ are symmetrised products over 
one dimensional functions $q_r$:
\begin{align}
  Q^{\MODALLSS{}}_n (x,y,z) \equiv q_{\{r}(x)q_{s}(y)q_{t\}}(z),
\end{align}
with $\{\dots\}$ representing symmetrisation over the indices $r,s,t$,
and each $n$ corresponds to a combination of $r,s,t$. $k_{max}$ is the
resolution of the tetrahedral domain defined above. The choice of $q_r$
is arbitrary and there are many sensible choices including $k$-bins (which are
localised in $k$-space), wavelets (which are localised in real space), 
Fourier modes, etc. We adopt polynomials since they offer efficient
compression of the data so fewer modes can be used without information 
loss. Note that the ${Q^{\MODALLSS{}}_n}$ form a complete basis for the 
expansion of $B^{th}$, but naturally we truncate the expansion at some $n_{max}$ 
depending on the accuracy required. For convenience in our discussion below we will 
assume that the truncation causes errors are tiny and assume that 
\Cref{expand} is exact.

It has been shown that the convergence of the sum in \Cref{expand} is independent of the choice of polynomials $q_r$. Different choices of polynomials only change the individual $\alpha^Q_n$ but not the sum. As such we choose our polynomials in order to ensure
numerical stability of the method on the tetrahedral domain $\mathcal{V}_B$. Currently we find shifted Legendre polynomials
$\tilde{P}_l(x)=P_l(2x-1)$ perform well and are adopted for $q_r$
as they demonstrate better orthogonality at low $n$ and
encapsulate the behaviour of the bispectrum at non-linear scales 
very well. Calculation of higher order polynomials also 
demonstrates good numerical stability when calculated recursively.

Another issue is the mapping between $n$ and $r,s,t$. The ordering
of this mapping is arbitrary, here we have adopted `slice ordering'
which orders the triples by the sum $r+s+t$. A sub-ordering is
introduced along each column in cases of degeneracy, i.e.
\begin{align}\label{eq:slicingorder}
  &\underline{0  \rightarrow 000}   \quad  4  \rightarrow 111  
    \quad~\,   8  \rightarrow 022\quad   12  \rightarrow 113 \nonumber \\
  & \underline{1\rightarrow 001}   \quad   5  \rightarrow 012 
    \quad~\,   9  \rightarrow 013  \quad  13  \rightarrow 023 \nonumber \\
  & 2 \rightarrow 011  \quad \underline{6  \rightarrow 003}\quad \underline{10  
    \rightarrow 004}  \quad  14  \rightarrow 014 \\   
  &\underline{ 3 \rightarrow 002}  \quad  7  \rightarrow 112 \quad   
    11  \rightarrow 122 \quad   \underline{15  \rightarrow 005} ~\cdots\,, \nonumber
\end{align}
where the lines mark the end of each overall polynomial order. 

Using the \MODALLSS{} expansion in \Cref{expand} we can rewrite
$\hat{f}_{nl}$ in \Cref{fnl} as:

\begin{align}
  \hat{f}_{nl}
  & =\frac{(2\pi)^6}{N_{th}} \int_{\mathbf{k}_1,\mathbf{k}_2,\mathbf{k}_3}
    \delta_D(\mathbf{k}_1+\mathbf{k}_2+\mathbf{k}_3)
    \nonumber \\
  & \qquad\qquad\times\frac{\sum_n \alpha^Q_n
    q_{\{r}(\frac{k_1}{k_{max}})q_s(\frac{k_2}{k_{max}})
    q_{t\}}(\frac{k_3}{k_{max}})}{\sqrt{k_1P(k_1)k_2P(k_2)k_3P(k_3)}}
    \nonumber \\
  & \qquad\qquad\times\left(\delta_{\mathbf{k}_1}\delta_{\mathbf{k}_2}\delta_{\mathbf{k}_3}-
    \langle\delta_{\mathbf{k}_1}\delta_{\mathbf{k}_2} 
    \rangle\delta_{\mathbf{k}_3}\right) \nonumber \\
  & =\frac{(2\pi)^3}{N_{th}} \sum_n \alpha^Q_n \int d^3 x
    \int\frac{\prod_i d^3k_i}{(2\pi)^9}
    e^{i(\mathbf{k}_1+\mathbf{k}_2+\mathbf{k}_3)\cdot\mathbf{x}}
    \nonumber \\
  & \qquad\qquad\times\frac{q_{\{r}(
    \frac{k_1}{k_{max}})q_s(\frac{k_2}{k_{max}})q_{t\}}(
    \frac{k_3}{k_{max}})}{\sqrt{k_1P(k_1)k_2P(k_2)k_3P(k_3)}}
    \nonumber \\
  & \qquad\qquad\times\left(\delta_{\mathbf{k}_1}\delta_{\mathbf{k}_2}\delta_{\mathbf{k}_3}-
    \langle\delta_{\mathbf{k}_1}\delta_{\mathbf{k}_2} 
    \rangle\delta_{\mathbf{k}_3}\right)
    \nonumber \\          
  & =\frac{(2\pi)^3}{N_{th}} \sum_n \alpha^Q_n \int d^3 x
    \bigg[M_r(\mathbf{x})M_s(\mathbf{x})M_t(\mathbf{x})
    \nonumber \\
  & \qquad\qquad-\langle M_{\{r}(\mathbf{x})M_s(\mathbf{x})
    \rangle M_{t\}}(\mathbf{x})\bigg],
    \label{eqn:fnl_sub}
\end{align}
where in the second line we have used the integral from of the delta function with variable
$\mathbf{x}$, and we defined 
\begin{align}
  M_r(\mathbf{x}) \equiv \int\frac{d^3k}{(2\pi)^3}\frac{\delta_{\mathbf{k}}q_r(k/k_{max})}
  {\sqrt{kP(k)}}e^{i\mathbf{k}\cdot\mathbf{x}},
  \label{Mfunc}
\end{align}
which is an inverse Fourier transform\footnote{Here the choice
  of the polynomials $q_r$ becomes important. For example, the integral in
  \Cref{Mfunc} convergences poorly for large $r$ if we choose monomials 
  $q_r=x^r$.}. Note that there is no
symmetrisation over $r,s,t$ in the first term inside the square brackets 
as the product is already symmetric. As we are only analysing simulation
data which approximately homogeneous and isotropic we can ignore the second term in the square brackets as it evaluates to zero. We then introduce
\begin{align}
  \beta^Q_n = (2\pi)^3\int d^3x \,M_r(\mathbf{x})M_s(\mathbf{x})M_t(\mathbf{x})
  \label{betaQ}
\end{align}
which allows us to express $\hat{f}_{nl}$ in a simple and elegant form:
\begin{align}
  \hat{f}_{nl} = \frac{1}{N_{th}}\sum_n \alpha^Q_n \beta^Q_n.
  \label{simple}
\end{align}
The beta coefficients $\beta^Q_n$ are approximately analogous (there is a subtly we will meet in the next section) to the alpha coefficients
$\alpha^Q_n$ but they are used in the expansion of observational/simulation
bispectra instead of theoretical ones.

In summary, we have reduced the complicated integral in \Cref{fnl} to a
the calculation of $\alpha^Q_n$ and $\beta^Q_n$ coefficients. The computation
of $\alpha^Q_n$ coefficients is a non-trivial problem but has been made
efficient by the authors of \citep{modal_HPC} whose implementation which we use here. The $\beta^Q_n$ coefficients
on the other hand only require a number of (inverse) Fourier transforms
(evident upon inspection of \Cref{Mfunc}) which can be evaluated efficiently
with the fast Fourier transform (FFT) algorithm\footnote{We use the 
  \texttt{FFTW3} \citep{FFTW05} implementation of the algorithm.},
together with an integral over the spatial extent of the data set (\Cref{betaQ})
which can highly parallelised with Open Multi-Processing (\texttt{OpenMP}).

\subsubsection{An orthogonal basis
  \label{sec:ortho_basis}}

Unlike the theoretical bispectrum the observational/simulation bispectrum is
a statistical quantity, and and it can only be estimated through different 
realisations of the density field $\delta$. We expand the estimated 
observational bispectrum $\hat{B}_\delta$ in the following way:
\begin{align}
  &\sqrt{\frac{k_1k_2k_3}{P(k_1)P(k_2)P(k_3)}}\hat{B}_\delta(k_1,k_2,k_3) 
    \nonumber \\
  &\quad=\sum_n^{n_{max}} \tilde{\beta}^Q_n 
    Q_n(k_1/k_{max},k_2/k_{max},k_3/k_{max}),
    \label{eqn:modal_est}
\end{align}
the expectation value of which is the true underlying observational bispectrum
$B_\delta\equiv\langle\hat{B}_\delta\rangle$:
\begin{align}
  &\sqrt{\frac{k_1k_2k_3}{P(k_1)P(k_2)P(k_3)}}B_\delta(k_1,k_2,k_3) \nonumber \\
  &\quad=\sum_n^{n_{max}} \langle\tilde{\beta}^Q_n\rangle 
    Q_n(k_1/k_{max},k_2/k_{max},k_3/k_{max}).
    \label{tilde}
\end{align}
We have introduced these new beta coefficients \footnote{We could have
  instead to reversed the placement of the tilde to make $\alpha^Q_n$
  and $\beta^Q_n$ more analogous, but we have adopted this notation as
  it more closely represents the computational flow of the method.}
$\tilde{\beta}^Q_n$. To relate 
$\tilde{\beta}^Q_n$ to $\beta^Q_n$ 
we substitute \Cref{tilde} into \Cref{expectation}:
\begin{align}
  \langle\hat{f}_{nl}\rangle
  &= \frac{1}{N_{th}}\frac{V}{\pi} 
    \int_{\mathcal{V}_B}dV_k\,
    \sum_{nm}\alpha_n^Q \langle\tilde{\beta}^Q_n\rangle
    Q_nQ_m \nonumber \\
  &= \frac{1}{N_{th}}\sum_{nm}\alpha_n^Q \langle\tilde{\beta}^Q_n\rangle
    \gamma_{nm},
    \label{fnl_Q}
\end{align}
where\footnote{Note
  that when a large number of modes are used, this integral
  evaluated with a regular grid on the tetrapyd domain and with FFTs differs
  greatly, especially in the limit of a low number of grid points.
  We conclude that discrete sampling has a different effect on
  direct integration compared to when FFTs are used, and to ensure
  internal consistency of the $\alpha$ and $\beta$ coefficients
  we evaluate $\gamma_{nm}$ separately by integration on the
  tetrapyd for $\alpha_n^Q$ and via FFTs for $\beta_n^Q$ to rotate
  them into the $\{R_n\}$ basis. For large grids
  $N_g>\mathcal{O}(1024)$ the memory requirements of computing
  $\gamma_{nm}$ with FFTs are too great, but we have verified that
  for such grids the two methods give consistent results and hence
  direct integration is used instead. See \Cref{appendix_C} for more
  details.
}
\begin{align}
  \gamma_{nm}\equiv\frac{V}{\pi}\int_{\mathcal{V}_B}dV_kQ_nQ_m
  \label{gamma}
\end{align} 
is the inner product between the $Q_n$ functions on the tetrapyd domain.
Generally speaking $\gamma_{nm}$ is not diagonal since the $Q_n$ functions
are not orthogonal to each other. Comparing this with the expectation
value of \Cref{simple} we obtain
\begin{align}
  \langle\beta^Q_n\rangle = \sum_m \gamma_{nm}\langle\tilde{\beta}^Q_m\rangle
  \Rightarrow \beta^Q_n = \sum_m \gamma_{nm}\tilde{\beta}^Q_m.
  \label{eqn:beta_basis}
\end{align}

While $\beta^Q_n$ may be straightforward to evaluate numerically through
\Cref{betaQ}, it often proves simpler to use an orthonormalised version we create
by diagonalising $\gamma_{nm}$. We therefore introduce a basis $\{R_n\}$ which is defined
relative to $\{Q_n\}$ by
\begin{align}
  R_n\equiv\lambda_{nm}Q_m \Leftrightarrow Q_p\equiv(\lambda^{-1})_{pq}R_q
  \label{eqn:RQconv}
\end{align}
such that it is orthonormal on the tetrapyd domain:
\begin{align}
  \frac{V}{\pi}\int_{\mathcal{V}_B}dV_kR_nR_m=\delta_{nm}.
  \label{orthonormal}
\end{align} 
From \Cref{gamma,orthonormal} we deduce that $\gamma=\lambda^{-1}(\lambda^{-1})^T$.  Choosing $R_n$ to have the same polynomial order as $Q_n$ forces this $\lambda$ to be the Cholesky decomposition.  This is equivalent to a performing a modified Gram-Schmitt orthonormalisation of the $Q_n$ directly.  We now apply the expansion in the
$\{R_n\}$ basis:
\begin{align}
  &\sqrt{\frac{k_1k_2k_3}{P(k_1)P(k_2)P(k_3)}}B^{th}(k_1,k_2,k_3) \nonumber \\
  &\quad=\sum_n^{n_{max}} \alpha^R_n R_n(k_1/k_{max},k_2/k_{max},k_3/k_{max}), \\
  &\sqrt{\frac{k_1k_2k_3}{P(k_1)P(k_2)P(k_3)}}B_\delta(k_1,k_2,k_3) \nonumber \\
  &\quad=\sum_n^{n_{max}} \langle\beta^R_n\rangle R_n(k_1/k_{max},k_2/k_{max},k_3/k_{max}).
\end{align}
Note that due to the orthonormality of the $R_n$ functions we do not
need two sets of $\beta$ coefficients in this basis. Since
$\sum_n\alpha^Q_nQ_n=\sum_n\alpha^R_nR_n$, one can derive the following
relationships between the coefficients in the $\{Q_n\}$ and $\{R_n\}$ 
bases:
\begin{align}
  \alpha^R_n=\sum_m(\lambda^{-1})^T_{nm}\alpha^Q_m,\quad
  \beta^R_n=\sum_m(\lambda^{-1})^T_{nm}\tilde{\beta}^Q_m,
\end{align}
which allows us to write 
\begin{align}
  \langle\hat{f}_{nl}\rangle = \frac{1}{N_{th}}\sum_n \alpha^R_n 
  \langle\beta^R_n\rangle.
\end{align}
One can very easily show this is consistent with \Cref{fnl_Q} above.
Using the \MODALLSS{} ansatz with \Cref{nth} above we find that
$N_{th}=\sum_n \alpha^R_n\alpha^R_n$. Therefore if the theoretical and
data bispectrum match perfectly, i.e. $B^{th}=B_\delta$ and hence
$\langle\hat{f}_{nl}\rangle = 1$, we deduce that
$\langle\beta^R_n\rangle = \alpha^R_n$.

\subsubsection{Numerical implementation\label{subsec:num}}

An implementation of the \MODALLSS{} method has already produced 
some good results \citep{MODAL-LSS}. The code has since been completely 
overhauled and parallelised with \texttt{OpenMP} and multi-threaded 
\texttt{FFTW} for a dramatic reduction in run time, allowing us to 
estimate the bispectra of much larger simulations and also using 
more modes. We are now able to estimate the bispectrum of 
$2048^3$ density grids with $n_{max}=\mathcal{O}(1000)$ modes in 
$\sim35$ minutes using 512 CPU-cores, a significant improvement in 
run time and resolution over the analysis of $512^3$ grids with 
$n_{max}=\mathcal{O}(50)$ in \citep{MODAL-LSS}. We would like to 
emphasise that the computational costs for bispectrum estimation with 
\MODALLSS{} scales with the size of the density grid and is a tiny 
fraction of the costs of N-body runs, and thus can be included in 
existing pipelines with little additional cost.

Another innovation to improve the performance of \MODALLSS{} is the
introduction of custom modes based on the separable bispectrum shapes
given in \Cref{sec:shap-non-gauss}. Explicitly we split the
SN-weighted versions of tree-level bispectrum (\Cref{eq:tree_shape})
and late-time local bispectrum (\Cref{eq:local_late_shape}) as follows
(Note that $P(k)$ represents the non-linear power spectrum of choice):
\begin{itemize}
\item The tree-level bispectrum requires 6 custom polynomials:
  \begin{itemize}
  \item $q^{\text{tree}}_0(k)=\sqrt{\frac{k}{P(k)}}\frac{5}{14}$
  \item $q^{\text{tree}}_1(k)=\sqrt{\frac{k}{P(k)}}P(k)$
  \item $q^{\text{tree}}_2(k)=-\sqrt{\frac{k}{P(k)}}P(k)k^2$
  \item $q^{\text{tree}}_3(k)=\sqrt{\frac{k}{P(k)}}\frac{P(k)}{k^2}$
  \item $q^{\text{tree}}_4(k)=\sqrt{\frac{k}{P(k)}}\frac{3}{14}k^2$
  \item $q^{\text{tree}}_5(k)=\sqrt{\frac{k}{P(k)}}\frac{1}{14}k^4$
  \end{itemize}
  which are combined into these 4 modes:
  \begin{itemize}
  \item $Q^{\text{tree}}_0 = q_{\{1}(x)q_{1}(y)q_{0\}}(z)$
  \item $Q^{\text{tree}}_1 = q_{\{2}(x)q_{3}(y)q_{0\}}(z)$
  \item $Q^{\text{tree}}_2 = q_{\{1}(x)q_{3}(y)q_{4\}}(z)$
  \item $Q^{\text{tree}}_3 = q_{\{3}(x)q_{3}(y)q_{5\}}(z)$ 
  \end{itemize}
\item The late-time local bispectrum requires 2 custom polynomials:
  \begin{itemize}
  \item $q^{\text{local,late}}_0(k)=\sqrt{\frac{k}{P(k)}}
    \sqrt{P_{\text{lin}}(k)}k^{n_s/2-2}$
  \item $q^{\text{local,late}}_1(k)=\sqrt{\frac{k}{P(k)}}
    \sqrt{P_{\text{lin}}(k)}k^{2-n_s/2}$
  \end{itemize}
  resulting in a single mode:
  \begin{itemize}
  \item $Q^{\text{local,late}}_0 = q_{\{0}(x)q_{0}(y)q_{1\}}(z)$
  \end{itemize}  
\end{itemize}
These custom modes help pick up general features in the matter
bispectra, which combined with the $Q_n$  functions ensures an
effective reconstruction of any dark matter bispectrum signal.

\begin{figure*}
  \begin{subfigure}[b]{0.32\textwidth}
    \includegraphics[width=\linewidth]{{theo_bis_6_1000_4_682_6400_55_z0p500_0.66857_408_bispectrum_tetrapyd.h5_cropped}.jpeg}
  \end{subfigure}
  ~
  \begin{subfigure}[b]{0.32\textwidth}
    \includegraphics[width=\linewidth]{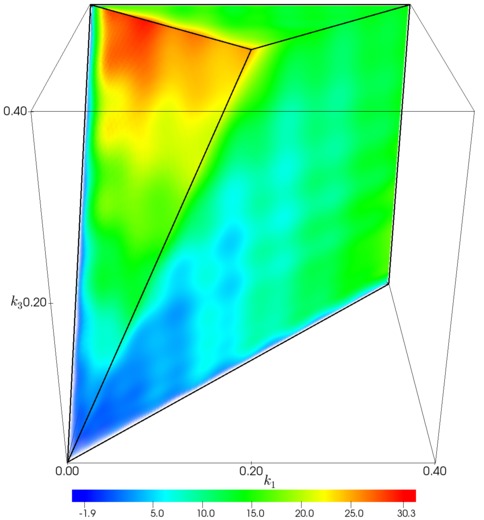}
  \end{subfigure}
  ~
  \begin{subfigure}[b]{0.32\textwidth} 
    \includegraphics[width=\linewidth]{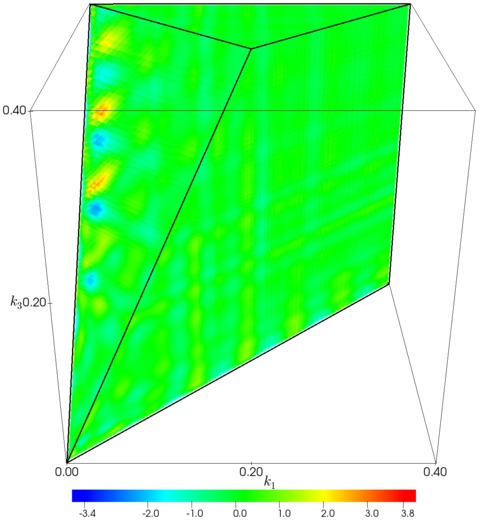}
  \end{subfigure}
  \caption{The nine-parameter up to $k_{max}=0.4\,h\,\text{Mpc}^{-1}$ by direct
    calculation (left), its reconstruction by \MODALLSS{} with 1000 modes
    (middle) and the residuals between them (right). Note the change of scale
    in the colour bars.
  }
  \label{fig:modal_theory_6400}
\end{figure*}

\begin{figure*}
  \begin{subfigure}[b]{0.32\textwidth}      \includegraphics[width=\linewidth]{{theo_bis_6_1000_4_682_1280_52_z0p500_3.34285_408_bispectrum_tetrapyd.h5_cropped}.jpeg}
  \end{subfigure}
  ~
  \begin{subfigure}[b]{0.32\textwidth}
    \includegraphics[width=\linewidth]{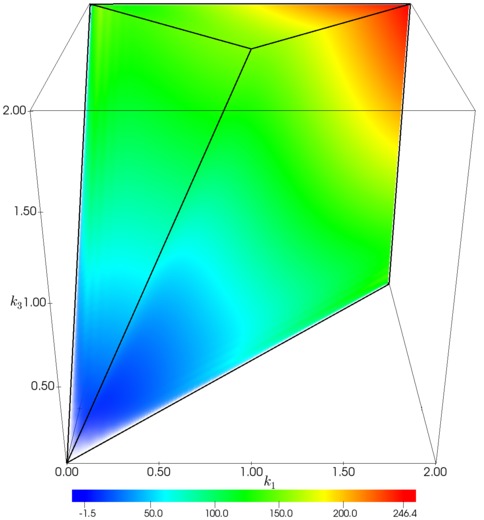}
  \end{subfigure}
  ~
  \begin{subfigure}[b]{0.32\textwidth} 
    \includegraphics[width=\linewidth]{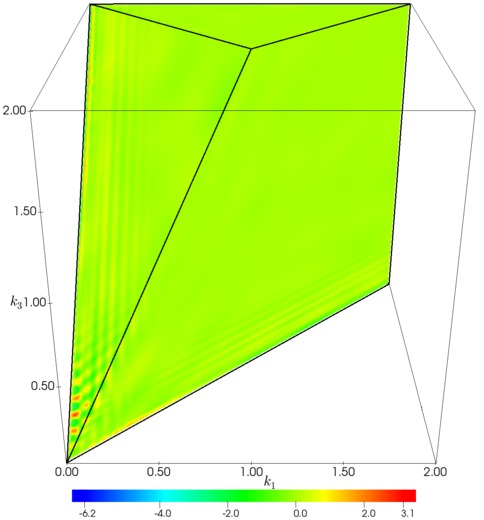}
  \end{subfigure}
  \caption{The 3-shape model up to $k_{max}=2.0\,h\,\text{Mpc}^{-1}$ by direct
    calculation (left), its reconstruction by \MODALLSS{} with 1000 modes
    (middle) and the residuals between them (right). Note the change of scale
    in the colour bars.
  }
  \label{fig:modal_theory_1280}
\end{figure*}

\begin{table*}
  \begin{tabular}{c|c|c|c|c|c|c}
    \multirow{2}{*}[-1ex]{Bispectrum shape}
    & \multirow{2}{*}[-1ex]{$n_{max}$} & \multicolumn{2}{c|}{$k_{max}=0.4\,h\,\text{Mpc}^{-1}$} & \multicolumn{2}{c|}{$k_{max}=2.0\,h\,\text{Mpc}^{-1}$} & Computational cost \\[1ex] \cline{3-6}
    \rule{0pt}{3ex} & & $1-\mathcal{S}_{\alpha,th}$ & $1-\mathcal{T}_{\alpha,th}$ & $1-\mathcal{S}_{\alpha,th}$ & $1-\mathcal{T}_{\alpha,th}$ & (CPU-minutes) \\[1ex] \hhline{=|=|=|=|=|=|=}
    \rule{0pt}{3ex}
    \multirow{4}{*}{Tree-level bispectrum}
    & $50^*$ & \num{6.7e-4} & \num{3.6e-2} & \num{1.3e-3} & \num{5.1e-2} & 160 \\
    & 10 & 0 & 0 & 0 & 0 & 90 \\
    & 50 & 0 & 0 & 0 & 0 & 160 \\
    & 200 & 0 & 0 & 0 & 0 & 370 \\
    & 1000 & 0 & 0 & 0 & 0 & 1600 \\[1ex] \hhline{-|-|-|-|-|-|-}
    \rule{0pt}{3ex} \multirow{4}{*}{Nine-parameter model}
    & $50^*$ & \num{6.6e-4} & \num{3.6e-2} & - & - & 450 \\
    & 10 & \num{3.3e-4} & \num{2.6e-2} & - & - & 390 \\
    & 50 & \num{2.2e-4} & \num{2.1e-2} & - & - & 450 \\
    & 200 & \num{7.9e-5} & \num{1.3e-2} & - & - & 660 \\
    & 1000 & \num{2.2e-5} & \num{6.7e-3} & - & - & 1870 \\[1ex] \hhline{-|-|-|-|-|-|-}
    \rule{0pt}{3ex} \multirow{4}{*}{3-shape model}
    & $50^*$ & \num{3.5e-4} & \num{2.6e-2} & \num{5.6e-5} & \num{1.1e-2} & 190 \\
    & 10 & \num{5.8e-4} & \num{3.4e-2} & \num{3.8e-4} & \num{2.8e-2} & 120 \\
    & 50 & \num{1.1e-4} & \num{1.5e-2} & \num{6.0e-5} & \num{1.1e-2} & 190 \\
    & 200 & \num{1.6e-5} & \num{5.7e-3} & \num{1.2e-5} & \num{4.9e-3} & 400 \\
    & 1000 & 0 & 0 & 0 & 0 & 1610 \\[1ex]
  \end{tabular}
  \caption{
    The performance of \MODALLSS{} at reconstructing different
    theoretical bispectrum shapes at different $k_{max}$ while varying
    the number of modes used in the reconstruction. $50^*$ indicates 
    only shifted Legendre polynomials and no custom modes were used, 
    highlighting the strength of the custom modes in capturing desired 
    bispectrum signals. We use the shape $\mathcal{S}_{\alpha,th}$ and 
    total correlator $\mathcal{T}_{\alpha,th}$ introduced in 
    \Cref{total_alpha} to assess the accuracy of the reconstructed 
    bispectra. It is clear that the total correlator is a much more 
    stringent test than the shape correlator. With 1000 modes we obtain 
    $\mathcal{T}_{\alpha,th}>0.99$ in all cases, giving us high confidence 
    in the validity of the \MODALLSS{} expansion. Note that we omit the 
    nine-parameter model at $k_{max}=2.0\,h\,\text{Mpc}^{-1}$ since it is 
    ill-defined at such non-linear scales. We give the computational 
    cost of the method by the CPU-minutes required to reconstruct the 
    theoretical bispectra on a $2048^3$ grid in pure \texttt{OpenMP} 
    mode. It demonstrates better than linear scaling with $n_{max}$ 
    which shows the highly optimised nature of the code. The performance 
    also scales with $N_{grid}^3$, where $N_{grid}$ is the number of grid 
    points, and will therefore run much faster for analyses that do not 
    require such high resolution.
  }
  \label{fig:total_alphas}
\end{table*}

We conclude this section by assessing the accuracy of the \MODALLSS{}
expansion. This is only possible with theoretical bispectra where we
know the true answer since statistical noise will always be present
in simulations\footnote{We have however made comprehensive tests of
  the \MODALLSS{} algorithm for estimating bispectrum of density
  fields, detailed in \Cref{appendix_C}.}. A qualitative comparison
is illustrated in \Cref{fig:modal_theory_6400,fig:modal_theory_1280}
where we plot the theoretical and reconstructed bispectra as well as
the residuals between them different $k_{max}$. Quantitatively we
evaluate both the shape and total correlator between a theoretical
bispectrum $B^{th}$ and its \MODALLSS{} counterpart
$\sum_n\alpha^R_nB^R_n$, where 
\begin{align}
  &B^R_n(k_1,k_2,k_3) = \nonumber \\
  &\quad\sqrt{\frac{P(k_1)P(k_2)P(k_3)}
    {k_1k_2k_3}}R_n(k_1/k_{max},k_2/k_{max},k_3/k_{max}).
\end{align}
Using \Cref{eqn:shape,total} we find that 
\begin{align}
  \mathcal{S}_{\alpha,th}&\equiv\mathcal{S}(\sum_n\alpha^R_nB^R_n,B^{th})
                           =\sqrt{\frac{\sum_n(\alpha^R_n)^2}{\left[B^{th},B^{th}\right]}},
                           \nonumber \\
  \mathcal{T}_{\alpha,th}&\equiv\mathcal{T}(\sum_n\alpha^R_nB^R_n,B^{th})
                           =1-\sqrt{1-\frac{\sum_n(\alpha^R_n)^2}{\left[B^{th},B^{th}\right]}},
                           \label{total_alpha}
\end{align}
where we have used the orthonormality of the ${R_n}$ basis functions
to obtain\footnote{Note that in principle
  $B^{th}=\sum^\infty_n\alpha^R_nB^R_n$.} $\left[\sum_n\alpha^R_nB^R_n
  ,B^{th}\right]=\sum_n(\alpha^R_n)^2$.

We tested \MODALLSS{} with a range of bispectrum shapes, including
the tree-level bispectrum (\Cref{treeNL}), nine-parameter model
(\Cref{9param}) and the 3-shape model (\Cref{eqn:3-shape}), at
different $k_{max}$ and number of modes up to $n_{max}=1000$
(\Cref{fig:total_alphas}). \MODALLSS{} is able to reconstruct all
bispectrum shapes with $\mathcal{T}_{\alpha,th}>99\%$ at different
$k$-ranges, and improvements can certainly be made by using more
modes. This result justifies our decision to take the approximation
in \Cref{expand} to be exact. This also gives us confidence that
\MODALLSS{} can very accurately estimate simulation and observational
bispectra. The computational cost of \MODALLSS{} is estimated by the
CPU-minutes used when reconstructing the various bispectrum. The code for
reconstructing theoretical bispectra is parallelised with hybrid
\texttt{MPI-OpenMP} but the tests here were ran with pure
\texttt{OpenMP} and 1 thread per CPU core. Note that this may not
be the optimal number of threads and further reductions in run time
may be possible.

\subsection{Sources of error in bispectrum estimation}
\label{sec:error}

In order to make meaningful comparisons between 
simulation/observational data with theoretical predictions
one must have a thorough understanding of the errors that 
occur in our measurements. Since the main focus of this paper
is on simulations we will not discuss observational effects 
such as survey geometry and redshift-space distortions (RSD).
The main contributions we consider here are Poisson shot noise,
covariance of the \MODALLSS{} estimator, and aliasing due to the 
use of FFTs, all of which are relevant for the analysis of 
observational data in the future.

\subsubsection{Shot noise contribution to the power 
  spectrum and bispectrum}

Since dark matter halos and galaxies are discrete tracers of their
respective density fields, measurements of their statistics are
biased relative to the true values that are of interest to us. This
is known as \emph{Poisson shot noise}. This effect
is well known for the power spectrum and bispectrum, and we quote
here the relationships between the statistics of the discrete
sample and the underlying continuous field:
\begin{align}
  P_n(k)&=P(k)+\frac{1}{\bar{n}}
          \label{eqn:ps_shot}\\
  B_n(k_1,k_2,k_3)&=B(k_1,k_2,k_3)
                    \nonumber \\  
        &\quad+\frac{1}{\bar{n}}[
          P(k_1)+P(k_2)+P(k_3)]+\frac{1}{\bar{n}^2},
\end{align}
where the subscript $n$ denotes the discrete number density
and $\bar{n}$ is the mean number density of the sample.
When making comparisons between theoretical and
simulation bispectra in \Cref{sub:sim_theory} one simply has to
subtract the shot noise contribution in the simulation bispectra
before calculating any correlators.

\subsubsection{Covariance of estimators}

The variance of an estimator is given by its 
covariance matrix $C_X$ which can be written schematically as:
\begin{align}
  C_X&\equiv\text{cov}(\hat{X}(\theta),\hat{X}(\theta')) 
       \nonumber \\
     &=\expval{\hat{X}(\theta)\hat{X}(\theta')} -
       \expval{\hat{X}(\theta)}\expval{\hat{X}(\theta')}.
\end{align}
In addition to calculating covariance matrices numerically through
simulations we also need a framework to calculate them 
(semi-)analytically as a consistency check. 

\vspace{2ex}

\paragraph{Power spectrum covariance}
We first give a brief introduction to matter power spectrum estimation 
and the calculation of its covariance as this has been widely discussed 
in the literature.
This will prepare us for the discussion on the bispectrum covariance
later. Consider for example estimating the power spectrum by binning
it in $k$-space and averaging over all modes within each bin
\cite{blot2,peacock}:
\begin{align}
  \hat{P}(k)=\frac{k_F^3}{(2\pi)^3}\int_k \frac{d^3p}{V_s(k)}
  \left|\delta(\mathbf{p})\right|^2,
\end{align}
where $k_F=2\pi/L=(1/\delta^D(\mathbf{0}))^{1/3}$ is the fundamental 
frequency of the simulation
box of length $L$, and the integral is performed over all modes
that lie in the spherical shell 
$\left|\mathbf{p}-k\right|\leq\Delta k/2$ which has width
$\Delta k$. The normalisation factor $V_s$ is the volume of the
shell: $V_s=\int_k d^3p=4\pi k^2\Delta k + \pi(\Delta k)^3/3$. This 
estimator is unbiased because 
\begin{align}
  \expval{\hat{P}(k)}
  &=\frac{k_F^3}{(2\pi)^3}\int_k \frac{d^3p}{V_s(k)}
    \expval{\left|\delta(\mathbf{p})\right|^2}
    =\frac{k_F^3}{(2\pi)^3}
    \expval{\left|\delta(k)\right|^2}
    \nonumber \\
  &=\frac{k_F^3}{(2\pi)^3}(2\pi)^3\delta(\mathbf{0})P(k)
    =P(k).
\end{align}
The covariance matrix for this estimator
is 
\begin{align}
  &C_P(k,k')
    \nonumber \\
  ={}&\frac{k_F^6}{(2\pi)^6}\int_k \frac{d^3p}{V_s(k)}
       \int_{k'} \frac{d^3q}{V_s(k')}
       \expval{\delta^*_p\delta_p\delta^*_q\delta_q}-P(k)P(k') 
       \nonumber \\
  ={}&\frac{2k_F^3}{V_s(k)}P^2(k)\delta_{k,k'}
       \nonumber \\
  &\qquad+
    \frac{k_F^6}{(2\pi)^6}\int_k \frac{d^3p}{V_s(k)}
    \int_{k'} \frac{d^3q}{V_s(k')}
    \expval{\delta^*_p\delta_p\delta^*_q\delta_q}_c,
    \nonumber \\
  ={}&\frac{2k_F^3}{V_s(k)}P^2(k)\delta_{k,k'}
       \nonumber \\
  &\qquad+
    \frac{k_F^3}{(2\pi)^3}\int_k \frac{d^3p}{V_s(k)}
    \int_{k'} \frac{d^3q}{V_s(k')}
    T(\mathbf{p},-\mathbf{p},\mathbf{q},-\mathbf{q}),
    \label{eqn:PS_cov}
\end{align}
where we have expanded the four-point correlator in 
terms of its connected pieces\footnote{Other contributions 
  vanish since $\expval{\delta}=0$ by definition.}:
$\expval{\delta^*_p\delta_p\delta^*_q\delta_q}=
\expval{\delta^*_p\delta_p}\expval{\delta^*_q\delta_q}+
\expval{\delta^*_p\delta^*_q}\expval{\delta_p\delta_q}+
\expval{\delta^*_p\delta_q}\expval{\delta^*_q\delta_p}+
\expval{\delta^*_p\delta_p\delta^*_q\delta_q}_c$, and
the trispectrum $T$ is defined by $\expval{\delta(\mathbf{k}_1) 
  \delta(\mathbf{k}_2) \delta(\mathbf{k}_3)\delta(\mathbf{k}_4)}_c
=(2\pi)^3 \delta_D (\mathbf{k}_1+\mathbf{k}_2+\mathbf{k}_3
+\mathbf{k}_4)T(\mathbf{k}_1,\mathbf{k}_2,\mathbf{k}_3,
\mathbf{k}_4)$ where the
subscript $c$ denotes \emph{connected}.
Connected $n$-point correlators with $n>2$ vanish if
$\delta$ is a Gaussian field, but e.g. gravitational
evolution induces mode coupling and hence non-Gaussianity
in the form of higher order correlators.

The first term in \Cref{eqn:PS_cov} is the \emph{Gaussian}
contribution to the power spectrum covariance and can be 
estimated with $\hat{P}$; the Kronecker delta $\delta_{k,k'}$ 
enforces the diagonality of the Gaussian covariance. The 
trispectrum term is the non-Gaussian covariance which is 
non-trivial to estimate directly from simulations or calculate 
theoretically. Crucially the non-Gaussian covariance does 
not scale inversely with the number of modes in each bin
unlike the Gaussian covariance \cite{blot2,cov}; this also
applies to the bispectrum. However they both scale inversely 
with the simulation box size through $k_F^3$, and clearly can both be 
suppressed by averaging over different simulation realisations.

\paragraph{Covariance of the \MODALLSS{} estimator}
Now we turn our attention to the covariance of the \MODALLSS{} 
bispectrum estimator (\Cref{eqn:modal_est}), which is unbiased
because
\begin{align}
  &\sqrt{\frac{k_1k_2k_3}{P(k_1)P(k_2)P(k_3)}}
    \expval{\hat{B}_\delta(k_1,k_2,k_3)} 
    \nonumber \\
  ={}&\sum_n^{n_{max}} \expval{\tilde{\beta}^Q_n} 
       Q_n(k_1/k_{max},k_2/k_{max},k_3/k_{max})
       \nonumber \\
  ={}&\sum_n^{n_{max}} \alpha^Q_n 
       Q_n(k_1/k_{max},k_2/k_{max},k_3/k_{max}),  
       \nonumber \\
  ={}&\sqrt{\frac{k_1k_2k_3}{P(k_1)P(k_2)P(k_3)}}
       B_\delta(k_1,k_2,k_3).
\end{align}
The covariance of $\hat{B}_\delta$, $C_B$, is given by:
\begin{align}
  &C_B(k_1,k_2,k_3,k_1',k_2',k_3')
    \nonumber \\
  ={}&\sqrt{\frac{P_1P_2P_3P_1'P_2'P_3'}
       {k_1k_2k_3k_1'k_2'k_3'}}\sum_{mn}^{n_{max}}
       \expval{\tilde{\beta}^Q_m\tilde{\beta}^Q_n}
       Q_mQ_n'
       \nonumber \\    
  &\quad-B(k_1,k_2,k_3)B(k_1',k_2',k_3') 
    \nonumber \\
  ={}&\sqrt{\frac{P_1P_2P_3P_1'P_2'P_3'}
       {k_1k_2k_3k_1'k_2'k_3'}}\sum_{mnop}^{n_{max}}
       (\gamma^{-1})_{om}(\gamma^{-1})_{pn}
       \expval{\beta^Q_m\beta^Q_n}Q_oQ_p'
       \nonumber \\    
  &\quad-B(k_1,k_2,k_3)B(k_1',k_2',k_3'),
    \label{eqn:bis_cov}
\end{align}
where $P_1=P(k_1)$ etc., and the arguments of the $Q_n$ basis 
functions have been suppressed for brevity. We have also
used \Cref{eqn:beta_basis} to convert from $\tilde{\beta}^Q_n$
to $\beta^Q_n$. In order to evaluate $\expval{\beta^Q_m\beta^Q_n}$
we write $\beta^Q_n$ as follows using \Cref{eqn:fnl_sub}:
\begin{align}
  &\beta^Q_n
    \nonumber \\
  ={}&(2\pi)^6\int_{\mathbf{k}_1,\mathbf{k}_2,\mathbf{k}_3}
       \frac{\delta_{\mathbf{k}_1}\delta_{\mathbf{k}_2}\delta_{\mathbf{k}_3}
       Q_n}{\sqrt{k_1k_2k_3P_1P_2P_3}}
       \delta_D(\mathbf{k}_1+\mathbf{k}_2+\mathbf{k}_3)
       \nonumber \\
  ={}&(2\pi)^3\int d^3x\int_{\mathbf{k}_1,\mathbf{k}_2,\mathbf{k}_3}
       \frac{\delta_{\mathbf{k}_1}\delta_{\mathbf{k}_2}\delta_{\mathbf{k}_3}
       Q_n}{\sqrt{k_1k_2k_3P_1P_2P_3}}
       e^{i(\mathbf{k}_1+\mathbf{k}_2+\mathbf{k}_3)\cdot\mathbf{x}},
       \label{eqn:beta_estimator}
\end{align}
which leads to this rather messy expression:
\begin{align}
  &\expval{\beta^Q_m\beta^Q_n}
    \nonumber \\
  ={}&(2\pi)^{12}\int_{1,2,3,1',2',3'}
       \frac{Q^{\vphantom{\prime}}_m}{\sqrt{k^{\vphantom{\prime}}_1k_2k_3P_1P_2P_3}}
       \frac{Q_n'}{\sqrt{k_1'k_2'k_3'P_1'P_2'P_3'}}
       \nonumber \\
  &\quad\times\delta_D(\mathbf{k}_1+\mathbf{k}_2+\mathbf{k}_3)
    \delta_D(\mathbf{k}_1'+\mathbf{k}_2'+\mathbf{k}_3')
    \nonumber \\
  &\quad\times\expval{\delta_{\mathbf{k}^{\vphantom{\prime}}_1}
    \delta_{\mathbf{k}^{\vphantom{\prime}}_2}
    \delta_{\mathbf{k}^{\vphantom{\prime}}_3}
    \delta_{\mathbf{k}_1'}\delta_{\mathbf{k}_2'}\delta_{\mathbf{k}_3'}},
    \label{eqn:six-point}
\end{align}
where we further abbreviate the integral over the 6 wavevectors to
$\int_{1,2,3,1',2',3'}\equiv\int\frac{\prod_{i=1}^3 d^3k_i}{(2\pi)^9}
\frac{\prod_{i=1}^3 d^3k_i'}{(2\pi)^9}$. With some difficulty this
can be rewritten as:
\begin{widetext}
  \begin{align}
    &\expval{\beta^Q_m\beta^Q_n}
      \nonumber \\
    ={}&6(2\pi)^{3}\gamma_{mn}+\alpha^Q_m\alpha^Q_n
         +V(2\pi)^{12}\int_{1,2,3,1',2',3'}
         \frac{Q^{\vphantom{\prime}}_m}{\sqrt{k^{\vphantom{\prime}}_1k_2k_3P_1P_2P_3}}
         \frac{Q_n'}{\sqrt{k_1'k_2'k_3'P_1'P_2'P_3'}}
         \delta_D(\mathbf{k}_1+\mathbf{k}_2+\mathbf{k}_3)
         \delta_D(\mathbf{k}_1'+\mathbf{k}_2'+\mathbf{k}_3')
         \nonumber \\
    &\quad\times
      \Bigg(
      (2\pi)^3\delta_D(\mathbf{k}_3-\mathbf{k}_3')
      B(k_1,k_2,k_3')B(k_1',k_2',k_3)
      +8\,\text{perms}
      \nonumber \\
    &\qquad+      
      (2\pi)^3
      \delta_D(\mathbf{k}_1+\mathbf{k}_1')
      T(\mathbf{k}_2,\mathbf{k}_3,\mathbf{k}_2',\mathbf{k}_3')P(k_1)
      +8\,\text{perms}+
      P_5(\mathbf{k}_1,\mathbf{k}_2,\mathbf{k}_3,\mathbf{k}_1',
      \mathbf{k}_2',\mathbf{k}_3')\Bigg).
      \label{betaQ2}
  \end{align}
  where the pentaspectrum $P_5$ is defined by $\expval{\delta(\mathbf{k}_1) 
    \delta(\mathbf{k}_2) \delta(\mathbf{k}_3)\delta(\mathbf{k}_4)
    \delta(\mathbf{k}_5)\delta(\mathbf{k}_6)}_c
  =(2\pi)^3 \delta_D (\mathbf{k}_1+\mathbf{k}_2+\mathbf{k}_3
  +\mathbf{k}_4+\mathbf{k}_5+\mathbf{k}_6)P_5(\mathbf{k}_1,\mathbf{k}_2,
  \mathbf{k}_3,\mathbf{k}_4,\mathbf{k}_5,\mathbf{k}_6)$.
\end{widetext}
While there is no easy way to evaluate the last two set of terms 
involving the trispectrum and pentaspectrum, the Gaussian covariance
of the $\beta^R_n$ is given trivially as
\begin{align}
  \label{eqn:beta_cov}
  C^{\beta}_{mn} \equiv \expval{\beta^R_m\beta^R_n} -
  \expval{\beta^R_m}\expval{\beta^R_n} \approx 6(2\pi)^{3}\delta_{mn},
\end{align}
which is diagonal. Unfortunately $C_B$ cannot be evaluated analytically,
even in the Gaussian limit, since \Cref{eqn:bis_cov} yields
\begin{align}
  &\sqrt{\frac{k_1k_2k_3k_1'k_2'k_3'}
    {P_1P_2P_3P_1'P_2'P_3'}}
    C_B(k_1,k_2,k_3,k_1',k_2',k_3')
    \nonumber \\
  \approx{}
  &6(2\pi)^3\sum_{mn}^{n_{max}}
    Q_m'(k_1',k_2',k_3')(\gamma^{-1})_{mn}Q_n(k_1,k_2,k_3)
    \nonumber \\    
  ={}&6(2\pi)^3\sum_{n}^{n_{max}}
       R_n'(k_1',k_2',k_3')R_n(k_1,k_2,k_3)
       \label{eqn:bis_cov_final}
\end{align}
where we have used \Cref{eqn:RQconv} to convert from the 
$\{Q_n\}$ basis to $\{R_n\}$. The last line cannot be
further simplified because in practice we can never use enough
modes to ensure $\{R_n\}$ forms a complete basis. Nevertheless
we can calculate the Gaussian covariance of $\hat{f}_{nl} =
\sum_n \alpha^R_n \beta^R_n/\sum_n \alpha^R_n\alpha^R_n$ here
which we will explore numerically in \Cref{sub:cov}:
\begin{align}
  C_{f_{nl}}
  & \equiv \expval{\hat{f}_{nl}^2} -
    \expval{\hat{f}_{nl}}^2  \nonumber \\
  & = \frac{\sum_{mn}
    \alpha^R_m\alpha^R_n\expval{\beta^R_m\beta^R_n} - 
    (\sum_n \alpha^R_n\expval{\beta^R_n})^2}
    {(\sum_n \alpha^R_n\alpha^R_n)^2} \nonumber \\
  & \approx \frac{1}{(\sum_n \alpha^R_n\alpha^R_n)^2}
    \Bigg(\sum_{mn}
    \alpha^R_m\alpha^R_n\left(6(2\pi)^3\delta_{mn}
    +\alpha^R_m\alpha^R_n\right) \nonumber \\
  & \qquad\qquad\qquad\qquad -
    (\sum_n \alpha^R_n\alpha^R_n)^2\Bigg)
    \nonumber \\
  & = \frac{6(2\pi)^3}{\sum_n \alpha^R_n\alpha^R_n}.
    \label{eqn:fnl_cov}
\end{align}

\paragraph{Suppression of large-scale variances}
Large variances are prominent at large scales due to the
finite volume of the simulation box or observational area leading to a
lack of Fourier modes for statistical calculations. These are typically
known as \emph{finite box} or \emph{cosmic variance} effects, although
in the former case there is the added complication of mode coupling induced
by non-linear gravitational evolution \citep{paired2}. These errors need to be controlled as to extract
cosmological parameters from galaxy surveys, and there is evidence to suggest
detection of new physics may require $\mathcal{O}(0.1\%)$ accuracy in
simulations \citep{tobias}. While cosmic variance, which is defined by the observational volume of a
given survey, is unavoldable, we could reduce \emph{finite box} errors in simulations by simply expanding the box or averaging multiple simulations. Unfortunatly both of these approaches are costly in terms of time and computational resources. For a more efficient way of obtaining ensemble
averaged quantities such as the power spectrum and bispectrum the the
authors of \citep{paired1,paired2} have proposed a method of pairing up
simulations which have opposite phases in their initial conditions.
The phase inversion has no affect on the statistical properties of the
simulation thus the pairing up process does not bias power spectra and
bispectra estimation. However, leading order contributions to the Gaussian
covariances, which are the dominant contribution to cosmic variance, will
cancel as they are out-of-phase with each other.

We will quickly review the method. First we expand the late-time non-linear density
field in standard perturbation theory (SPT) \citep{Bernardeau}:
\begin{align}
  \delta(\mathbf{k},z)=\sum^\infty_{n=1}\delta_n(\mathbf{k},z),
\end{align}
where $\delta_1$ represents linear growth of the initial conditions,
an $\delta_n$ are $n$ copies of $\delta_1$ convolved with the
SPT kernels $F_n$. We can calculate the power spectrum in this formalism,
expanding to 4th order in products of $\delta_1$ we obtain:
\begin{align}
  P=P_{11}+P_{12}+P_{21}+P_{13}+P_{22}+P_{31}+\cdots,
\end{align}
where $P=\left<\delta\delta\right>$ and $P_{nm}$ denotes
$P=\left<\delta_n\delta_m\right>$. Assuming Gaussian initial conditions
so that $\delta_1$ is also Gaussian, we can use Wick's theorem to
eliminate terms containing odd multiples of $\delta_1$, thus giving:
\begin{align}
  P^{\text{Gaussian IC}}=P_{11}+P_{13}+P_{22}+P_{31}+\cdots.
\end{align}
The effect of phase inversion is to reverse the sign of $\delta_1$,
and the pairing up procedure serves to annihilate the same odd-parity
terms that are expected to vanish in the ensemble average, while
leaving the signal terms, which have even parity, intact. On the other
hand since the non-Gaussian covariances also have even parity they
remain unaffected.

The same applies for the bispectrum. The expansion in SPT is now
(neglecting permutations)
\begin{align}
  B&=B_{111}+B_{112}+B_{113}+B_{122}
     \nonumber \\
   &\qquad+B_{114}+B_{123}+B_{222}+\cdots,
\end{align}
so that for Gaussian initial conditions we have
\begin{align}
  B^{\text{Gaussian IC}}=B_{112}+B_{122}+B_{114}+B_{123}+B_{222}+\cdots.
\end{align}
Again we see that terms containing an odd number of $\delta_1$
vanish which coincides with the effect of pairing up phase inverted
simulations. While the suppression of variance in power spectra
estimation was explored in great detail in \citep{paired2} no
equivalent test have been performed with the bispectrum, which
we leave to future work.

\subsubsection{Systematic offsets due to aliasing contributions
  \label{sec:aliasing}}

\begin{figure*}
  \begin{subfigure}[b]{\textwidth}
    \begin{align}
      \vcenter{\hbox{\includegraphics[width=0.31\linewidth]{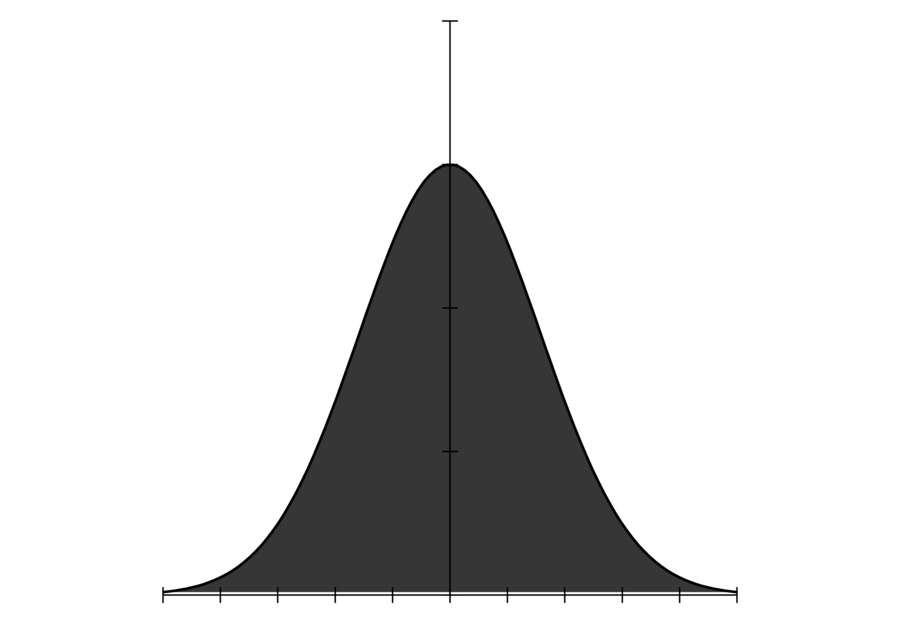}}}
      \,\times
      \vcenter{\hbox{\includegraphics[width=0.31\linewidth]{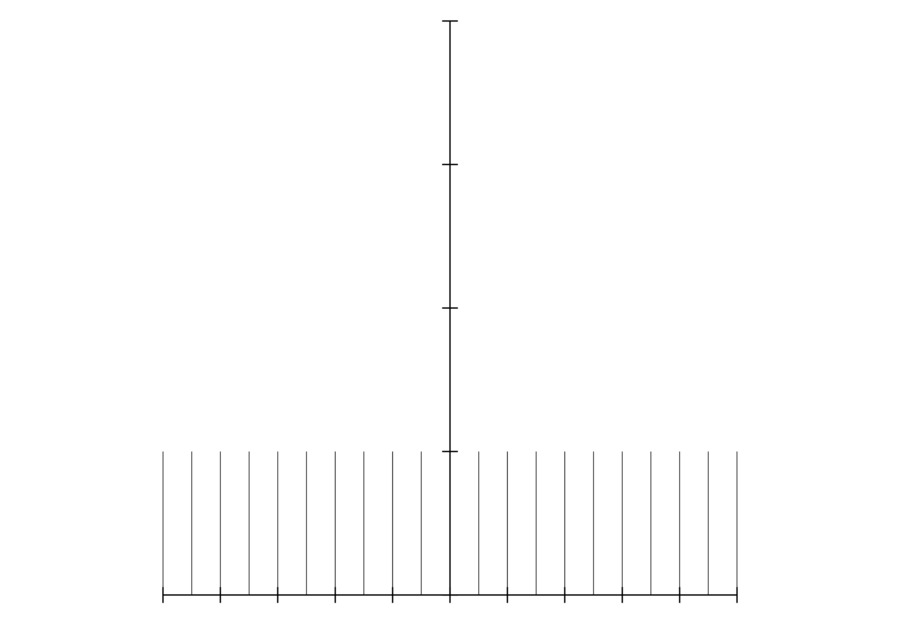}}}
      \,=
      \vcenter{\hbox{\includegraphics[width=0.31\linewidth]{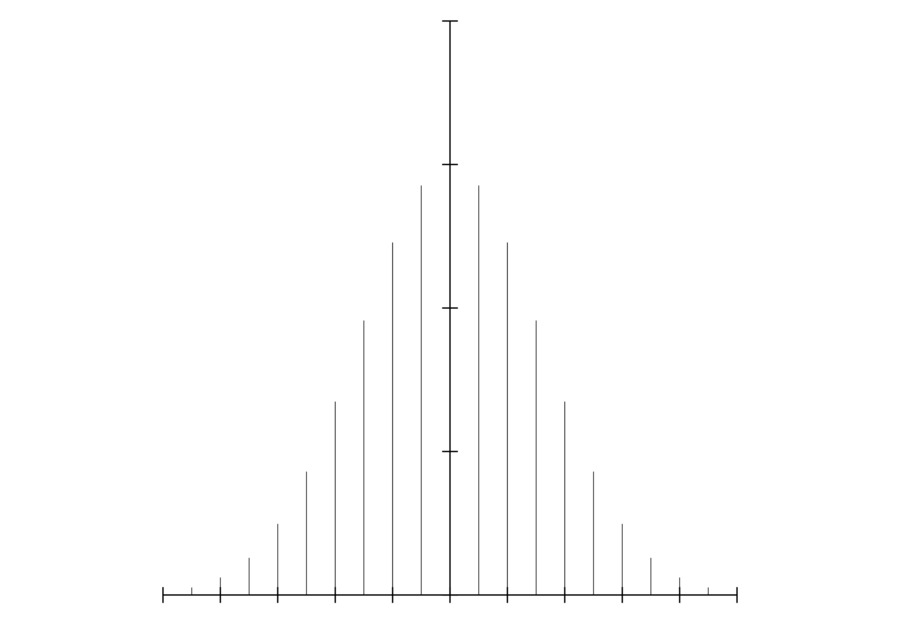}}}
      \nonumber
    \end{align}
    \caption{Sampling in real space is a multiplication of the signal
      with a Dirac comb.} 
    \label{fig:diraccomb_mult}
  \end{subfigure}
  
  \begin{subfigure}[b]{\textwidth}
    \begin{align}
      \vcenter{\hbox{\includegraphics[width=0.31\linewidth]{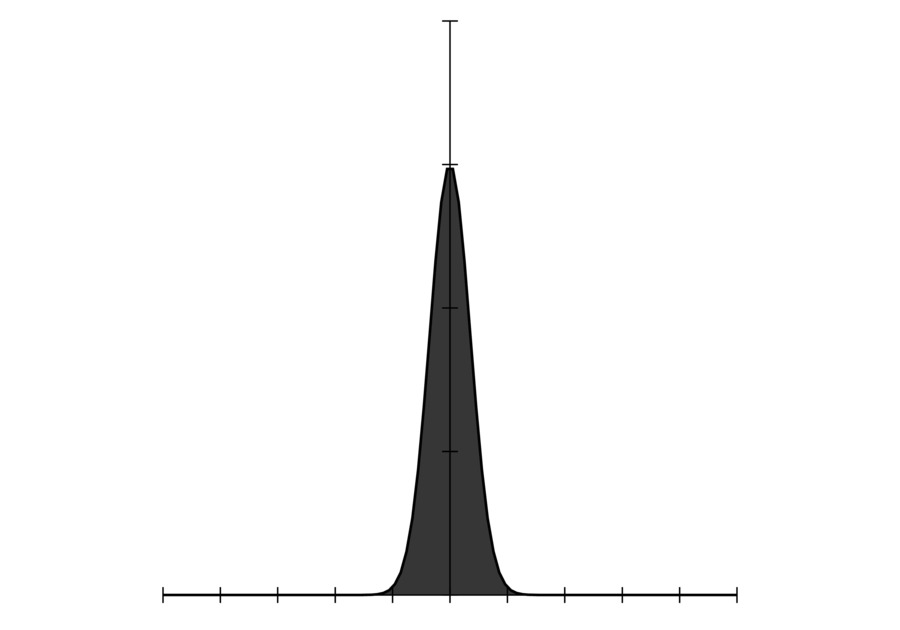}}}
      \,*
      \vcenter{\hbox{\includegraphics[width=0.31\linewidth]{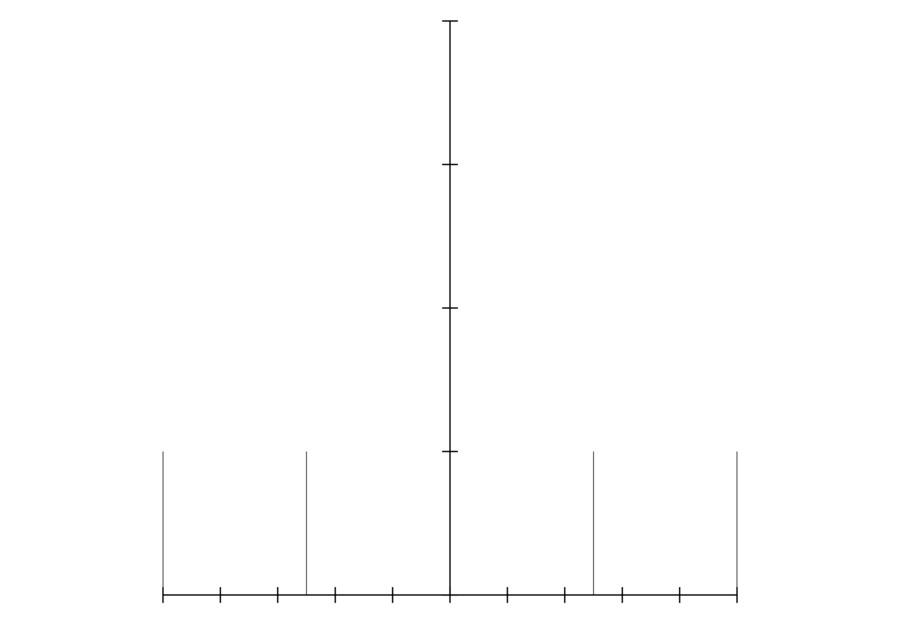}}}
      \,=
      \vcenter{\hbox{\includegraphics[width=0.31\linewidth]{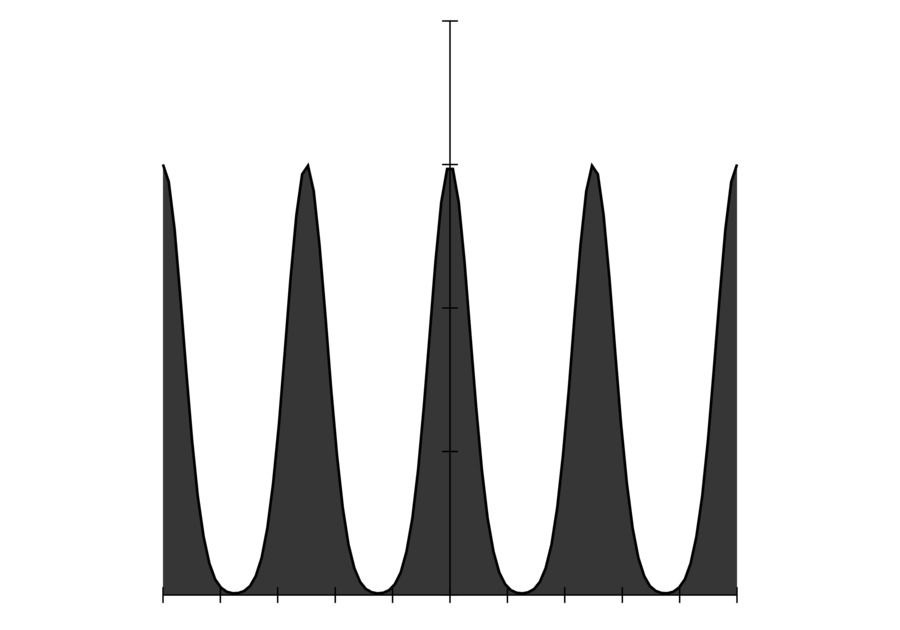}}}
      \nonumber
    \end{align}
    \caption{In Fourier space this becomes a convolution between the 
      signal and a Dirac comb, resulting in multiple, aliased copies 
      of the signal.} 
    \label{fig:diraccomb_conv}
  \end{subfigure}
  \caption{Sampling in real and Fourier space (Figure 1 from \citep{digital}).} 
\end{figure*}

\begin{figure*}
  \begin{subfigure}[b]{0.45\textwidth}
    \includegraphics[width=\linewidth]{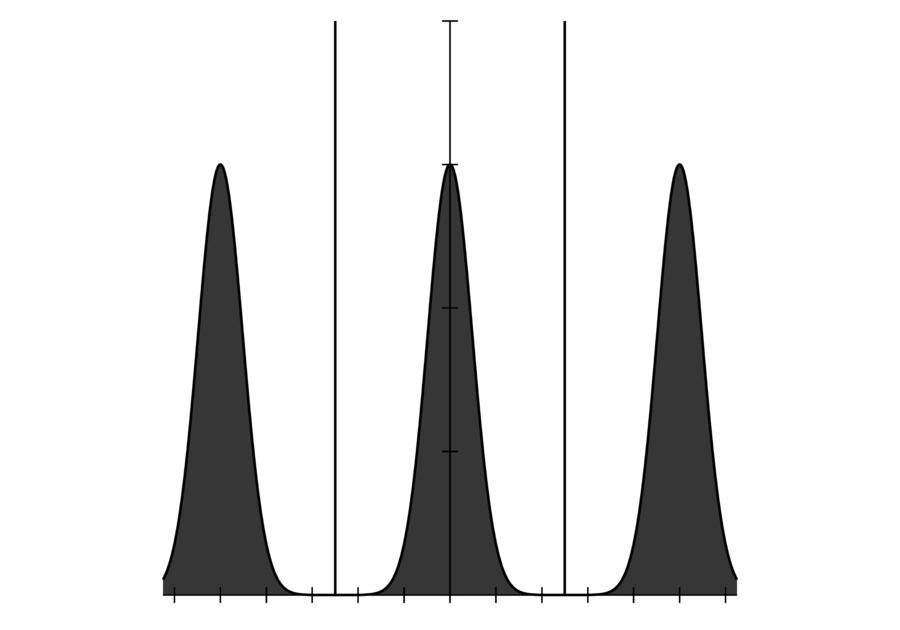}
    \caption{
      If the sampling frequency is more than twice
      the highest frequency in the signal, then the aliased 
      images that appear after convolving the signal with
      the Dirac comb do not overlap. In this case the signal
      is undistorted and can be uniquely restored.
    }
    \label{fig:no_alias}
  \end{subfigure}
  ~
  \begin{subfigure}[b]{0.45\textwidth}
    \includegraphics[width=\linewidth]{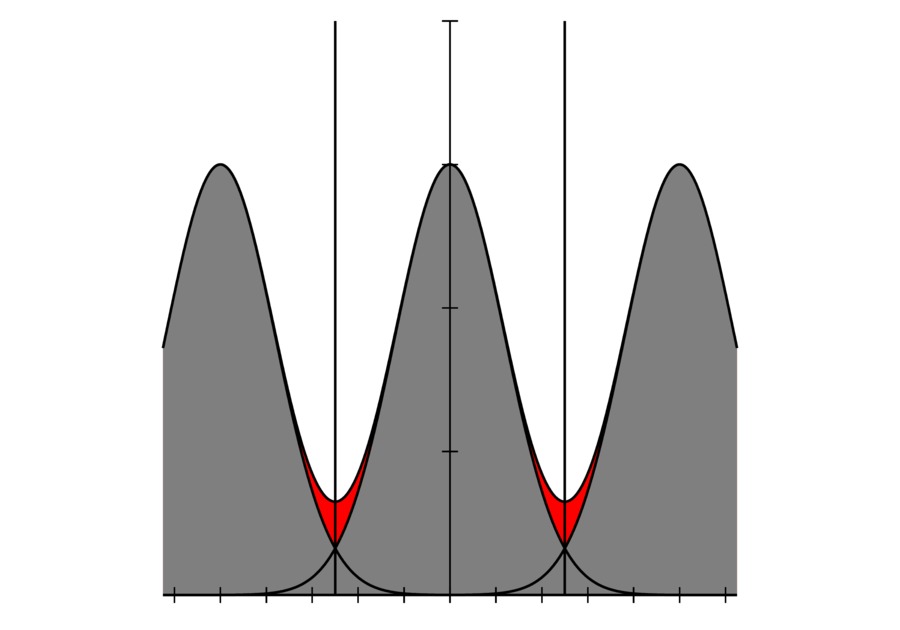}
    \caption{
      On the other hand if the Nyquist criterion is not met,
      the images will then overlap with each other due to 
      contributions from the higher frequencies, leading to
      significant distortions near the Nyquist frequency. There 
      is no easy way to recover the original signal.
    }
    \label{fig:alias}
  \end{subfigure}
  \caption{If the sampling frequency is too low, aliasing occurs (Figure 2 from \citep{digital}).}
\end{figure*}

Virtually all power spectra and bispectra 
analyses are done with FFTs due to the efficiency of calculating 
Fourier transforms versus direct calculation of correlation functions in real space \citep{jing2005}. 
The first step in using FFTs is to put the particles on a regular
grid. This involves a mass assignment scheme which dictates the
weighting with which each particle is distributed across its
surrounding grid points. Many of these schemes are well known in the
literature, e.g. Nearest Grid Point (NGP), Cloud in cell (CIC) and
Triangular Shaped Clouds (TSC) \citep{jing2005}, as well as higher
order interpolation schemes such as Piecewise Cubic Spline (PCS)
\citep{interlacing} and Daubechies wavelet transformations \citep{MAS}.
The effect of this assignment manifests as a convolution with the 
density field which becomes a product with the corresponding window
function $W(\mathbf{k})$ in Fourier space. In principle this can be 
corrected for easily by dividing out the window function in Fourier 
space.  However even in this case the use of discrete FFTs inevitably 
leads to information loss \citep{digital}. By the Shannon sampling theorem 
\citep{shannon} all the information in a signal can be recovered if
the sampling frequency is twice that of the highest frequency in the
signal, i.e. with a sufficiently high sampling frequency a 
\emph{band-limited} signal can be reproduced without information loss. 
This is known as the \emph{Nyquist criterion}.  The sampling theorem states that this limit is the 
Nyquist frequency $k_{Ny}=k_{max}/2=\pi/H$, where $k_{max}$ is the sampling 
frequency of the grid and $H$ is the grid spacing. For the purpose of
estimating correlation functions with FFTs it is known than the cutoff frequency for
the power spectrum is the Nyquist frequency $k_{Ny}$ \citep{jing2005,MAS,digital,interlacing}. For the bispectrum \citep{jeong} and \citep{interlacing} propose the 
limit for the bispectrum should be $2k_{Ny}/3$.

There is a second serious problem associated with discrete grids which
is the introduction of sampling artefacts near the Nyquist frequency. 
As explained in further detail in \citep{digital}, discrete sampling in
real space is effectively a multiplication of the signal with a Dirac
comb (\Cref{fig:diraccomb_mult}). In Fourier space this multiplication
becomes a convolution operation, resulting in multiple images of the 
signal evenly spaced at the sampling frequency of the grid 
(\Cref{fig:diraccomb_conv}). In the case that the sampling frequency 
is more than twice the maximum frequency of the signal, as in 
\Cref{fig:no_alias}, then the images of the signal do not overlap each 
other and no artefacts are induced. Otherwise if higher frequencies are 
indeed present (\Cref{fig:alias}), which certainly holds true in 
cosmological contexts, then the copies of the replicated signal will 
overlap and distort the sampled signal near the Nyquist frequency. 
We demonstrate this effect with \GADGET{} power spectra and bispectra in
\Cref{fig:alias_ps_bis} (for details of the simulations see \Cref{sim}
below).  Here we find that the cutoff frequency for the bispectrum is the same as the power spectrum, $k_{Ny}$ in disagreement with the predictions of \citep{jing2005,MAS,digital,interlacing}.

\begin{figure*}
  \begin{subfigure}[b]{0.49\textwidth}
    \includegraphics[width=\linewidth]{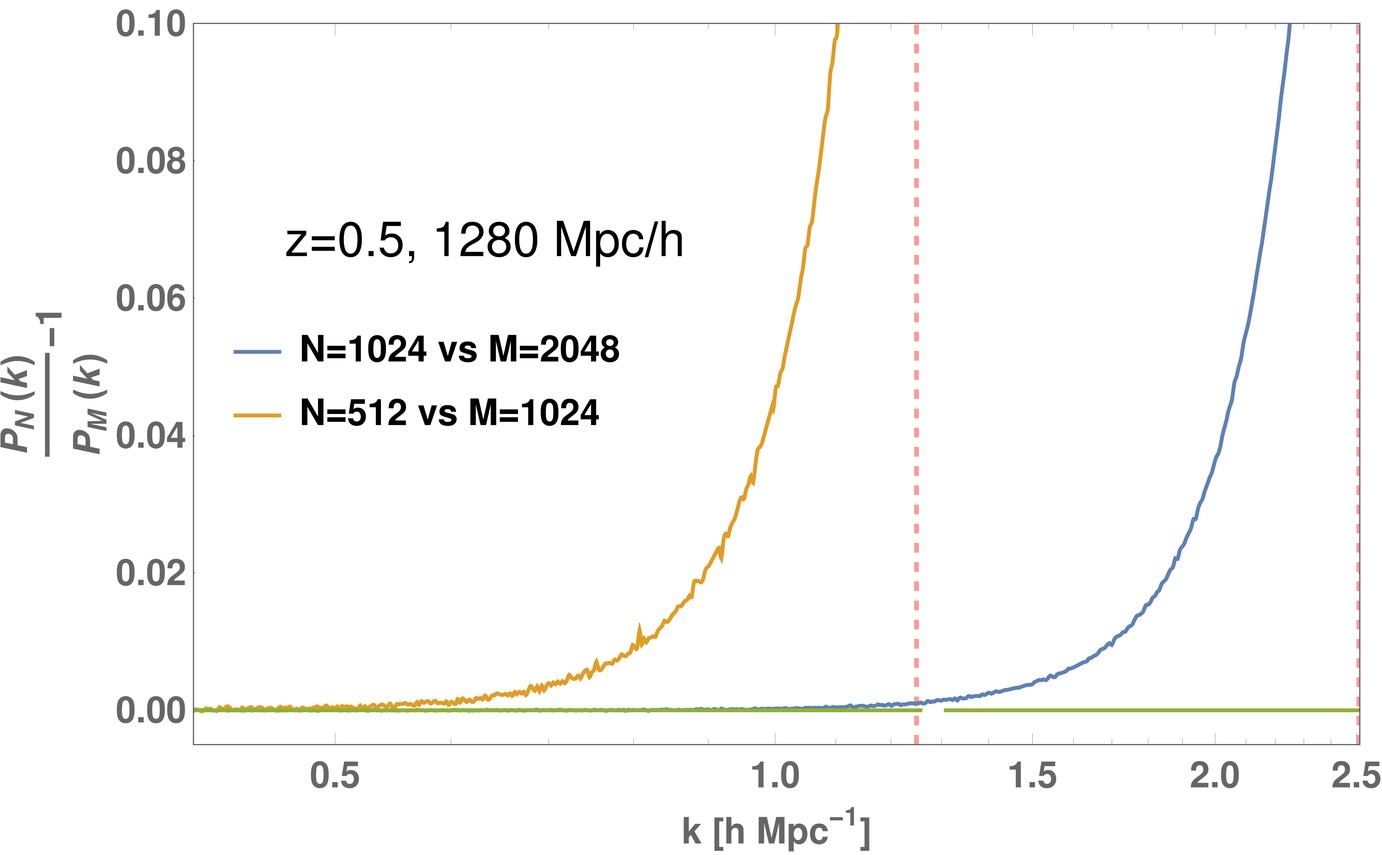}
    \caption{
      Ratio between \GADGET{} power spectra estimated
      with FFT grids of different sizes. The baseline is the larger
      of the two CIC grids, and the pink, dashed lines indicate the
      Nyquist frequencies $k_{Ny}$ for the $512^3$ and
      $1024^3$ CIC grids. It is clear how aliasing contributions 
      lead to overestimation of the power spectra near 
      $k=k_{Ny}$, but the functional form of this overshoot
      cannot be calculated analytically.
    }
    \label{fig:alias_ps}
  \end{subfigure}
  ~
  \begin{subfigure}[b]{0.47\textwidth}
    \includegraphics[width=\linewidth]{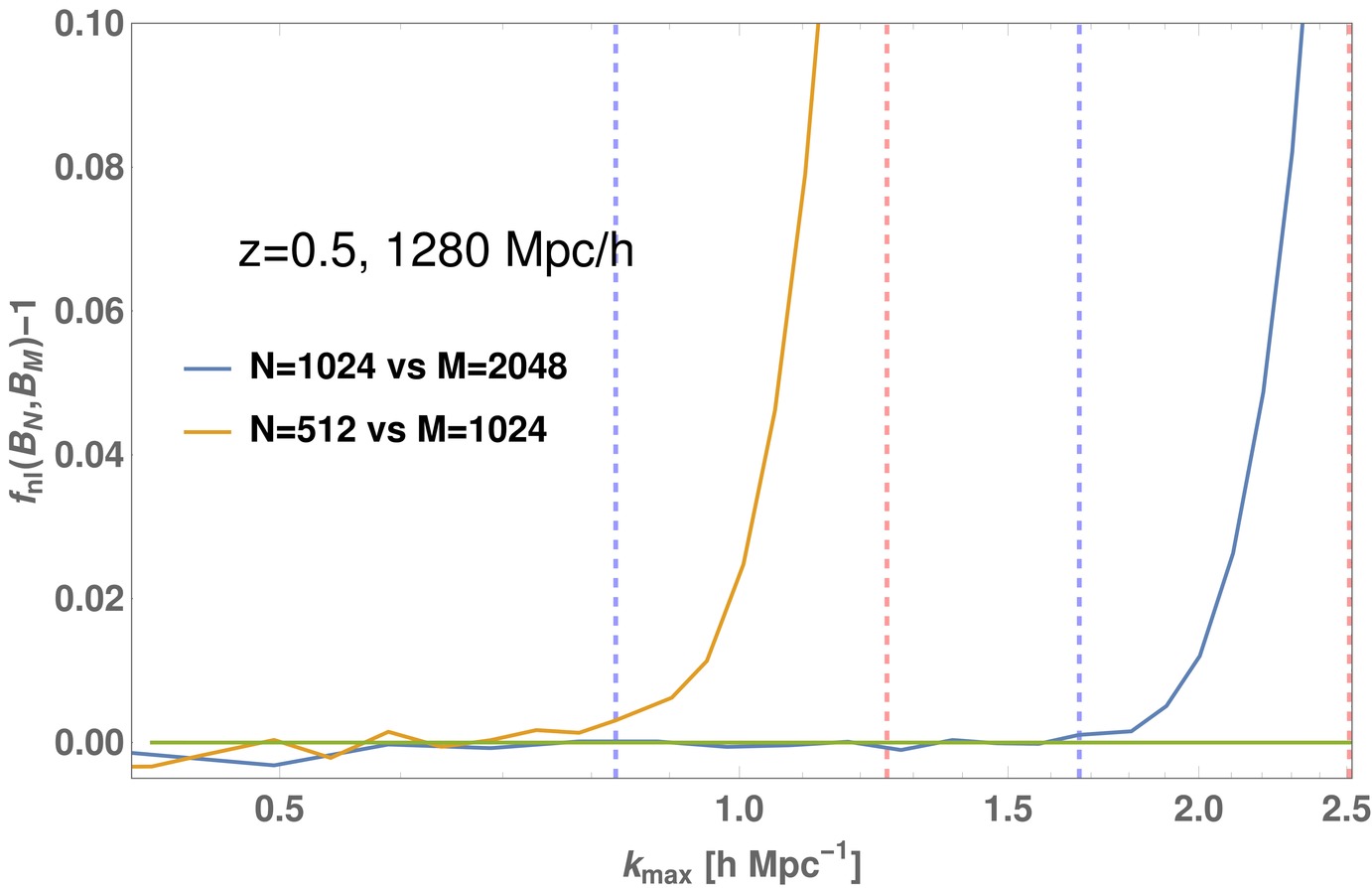}
    \caption{
      $f_{nl}$ correlators between \GADGET{} bispectra 
      estimated with the same FFT grids in \Cref{fig:alias_ps}. 
      Again pink, dashed lines indicate $k_{Ny}$ for the various 
      grids, but we additionally label $k=\frac{2}{3}k_{Ny}$
      with blue, dashed lines to find the correct cutoff 
      frequency. Contrary to \citep{jeong,interlacing} there is
      little to suggest that bispectrum estimation breaks down
      at $\frac{2}{3}k_{Ny}$, but rather at $k_{Ny}$ as for the
      power spectrum.
    }
    \label{fig:alias_bis}
  \end{subfigure}
  \caption{A demonstration of aliasing in the power spectrum and bispectrum
  for \GADGET{} simulations.}
  \label{fig:alias_ps_bis}
\end{figure*}

\begin{figure*}
  \begin{subfigure}[b]{0.45\textwidth}
    \includegraphics[width=\linewidth]{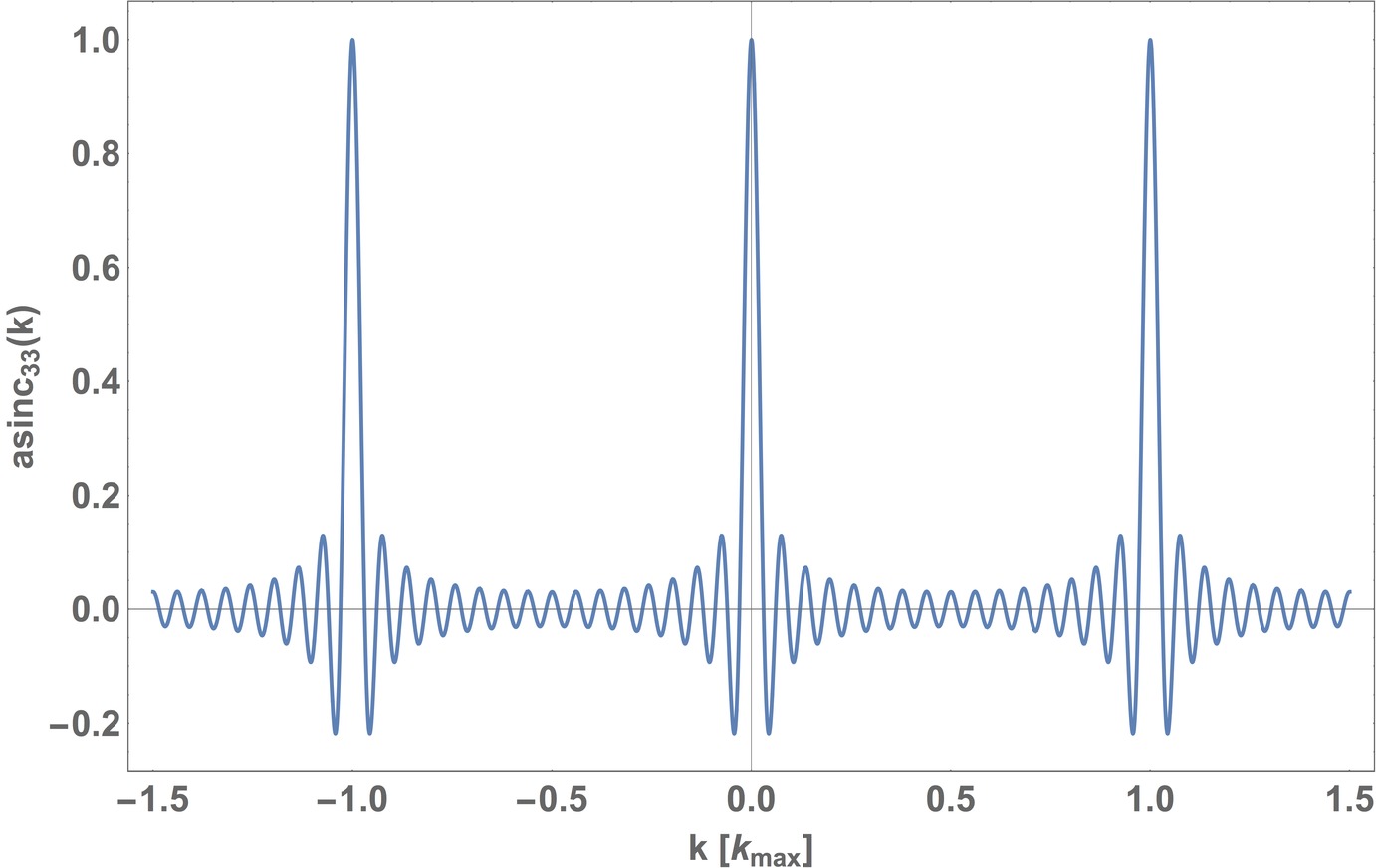}
  \end{subfigure}
  ~
  \begin{subfigure}[b]{0.45\textwidth}
    \includegraphics[width=\linewidth]{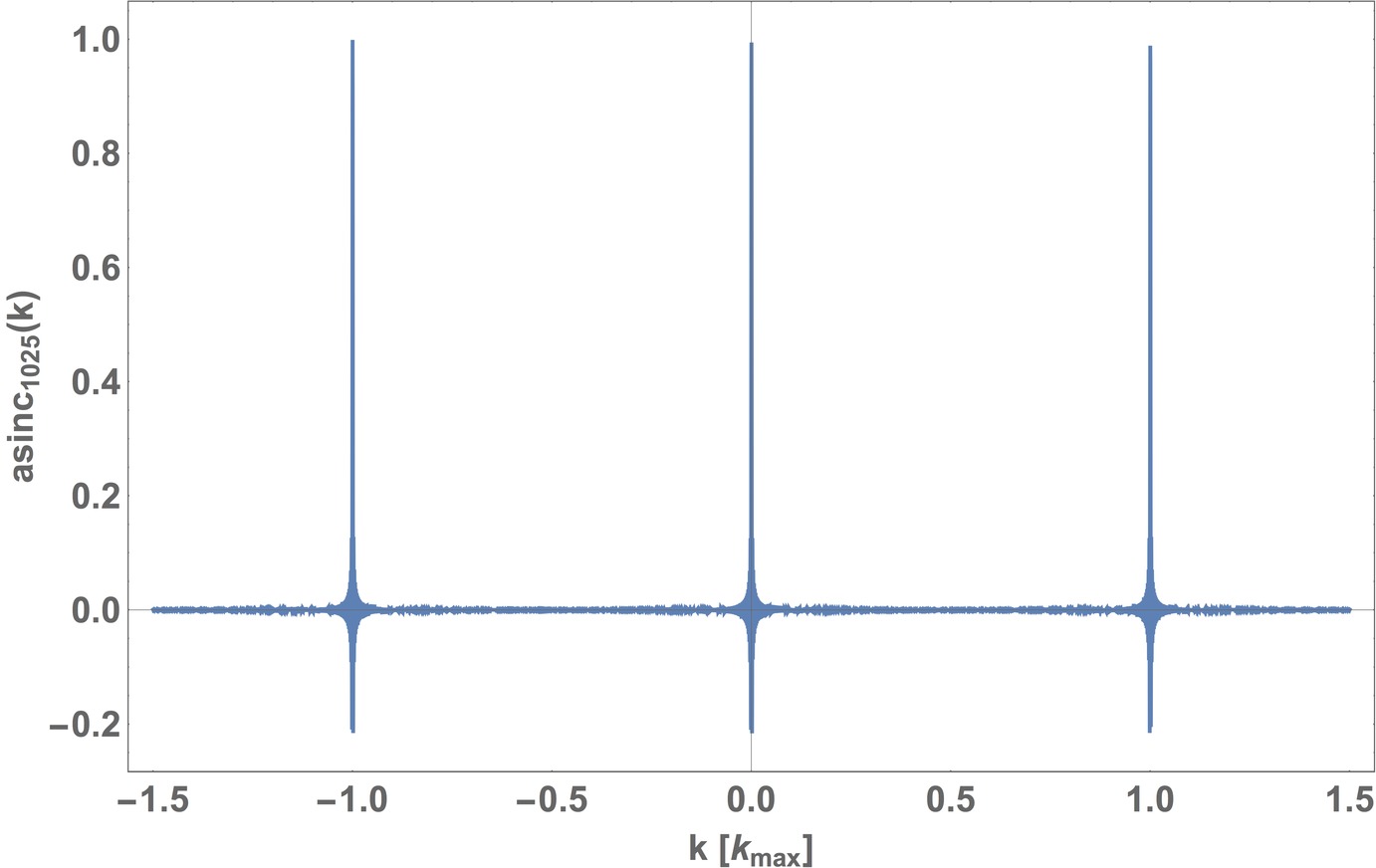}
  \end{subfigure}
  \caption{
    The aliased $\sinc$ function with $\asinc_M(k)$ with
    $M=33$ and 1025 plotted in units of the sampling
    frequency of the grid $k_{max}$. Unlike the Dirac comb
    $\asinc_M(k)$ is non-local and oscillatory between the
    peaks, leading to distortions and aliasing effects even
    for band-limited signals. As is evident in the $M=1025$
    case, both of these effects can be mitigated by using 
    finer sampling grids since the width of the primary peaks
    at its base is $2/M$, and the value of the function at
    $k=k_{Ny}=k_{max}/2$ is $1/M$.
  }
  \label{fig:asinc}
\end{figure*}

To derive this more rigorously
we begin by denoting the FFT density
grid in real space as 
\begin{align}
  \delta_n^f(\mathbf{r})=\Sh_r\left(\frac{\mathbf{r}}{H}\right)
  (\delta_n*W)(\mathbf{r})
\end{align}
where the superscript $f$ labels an FFT quantity and the subscript
$n$ indicates sampling with discrete objects as before. This is
equivalent to the statement that the $\delta_n^f(\mathbf{r})$ is a
multiplication of the sampling grid, i.e. the Dirac comb 
$\Sh_r(\mathbf{r})=\sum_{\mathbf{r}_g}\delta_D(\mathbf{r}-\mathbf{r}_g)
=\sum_{\mathbf{n}}\delta_D(\mathbf{r}-H\mathbf{n})$ where $\mathbf{r}_g$ 
are the grid points and $\mathbf{n}\in\mathbb{Z}^3$ is a vector 
composed of integers, with the convolution between the density field 
sampled by discrete objects $\delta_n(\mathbf{r})$ and the window 
function $W(\mathbf{r})$ due to mass assignment. The Fourier Transform 
of this grid is 
$\delta_n^f(\mathbf{k})=\mathcal{F}[\delta^f_n(\mathbf{r})]$, but one
should bear in mind that to obtain the FFT output one needs to further
multiply this by the Dirac comb in $\mathbf{k}$-space,
$\Sh_k(\mathbf{k})=\sum_{\mathbf{n}}\delta_D(\mathbf{k}-k_F\mathbf{n})$.
The aliasing effects discussed in the previous paragraph becomes
immediately apparent when one evaluates $\delta_n^f(\mathbf{k})$
explicitly which produces:
\begin{align}
  \delta_n^f(\mathbf{k})=\sum_{\mathbf{n}}
  \delta_n(\mathbf{k}-k_{max}\mathbf{n})W(\mathbf{k}-k_{max}\mathbf{n}).
\end{align}
This is merely a restatement of \Cref{fig:diraccomb_conv}: sampling 
with a Dirac comb leads to aliased images spaced at intervals of
$k_{max}$ in Fourier space. If the Nyquist criterion is satisfied, i.e.
all frequencies in the signal satisfy $k<k_{max}/2=k_{Ny}$, then the
images will not overlap and the signal remains undistorted
(\Cref{fig:no_alias}). Otherwise aliasing artefacts will occur
(\Cref{fig:alias}). The power spectrum we obtain via FFT, $P^f_n(k)$,
is thus
\begin{align}
  P^f_n(k)=\sum_{\mathbf{n}}
  \left(P\left(\left|\mathbf{k}-k_{max}\mathbf{n}\right|\right)
  +\frac{1}{\bar{n}}\right)
  \left|W\left(\mathbf{k}-k_{max}\mathbf{n}\right)\right|^2
  \label{eqn:ps_alias}
\end{align}
where we have included the effects of Poisson shot noise. We can
see that the aliasing contributions are most prominent near the
Nyquist frequency $k_{Ny}$ as was the case for the density field.
Finally we note that to obtain the true FFT output one must
multiply the expression in \Cref{eqn:ps_alias} by
$\Sh_f\left(\frac{\mathbf{k}}{k_F}\right)
\Sh_f\left(\frac{\mathbf{-k}}{k_F}\right)
=\Sh_f\left(\frac{\mathbf{k}}{k_F}\right)$.
The equivalent expression for the FFT bispectrum is 
\begin{align}
  &B^f_n(k_1,k_2,k_3)
    \nonumber \\
  ={}&\sum_{\mathbf{n}_1\mathbf{n}_2}\Bigg(B\left(q_1,q_2,
       \left|\mathbf{q}_1+\mathbf{q}_2\right|\right)
       \nonumber \\
  &\qquad+\frac{1}{\bar{n}}[P(q_1)+P(q_2)+
    P(\left|\mathbf{q}_1+\mathbf{q}_2\right|)]
    +\frac{1}{\bar{n}^2}\Bigg)
    \nonumber \\
  &\quad\times W(\mathbf{q}_1)W(\mathbf{q}_2)
    W(-\mathbf{q}_1-\mathbf{q}_2),
    \label{eqn:bis_alias}
\end{align}
where $\mathbf{q}_i=\mathbf{k}_i-k_{max}\mathbf{n}_i$, and the
multiplicative factor that gives the true FFT output becomes
\begin{align}
  &\Sh_f\left(\frac{\mathbf{k}_1}{k_F}\right)
    \Sh_f\left(\frac{\mathbf{k}_2}{k_F}\right)
    \Sh_f\left(\frac{-\mathbf{k}_1-\mathbf{k}_2}{k_F}\right)
    \nonumber \\
  ={}&\Sh_f\left(\frac{\mathbf{k}_1}{k_F}\right)
       \Sh_f\left(\frac{\mathbf{k}_2}{k_F}\right).
\end{align}

In principle this aliasing effect can be completely avoided by low-pass 
filtering the signal to remove the high-frequency contributions. This 
is equivalent to convolving the real-space signal with a $\sinc$ 
function \citep{digital}. However the $\sinc$ function is highly 
non-local and such an operation is computationally expensive since we 
would have to distribute all particles to every grid point. In addition
we have assumed so far that our sampling operation in real space, i.e.
$\Sh_r(\mathbf{r})$, has infinite extent, so that its Fourier transform 
is also an infinite Dirac comb. This cannot be achieved for practical
reasons, and the Fourier transform of a truncated one-dimensional Dirac
comb is the aliased $\sinc$ function 
$\asinc_M(k)$: 
\begin{align}
  \mathcal{F}\left[\frac{1}{M}\sum_{n=-\frac{M-1}{2}}^{\frac{M-1}{2}}
  \delta_D(r-Hn)\right]
  &=\frac{\sin(\pi kM/k_{max})}{M\sin(\pi k/k_{max})}
    \nonumber \\
  &\equiv\asinc_M(k),
\end{align}
where we have introduced the normalisation factor $1/M$. We plot
$\asinc_M(k)$ for $M=33$ and 1025 in \Cref{fig:asinc}, which correspond
to sampling with FFT grids of size $32^3$ and $1024^3$ respectively.
The aliased $\sinc$ function differ from the infinite Dirac comb
in a very important way, i.e. its non-locality. When convolved with
$\delta_n(\mathbf{r})$ the oscillatory features will distort the
signal, and aliased images will always overlap even if the signal
is band-limited. These aliasing contributions can be alleviated by
low-pass filtering the signal, but one can not eradicate them nor
uniquely restore the original signal \citep{digital}. However it should
be noted that with sufficiently large $M$ one can typically neglect
these contributions: the base width of the primary peaks is $2/M$
and the value of $\asinc_M(k)$ at the Nyquist frequency is $1/M$.
Finally we remark that these finite, discrete sampling effects are
exacerbated by the mass assignment procedure as the window function
$W(\mathbf{k})$ also enters the aliased sum. This is a mild complication
for the shot noise terms in \Cref{eqn:ps_alias,eqn:bis_alias} as
$W\left(\mathbf{k}-k_{max}\mathbf{n}\right)$ are typically simple
analytical expressions \citep{jing2005}. As for the product between
the power spectrum and window function \citep{jing2005} proposed a
procedure to cure these sampling effects iteratively by assuming the
power spectrum $P(k)$ behaves like a power-law near the Nyquist frequency
$k\sim k_{Ny}$. While this approximation seemed to work effectively for
the power spectrum, it is not clear how one would similarly construct a
simple analytical formula that captures the local behaviour of the
bispectrum and higher order correlators effectively.

While no method has been found to fully recover the bispectrum near the
Nyquist frequency, various solutions have been put forward to diminish
the effects of aliasing. A straightforward approach is using higher order
interpolation kernels such as PCS or Daubechies wavelets which are closer
approximations to the ideal low-pass filter. In particular the authors of
\citep{MAS} claim that even with deconvolution of the corresponding window
function, the power spectrum can be measured with the wavelets to an accuracy 
level of 2\% in for wavenumbers up to $0.7k_{Ny}$. Since particle-mesh
simulation codes rely on FFTs for rapid calculations of the gravitational
potential, the Daubechies wavelets may prove useful as an inexpensive yet
accurate way of representing particles on a grid. An alternative method
is to push the aliasing effects to higher $k$ by first `supersampling' the
density field at some higher resolution than the one desired \citep{digital}.
The super-sampled grid naturally has a higher Nyquist frequency thus we
expect the aliasing effects at the target resolution to be much reduced.
Finally we down-sample the super-sampled grid by deconvolving the relevant
window function and removing all unwanted $k$-modes to obtain the signal
sampled at the frequency of interest. The advantages of `supersampling' over
other methods are its effectiveness at removing undesirable aliasing
distortions \emph{at the target frequency}, and since low order mass
assignment schemes such as CIC and TSC can be used for supersampling it is
also computationally fast. However to super-sample at $n$ times the required
resolution demands $n^3$ the amount of memory which can be a big limiting
factor. A third method, propounded by \citep{interlacing}, sets out to
remove the dominant aliasing contributions from odd images (cf.
\Cref{fig:alias}) by interlacing two density grids that are shifted by half
the grid spacing with respect to each other. The authors claim that the
method, combined with a high order interpolation scheme such as PCS, can
reduce systematic biases from aliasing to levels below 0.01\% all the way
up to the Nyquist frequency for both power spectra and bispectra estimates.

Investigation of these effects in the case opf the bispectrum is beyond the current scope of this paper and we leave it to future work. For the remainder of the paper we will instead avoid the issues mentioned above by simply limiting ourselves to $k<k_{max}/3 = \frac{2}{3}k_{Ny}$.

\section{Results\label{work}}

\subsection{Comparison between Dark Matter Simulation Codes}
\label{sec:dark-matt-bisp}

As we enter the age of precision cosmology we are ever more reliant
on cosmological simulations to understand the dynamics of dark matter
and baryons. Numerical simulations act as a buffer between theory and
observation: we test cosmological models by matching simulation results
to observational data, and hence obtain constraints on cosmological
parameters. On the other hand since we only observe one universe we must
turn to simulations to understand the statistical significance of our
measurements. This is especially important with large galaxy data sets
coming from current and near-future surveys such as DES, LSST, Euclid
and DESI. While it would be ideal to use full N-body simulations to
generate these so-called mock catalogues for statistical analysis, their
huge demand for computational resources is prohibitive for generating
the large number of simulations required for accurate estimates of
covariances \citep{l-picola}. This has led to a proliferation of fast
dark matter simulation tools, such as PINOCCHIO
\citep{PINOCCHIO1,PINOCCHIO2}, Quick Particle Mesh (QPM) \citep{QPM},
Augmented Lagrangian Perturbation Theory (ALPT) \citep{ALPT} and the
Comoving Lagrangian Acceleration method (COLA) \citep{tassev}. While
the algorithms employed in all these methods are different, they all
share the common aim of speeding up the simulation process at the
expense of reduced accuracy at small scales.

These fast methods are typically bench-marked against N-body
codes with the power spectrum and other two-point clustering statistics,
as well as some form of three-point correlation, e.g. the reduced
bispectrum
\begin{align}
  Q(k_1,k_2,k_3) = \frac{B(k_1,k_2,k_3)}{P(k_1)P(k_2)+P(k_2)P(k_3)+P(k_3)P(k_1)}
\end{align}
in some restricted domain. With \MODALLSS{} we can incorporate full 
bispectrum estimation into the validation testing for these methods. The importance of these tests cannot
be underestimated: the analysis in \citep{tobias} has shown
that theoretical and numerical uncertainties can strongly influence the
extent to which observational data can be used to put constraints on
cosmological parameters and hence possibilities of detecting new physics.

As a proof of concept we have elected to test the bispectra of three
different fast dark matter methods, i.e. COLA, Particle-Mesh (PM) and
second-order Lagrangian perturbation theory (2LPT) \citep{2LPT}, against the
Tree-PM N-body code \GADGET{} at various redshifts. \LPICOLA{}
\citep{l-picola,scoccimarro} was used to generate the COLA, PM and 2LPT data
due to its versatility and massively parallel performance, and its ability to
generate and evolve the same 2LPT initial conditions used in our \GADGET{} runs.
This means that all final outputs share the same initial seed and random
phases, thus eliminating the need for cosmic variance considerations when
comparing them.

\subsubsection{Fast dark matter algorithms}
\label{sec:fast-dark-matter}

Here we briefly summarise the three algorithms we test in this paper.
For further details we refer the reader to relevant literature for
2LPT \citep{2LPT}, PM \citep{PM} and COLA \citep{l-picola,tassev}.

\paragraph{2LPT}

In Lagrangian perturbation theory (LPT) we track particles by their 
displacement $\bm{\psi}(\bm{q},t)$ from their initial position 
$\bm{q}$, i.e. $\bm{x}(t)=\bm{q}+\bm{\psi}(\bm{q},t)$, where $\bm{x}$ 
is the Eulerian position. First order in LPT leads to the well-known
Zeldovich Approximation (ZA), which is particularly useful due to its
analytical simplicity, and is often used to generate initial conditions
for numerical simulations. However as shown in \citep{2lpt2} 
2LPT is a superior method at limited additional computational cost,
and has since replaced ZA as the standard.

\paragraph{PM}

The PM algorithm speeds up the calculation of gravitational forces
though the use of a mesh: instead of summing all interactions between
all the particles, we calculate the density field on a grid and use
the Poisson equation to derive the gravitational potential in Fourier
space. This computation is sped up greatly with FFTs, and it is
straightforward to calculate the forces in real space at each grid
point with the gradient of the potential and an inverse-FFT. The force
on each particle is found by reversing the interpolation scheme used
to place the particles on the grid. Here we use \LPICOLA{}'s
implementation of the PM algorithm which is based on \texttt{PMCODE}
\citep{pmcode}.

\paragraph{COLA}

While the 2LPT produces excellent results at large scales, it quickly
becomes deficient going into smaller scales as it fails to capture
the full non-linearity of the system. The COLA algorithm is an
efficient extension of 2LPT, boasting both speed and accuracy by
trying to recover the residual Lagrangian displacement $\bm{\psi}_{res}$
between the 2LPT displacement and the full non-linear counterpart.
The extra computations rely on variables already calculated and stored,
such as the LPT and 2LPT displacements and the gravitational potential,
the last of which is provided by the PM method.

\subsubsection{Simulation Data}
\label{sim}

In order to probe a range of scales we have chosen two simulation
box sizes of $1280\,h^{-1}$ Mpc and $640\,h^{-1}$
Mpc\footnote{Corresponding to $k_F=0.005\,h\,\text{Mpc}^{-1}$
  and $k_{Ny}=5.0\,h\,\text{Mpc}^{-1}$, and
  $k_F=0.01\,h\,\text{Mpc}^{-1}$  and
  $k_{Ny}=10.0\,h\,\text{Mpc}^{-1}$ respectively}. The 2LPT Gaussian
initial conditions were generated using \LPICOLA{} at redshift $z_i=99$
to ensure the suppression of transients in power spectra and bispectra
estimates of our simulations \citep{transients}, with an input linear
power spectrum at redshift $z=0$ produced by \texttt{CAMB} \citep{CAMB}.
A PM grid size of $2048^3$ was then used to evolve the $2048^3$ particles
in each run where applicable. The fiducial cosmology is flat $\Lambda$CDM
with extended Planck 2015 cosmological parameters (TT,TE,EE+lowP+lensing+ext,
see \Cref{planck}). The expensive \GADGET{} run was completed on the
\texttt{COSMA} facility at Durham while the other codes and all subsequent
analysis was finished with the \texttt{COSMOS} supercomputer at Cambridge.
The small deviations in output redshifts between \GADGET{} and \LPICOLA{}
were corrected with the appropriate linear growth factor
\begin{align}
  \label{eq:D_1}
  D_1(a)=\frac{E(a)}{D_{1,0}}\int^a_0\frac{da'}{a'^3E^3(a')}
\end{align}
where 
\begin{align}
  \label{eq:E}
  E(a)=\frac{H(a)}{H_0}=\sqrt{\Omega_ma^{-3}+\Omega_\Lambda}
\end{align}
for a flat cosmology, and 
\begin{align}
  \label{eq:D_10}
  D_{1,0}=\int^1_0\frac{da'}{a'^3E^3(a')}
\end{align}
is introduced to normalise $D_1(z=0) = 1$.

\begin{table*}[!htb]
  \small
  \begin{subtable}{0.48\textwidth}
    \begin{tabular}{c|c|c}
      Description & Symbol  & Value \\[1ex] \hhline{=|=|=}
      \rule{0pt}{3ex}
      Hubble constant & $H_0$  & 67.74 $\text{km}\,\text{s}^{-1}$  \\
      Physical baryon density parameter & $\Omega_b h^2$  &  0.02230 \\
      Matter density parameter & $\Omega_m$   & 0.3089 \\
      Dark energy density parameter & $\Omega_{\Lambda}$   & 0.6911 \\
      Fluctuation amplitude at $8h^{-1}$ Mpc & $\sigma_8$  & 0.8196 \\ 
      Scalar spectral index & $n_s$  & 0.9667 \\
      Primordial amplitude & $10^9A_s$ & 2.142
    \end{tabular}
    \caption{Planck 2015 cosmological parameters (rightmost column of
      Table 4 in \citep{planck2015})}
  \end{subtable}
  \hspace{3ex}
  \begin{subtable}{0.45\textwidth}
    \begin{tabular}{c|c|c}
      Description & Symbol  & Value \\[1ex] \hhline{=|=|=}
      \rule{0pt}{3ex}
      Physical neutrino density parameter & $\Omega_{\nu} h^2$   & 0.000642 \\
      Number of effective neutrino species & $N_{eff}$   & 3.046 \\
      Curvature density parameter & $\Omega_{k}$   & 0.0000 \\
    \end{tabular}
    \caption{Extensions to base $\Lambda$CDM parameters (rightmost column of
      Table 5 in \citep{planck2015})}
  \end{subtable}
  \caption{For consistency between the Planck parameters and the
    \texttt{CAMB} output we incorporated one massive neutrino species
    with a small energy density. The lack of radiation and neutrino
    evolution in \LPICOLA{} and \GADGET{} has led us to define the
    matter power spectrum to consist only of cold dark matter and
    baryons, hence the raised value of $\sigma_8$ instead of the Planck
    value of 0.8159. The pivot scale for $n_s$ is 0.05
    $\text{Mpc}^{-1}$.}
  \label{planck}
\end{table*}

In addition to \Cref{planck}, the following are the key parameters we
used to generate the initial power spectrum and evolve the initial
conditions:

\paragraph{\texttt{CAMB}}
We use only cold dark matter (CDM) and baryons to define the matter
power spectrum and $\sigma_8$, i.e. {transfer\_power\_var = 8}.
The relevant neutrino parameters are
{massless\_neutrinos = 2.046} and {massive\_neutrinos = 1}.

\paragraph{\LPICOLA{}}
Three different logarithmic time steppings in $a$ were used to test
the accuracy of COLA: $\Delta(\ln a) = 0.01$ (the same time-stepping
we use for \GADGET{}), 0.046 and 0.23. They correspond to 460, 100
and 20 time-steps from $z=99$ to $z=0$ respectively.

\paragraph{\GADGET{}}
We used \citep{transients,crocce2006} as guides in setting the
parameters to ensure high numerical accuracy in our
simulations: {
  MaxRMSDisplacementFac = 0.1,
  ErrTolIntAccuracy = 0.01,
  MaxSizeTimestep = 0.01,
  ErrTolTheta = 0.2} and
{ErrTolForceAcc = 0.002}. A smoothing length of $0.05L/N$ where
$L$ is the simulation box size and $N=2048$ is the number of
particles per dimension was used.

\subsubsection{Simulation Power Spectra}

\begin{figure*}
  \begin{subfigure}[b]{0.49\textwidth}
    \includegraphics[width=\linewidth]{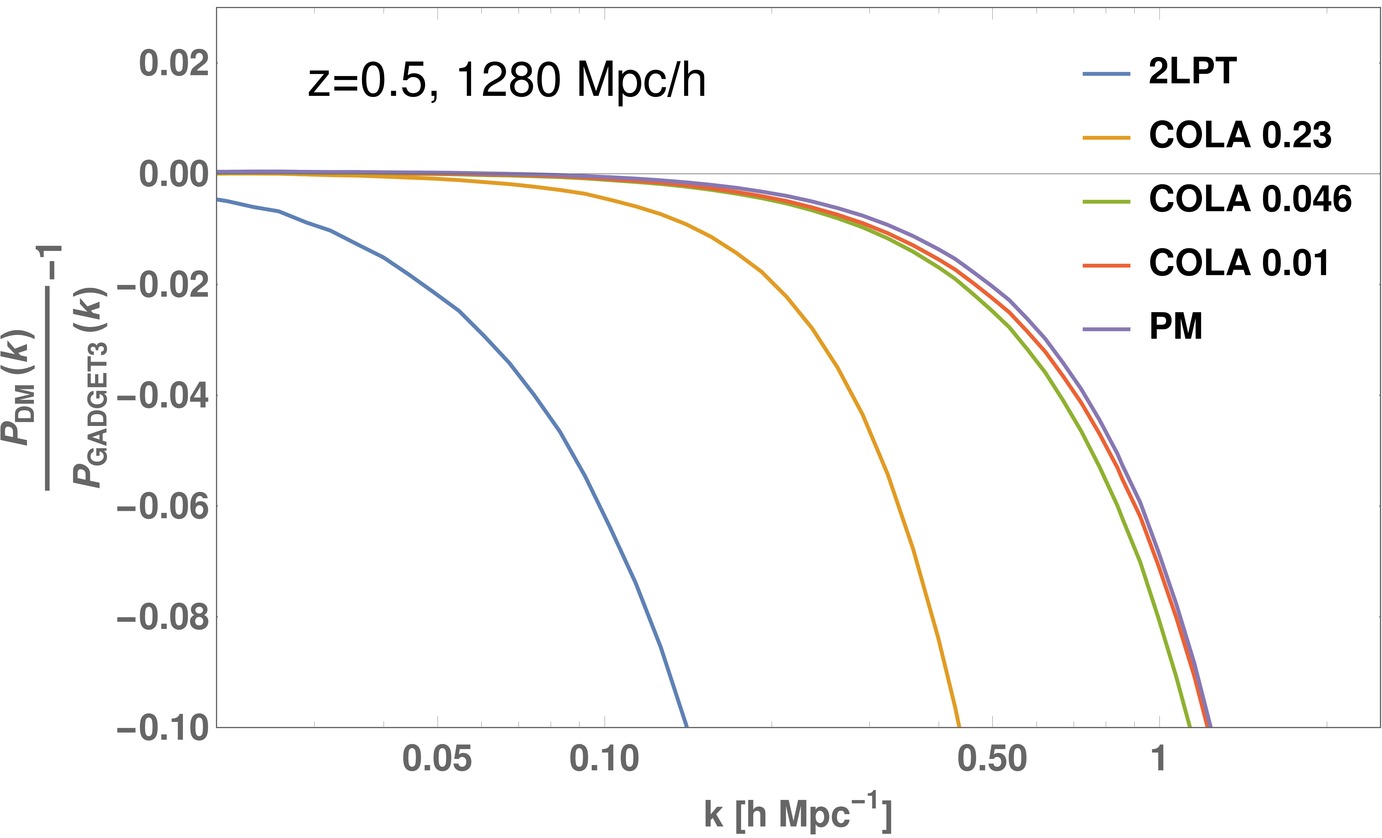}
  \end{subfigure}
  ~
  \begin{subfigure}[b]{0.49\textwidth}
    \includegraphics[width=\linewidth]{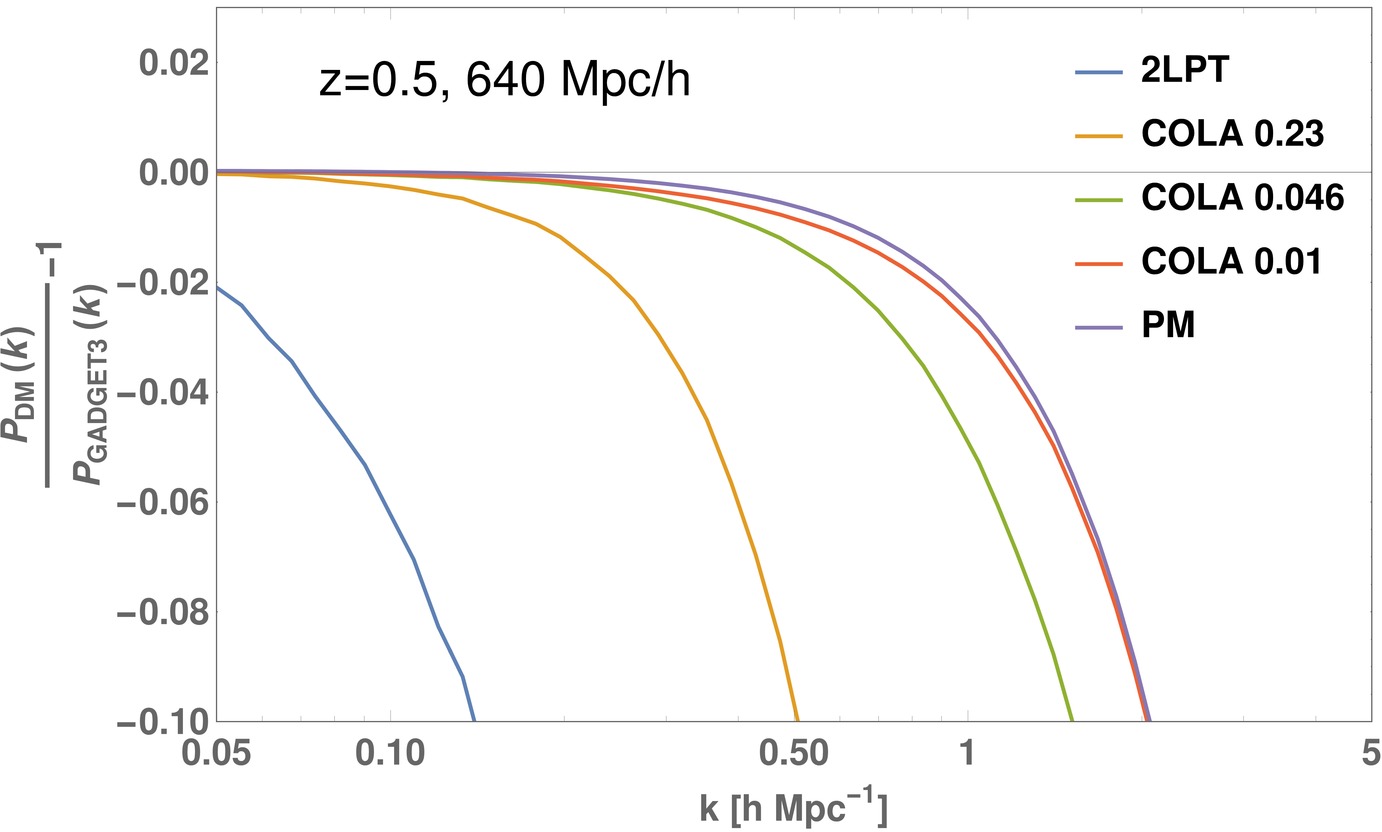}
  \end{subfigure}
  \caption{
    Ratio between the power spectra of the various fast dark matter
    codes and \GADGET{} for the 2 simulation boxes. All the power
    spectrum estimates were performed with \GADGET{}. The sub-par
    performance of 2LPT and COLA with a coarse time-stepping of
    $\Delta(\ln a) = 0.23$ is unsurprising, but the $\Delta(\ln a)_{0.046}$
    COLA simulation with compares quite favourably with PM and the
    $\Delta(\ln a)_{0.01}$ COLA with at a fraction of the computational
    cost. As noted by its authors the ability to reproduce the matter
    power spectrum at a reasonable accuracy but with reduced computational
    resources compared to conventional PM methods is the strength of
    the COLA method \citep{l-picola}.
  }
  \label{fig:powerspec}
\end{figure*}

We estimated the power spectra of our simulations with \GADGET{}.
To minimise errors coming from aliasing effects the power spectra
of each simulation was estimated three times: once with a $2048^3$
PM grid and two further times by `folding' \citep{fold} that grid
onto itself by factors of 2 and 4 respectively. The disadvantage
of this folding method is the reduction in the number of modes at
large scales leading to greater cosmic variance. We therefore
combine these three power spectra together to guarantee precision over the
entire $k$-ranges considered here. We did not observe shot noise in
the power spectra of the initial conditions, and due to large number
densities used did not find it necessary to correct for shot noise
in the simulation outputs (cf. \Cref{eqn:ps_shot}).

\Cref{fig:powerspec} shows the ratio between the power spectra of
the fast codes and \GADGET{} at redshift $z=0.5$. While 2LPT and
$\Delta(\ln a)_{0.23}$ COLA compare poorly to \GADGET{} as
expected, the power of the COLA algorithm to imitate the performance
of PM in fewer time-steps is shown by the $\Delta(\ln a)_{0.046}$
case. It should be noted that PM does perform slightly better than
COLA when the same number of time-steps are used.

\subsubsection{Simulation Bispectra}

\begin{figure*}
  \begin{subfigure}[b]{0.32\textwidth}
    \includegraphics[width=\linewidth]{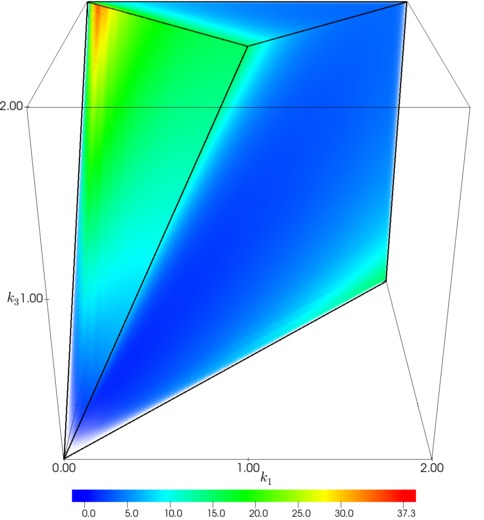}
    \caption{Redshift $z=9$}
  \end{subfigure}
  ~
  \begin{subfigure}[b]{0.32\textwidth}
    \includegraphics[width=\linewidth]{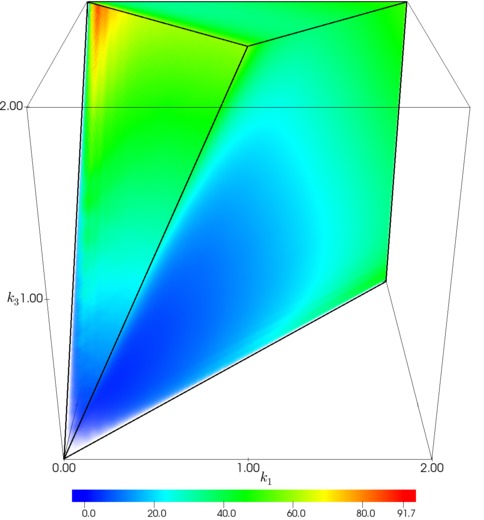}
    \caption{Redshift $z=3$}
  \end{subfigure}
  ~
  \begin{subfigure}[b]{0.32\textwidth}
    \includegraphics[width=\linewidth]{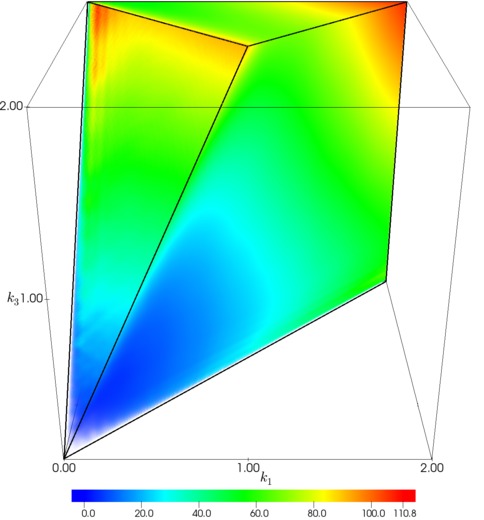}
    \caption{Redshift $z=2$}
  \end{subfigure}

  \begin{subfigure}[b]{0.32\textwidth}
    \includegraphics[width=\linewidth]{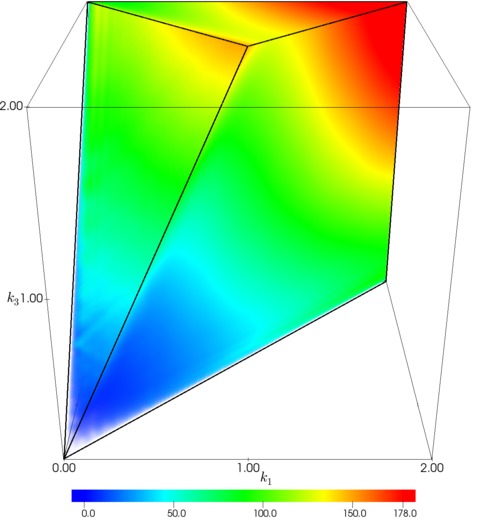}
    \caption{Redshift $z=1$}
  \end{subfigure}
  ~
  \begin{subfigure}[b]{0.32\textwidth}
    \includegraphics[width=\linewidth]{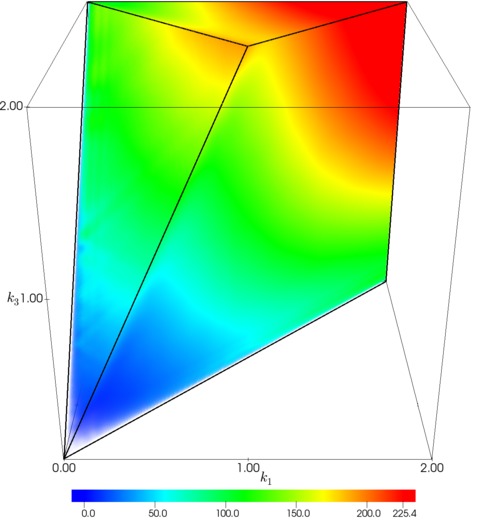}
    \caption{Redshift $z=0.5$}
  \end{subfigure}
  ~
  \begin{subfigure}[b]{0.32\textwidth}
    \includegraphics[width=\linewidth]{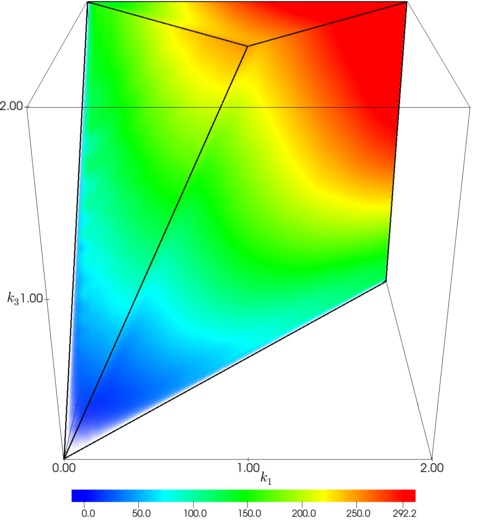}
    \caption{Redshift $z=0$}
  \end{subfigure}
  \caption{
    Redshift evolution of the estimated bispectra from a $1280\,h^{-1}$ Mpc
    \GADGET{} simulation, plotted up to $k_{max}=2.0\,h\,\text{Mpc}^{-1}$. This shows
    clearly how the flattened tree-level signal dominates the early time bispectra,
    but the constant shape brought about by the aggregation of matter takes over
    at late times. To emphasise this point we have scaled the maxima of the colour
    bars for redshifts $z=3\rightarrow0$ relative to redshift $z=9$ by the appropriate
    linear growth factor, $D_1(z)/D_1(z=9)$. The SN-weighted tree-level bispectrum grows
    as $D_1(z)$, and the saturation of the signal for redshifts $z=1,0.5,0$ demonstrate
    faster growth than that dictated by perturbation theory in the non-linear regime.
    It is remarkable that the only shape generated by the collapse of dark matter
    into halos is the constant shape. Therefore after $z~2$ we observe
    a steady growth in the strength of the signal but very little change in the
    bispectrum morphology.
  }
  \label{fig:gadget20}
\end{figure*}

\begin{figure*}
  \begin{subfigure}[b]{0.32\textwidth}
    \includegraphics[width=\linewidth]{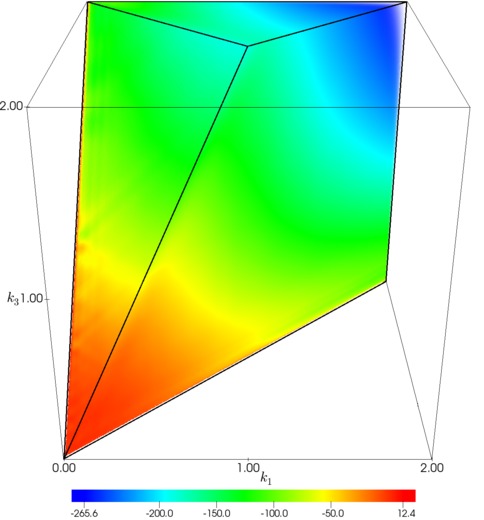}
    \caption{2LPT}
  \end{subfigure}
  ~
  \begin{subfigure}[b]{0.32\textwidth}
    \includegraphics[width=\linewidth]{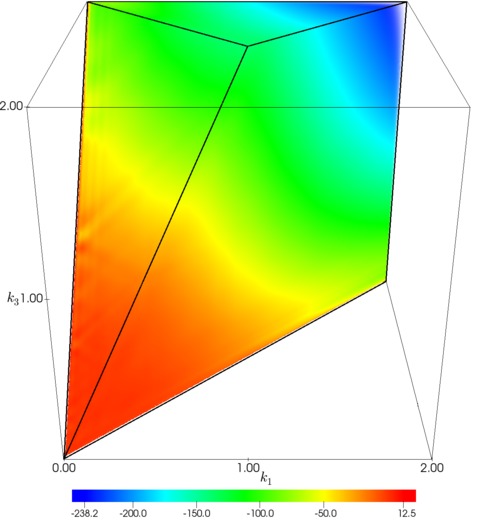}
    \caption{$\Delta(\ln a)_{0.23}$ COLA}
  \end{subfigure}
  ~
  \begin{subfigure}[b]{0.32\textwidth}
    \includegraphics[width=\linewidth]{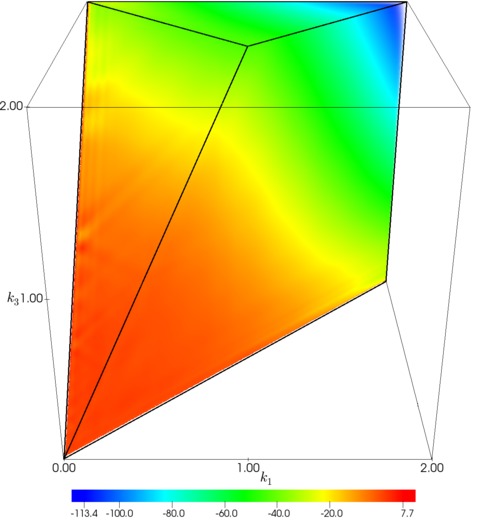}
    \caption{$\Delta(\ln a)_{0.01}$ COLA}
  \end{subfigure}

  \begin{subfigure}[b]{0.32\textwidth}
    \includegraphics[width=\linewidth]{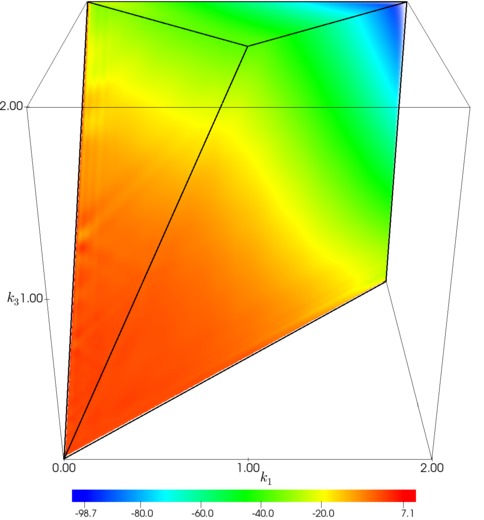}
    \caption{$\Delta(\ln a)_{0.046}$ COLA}
  \end{subfigure}
  ~
  \begin{subfigure}[b]{0.32\textwidth}
    \includegraphics[width=\linewidth]{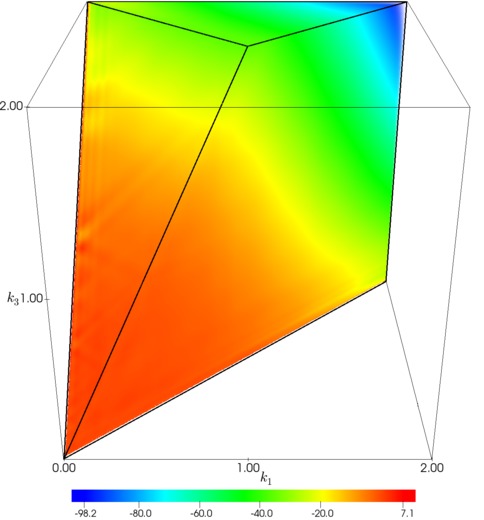}
    \caption{PM}
  \end{subfigure}
  ~
  \begin{subfigure}[b]{0.32\textwidth}
    \includegraphics[width=\linewidth]{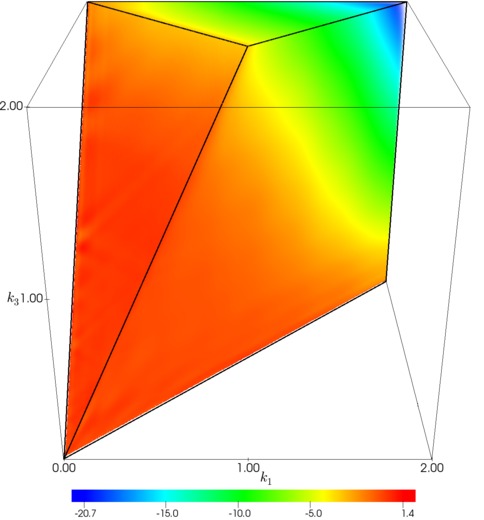}
    \caption{Boosted $\Delta(\ln a)_{0.01}$ COLA}
  \end{subfigure}
  \caption{
    Bispectrum residuals at redshift $z=0.5$ between the $1280\,h^{-1}$ Mpc fast
    dark matter and \GADGET{} simulations, plotted up to
    $k_{max}=2.0\,h\,\text{Mpc}^{-1}$. The lack of non-linear signal in the fast dark
    matter simulations is evident, leading to a deficient constant shape in their
    bispectra. 
  }
  \label{fig:residuals}
\end{figure*}

The density field of the simulations were first obtained via a
CIC mass assignment. A smoothed \GADGET{} power
spectrum\footnote{Smoothing is necessary at large scales
  where the lack of modes creates large variance in the estimated
  power spectrum, and was achieved by `dividing' out the variance:
  \begin{align}
    \hat{P}_{smooth}(k,z)=\hat{P}_{var}(k,z)\frac{P_L(k,z)}{\hat{P}_{IC}(k,z_i)}
    \times \frac{D_1^2(z_i)}{D_1^2(z)}
  \end{align}
  where $\hat{P}_{var}(k,z)$ is the original, variance-contaminated, power
  spectrum estimate, $P_L(k,z)$ is the linear power spectrum computed by
  \texttt{CAMB} at the same redshift and $\hat{P}_{IC}(k,z_i)$ is the
  estimated power spectrum of the initial conditions. This step is crucial
  for producing a smooth theoretical bispectrum since they often take the
  non-linear power spectrum as input, and a simulation power spectrum is
  usually chosen for that purpose to ensure fair comparison between simulation
  and theory (see \Cref{sub:sim_theory}).} at the appropriate redshifts were
used in the signal-to-noise weighting of the bispectrum (\Cref{SN}).

In \Cref{fig:gadget20} we show the estimated bispectra for the
$1280\,h^{-1}$ Mpc \GADGET{} simulations described in \Cref{sim} up to
$k_{max}=2.0\,h\,\text{Mpc}^{-1}$. We choose this resolution to best
highlight the transition from the tree-level dominant signal seen
in early redshifts to the strong constant shape presence induced by
non-linear gravitational evolution at late times. In particular we
see that this happens most prominently from redshift $z=3$, where
there is still some competition between the flattened and equilateral
signals, to redshift $z=2$, in which the constant shape has taken
over. This is one of the many advantages of estimating the full
bispectrum, as its morphology typically offers unique information
regarding structure formation that cannot be gained from the power
spectrum. Another point of note is that
the formation of dark matter halos through virialisation generates
only one bispectrum shape which is the constant shape, as evidenced
by the lack of change in the bispectrum past $z=2$ bar a growth in
signal strength. We also show the bispectrum residuals between the
fast dark matter codes and \GADGET{} in \Cref{fig:residuals}. The
inability of the fast codes to resolve small scale structure is
illustrated by the lack of constant shape signal in their bispectra.
These pictures agrees qualitatively with the power spectra results
in \Cref{fig:powerspec}.

\begin{figure*}
  \begin{subfigure}[b]{0.49\textwidth}
    \includegraphics[width=\linewidth]{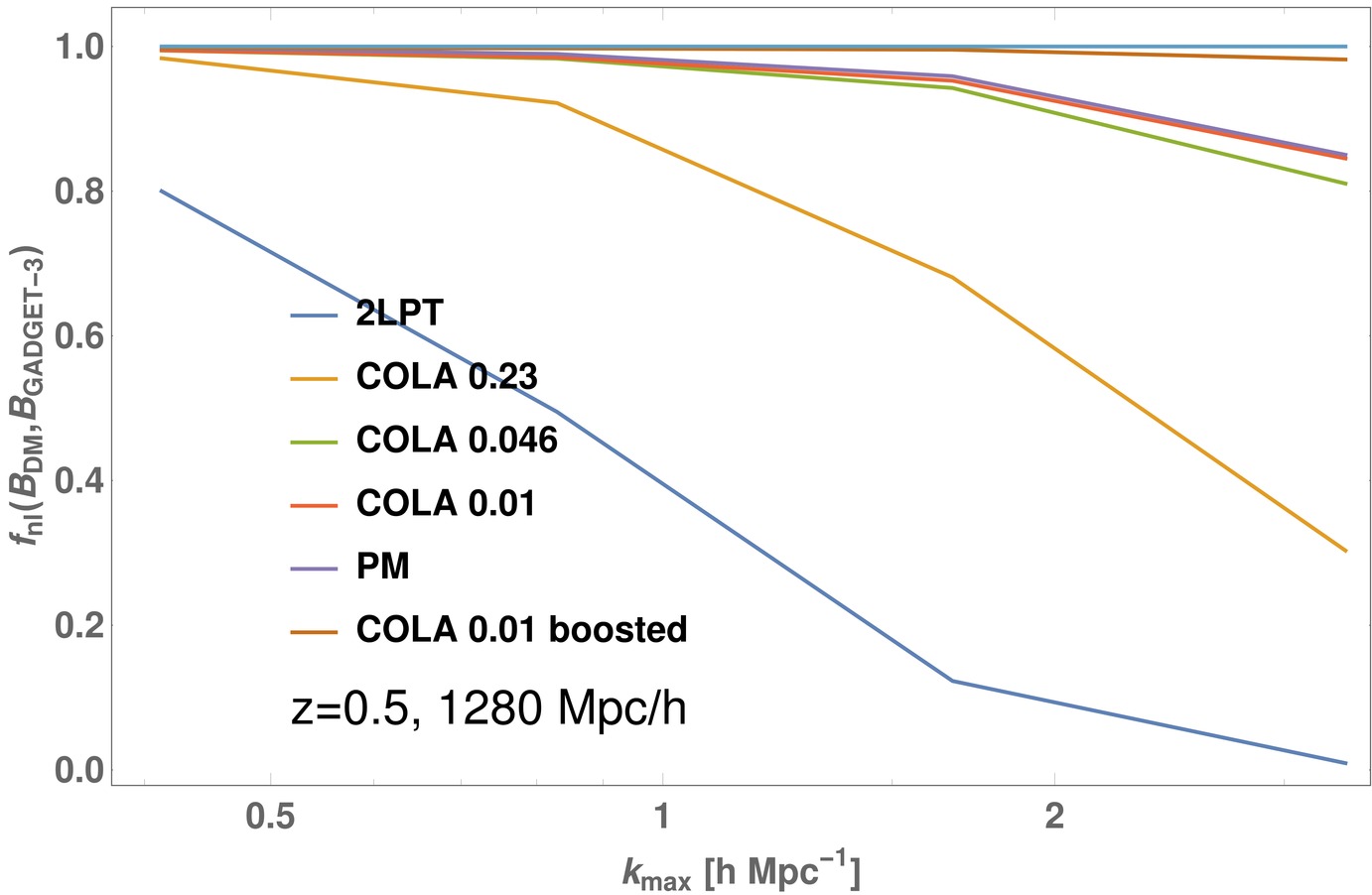}
    \caption{}
  \end{subfigure}
  ~
  \begin{subfigure}[b]{0.49\textwidth}
    \includegraphics[width=\linewidth]{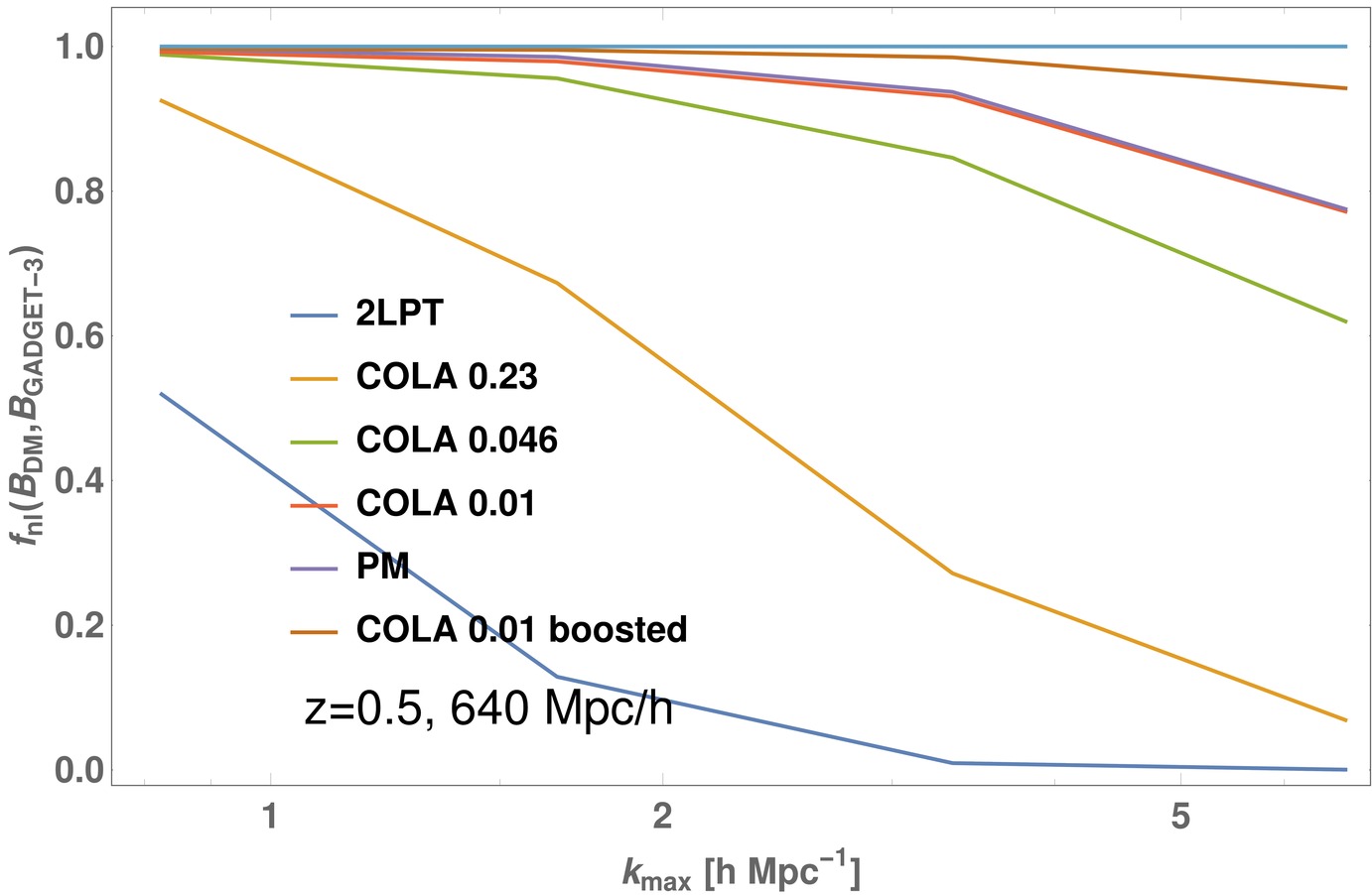}
    \caption{}
  \end{subfigure}
  \caption{
    $f_{nl}$ correlators between the bispectra of fast dark matter
    codes and \GADGET{}. The similarities of these plots to those in
    \Cref{fig:powerspec} is due to the power spectrum weighting
    present in inner products between bispectra (\Cref{inner_product}),
    thus a mismatch in power spectra naturally leads to discrepancies
    in bispectrum comparisons. This may suggest that the differences
    we see here are due to the power spectrum alone, but clearly this
    is not the case since the `boosted' COLA simulation has an identical
    power spectrum to \GADGET{} yet still suffers from a lack of bispectrum
    signal at small scales. However, the improved performance of the
    boosted COLA bispectrum demonstrates the effectiveness of the `boosting'
    method.
  } 
  \label{fig:bis_dm}
\end{figure*}

To make quantitative comparisons we invoke the correlators introduced
in \Cref{corr}. The $f_{nl}$ correlators of the fast dark matter codes
with \GADGET{}:
\begin{align}
  f_{nl}(\hat{B}_{\text{DM}},\hat{B}_{\GADGET{}})
  &=\frac{\sum_n\beta^R_{\text{DM},n}\beta^R_{\GADGET{},n}}
    {\sum_n(\beta^R_{\GADGET{},n})^2}
\end{align}
are shown in \Cref{fig:bis_dm}; we do not plot the
shape correlators as they only provide redundant information. The first
thing to note is a striking resemblance to the power spectra plots in
\Cref{fig:powerspec}, as the power spectrum enters the $f_{nl}$ correlator
through the weighted inner products between bispectra (\Cref{inner_product}).
Since we use the \GADGET{} power spectrum for the weighting, bispectra
comparisons will inevitably be biased by the lack of power in the fast
dark matter power spectra. To address this issue and show the differences
due to the bispectrum alone we propose boosting the power spectrum of the
fast code in Fourier space:
\begin{align}
  \label{eq:boost}
  \delta_{\text{DM}}(\mathbf{k})\rightarrow 
  \sqrt{\frac{\hat{P}_{\GADGET{}}(k)}{\hat{P}_{\text{DM}}(k)}}
  \delta_{\text{DM}}(\mathbf{k}).
\end{align}
The residuals between the boosted $1280\,h^{-1}$ Mpc $\Delta(\ln a)_{0.01}$ COLA
simulation and \GADGET{} is shown in \Cref{fig:residuals}, demonstrating
more than a 3x reduction in magnitude compared to the unboosted COLA and PM
runs. More quantitatively the boosted $\Delta(\ln a)_{0.01}$ COLA bispectra
also show much improved $f_{nl}$ correlation with \GADGET{} as seen in
\Cref{fig:bis_dm}. We therefore conclude this is an effective yet relatively
inexpensive\footnote{To obtain a smooth boosting factor in \Cref{eq:boost}
  we require one \GADGET{} and one fast code run that share the same initial
  conditions. This only has to be done once as the boosting factor should
  be reasonably realisation-independent.\label{smooth}} method to
improve the performance of fast simulation codes. Nevertheless a dip in
correlation at small scales remain after boosting which reflects that there is bispectrum information lost which is independent of the power spectrum. 

\subsection{Gaussian vs Non-Gaussian covariances
  \label{sub:cov}}

\begin{figure*}
  \includegraphics[width=0.55\textwidth]{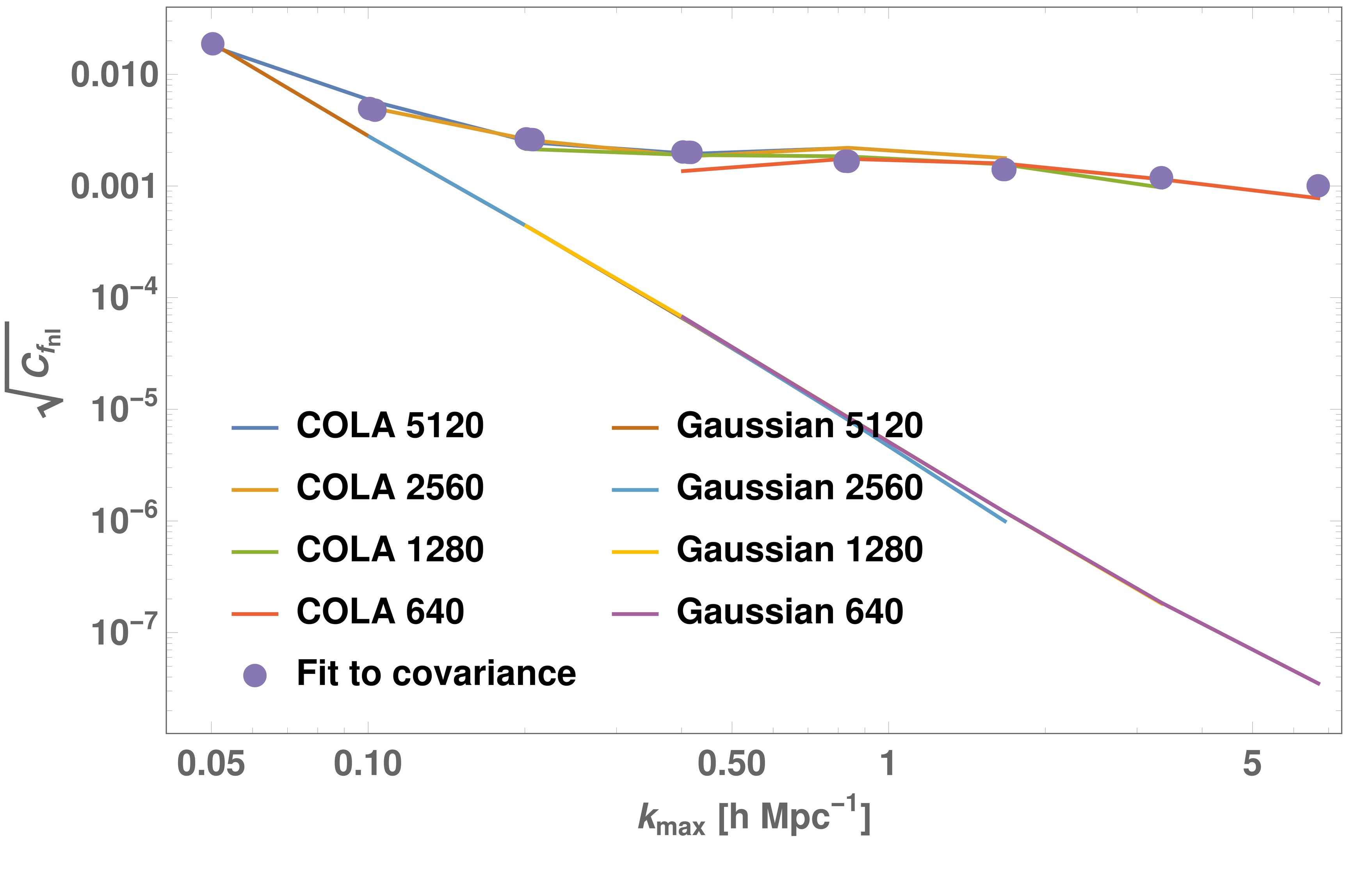}
  \caption{
    The full covariance of the $f_{nl}$ correlator estimated using 10
    COLA runs compared to the Gaussian contribution calculated
    using \Cref{eqn:fnl_cov} with the 3-shape model. The two begin
    to diverge significantly at $k\sim0.1\,h\,\text{Mpc}^{-1}$, signalling the
    dominance of non-Gaussian covariances. Since the covariance scales
    inversely as the cube of the box size, in order to combine the
    estimates from the different simulations we have re-scaled them
    accordingly against the $5120\,h^{-1}$ Mpc runs. The purple points are
    the best-fit to the full covariance with the function
    $f=A k^{-a}+B k^{-b}$ and the parameters
    $A=3.2477\times10^{-6},B=1.5871\times10^{-3},a=2.8339,b=0.2409$.
  }
  \label{fig:cov}
\end{figure*}

The extent to which we can put constraints on cosmological
parameters through the bispectrum is dependent on the covariance
of \MODALLSS{} estimator. To find the full covariance we first
average over 10 boosted COLA realisations for an estimate of the mean
bispectrum $\bar{\beta}$, then calculate the variance in
$f_{nl}(\beta,\bar{\beta})$ as an estimate for $C_{f_{nl}}$
(\Cref{eqn:fnl_cov}). The computational cost of COLA runs are
sufficiently low that additional to the $1280\,h^{-1}$ Mpc and
$640\,h^{-1}$ Mpc boxes we have also completed runs with $5120\,h^{-1}$ Mpc
and $2560\,h^{-1}$ Mpc box sizes\footnote{Since we
  do not have \GADGET{} simulations for the $5120\,h^{-1}$ Mpc and
  $2560\,h^{-1}$ Mpc boxes we estimate the dark matter power
  spectrum by boosting a COLA run as follows. First we repeat the
  smoothing procedure detailed in \Cref{smooth} to obtain a smoothed
  COLA power spectrum, then estimate the appropriate boosting factor
  with the $1280\,h^{-1}$ Mpc one.}, so that we can
explore the regime where Gaussian covariances dominate. We have made
a least-squares fit of the full covariance $\sqrt{C_{f_{nl}}}$ with the
\texttt{curve\_fit} algorithm in \texttt{Scipy}, using the default
Levenberg-Marquardt method \citep{lm}. We model the full covariance
a sum of two power laws: $f=A k^{-a}+B k^{-b}$, which represents the
Gaussian and non-Gaussian contributions respectively. The best-fit
is obtained using the following values for these parameters:
$A=4.6480\times10^{-6},B=1.0900\times10^{-3},a=2.5978,b=0.2315$.

Our estimates are shown in \Cref{fig:cov} where we also plot the
Gaussian covariances calculated using \Cref{eqn:fnl_cov} with the
3-shape model $\alpha^R_n$ coefficients. It is clear that while the
Gaussian covariance continues to diminish in the non-linear regime
due to more modes being available, the non-Gaussian covariance
starts to dominate at $k\sim0.1\,h\,\text{Mpc}^{-1}$ and then asymptotes
towards$\sim0.1\%$. This has important consequences on e.g. Fisher matrix
forecasts, especially if non-Gaussian covariances are not taken
in account which could strongly skew theoretical error estimates.
While the combination of power spectrum and bispectrum is superior
to using the power spectrum alone, the improvement may not be as
significant as one might have hoped due to this plateauing in the
bispectrum covariance.

\subsection{Comparison between Dark Matter Simulations and Theory
  \label{sub:sim_theory}}

\begin{figure*}
  \begin{subfigure}[b]{0.36\textwidth}
    \includegraphics[width=\linewidth]{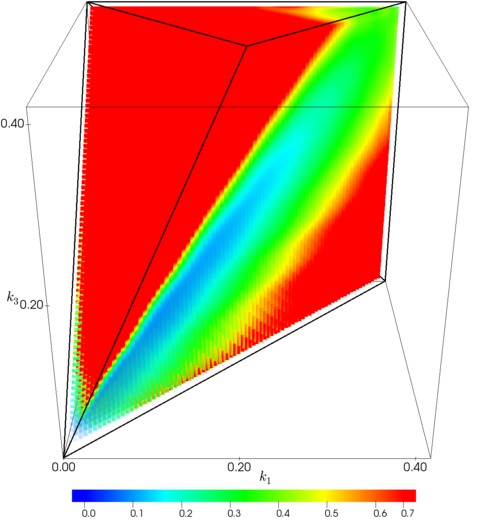}
    \caption{10000 averaged $256^3$ COLA runs}
  \end{subfigure}
  ~
  \begin{subfigure}[b]{0.36\textwidth}
    \includegraphics[width=\linewidth]{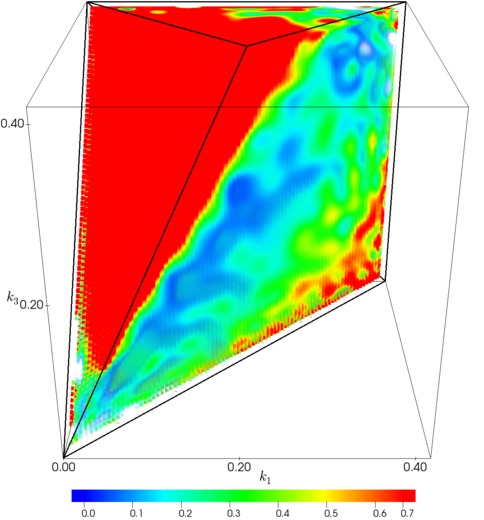}
    \caption{10 averaged $512^3$ COLA runs}
  \end{subfigure}

  \begin{subfigure}[b]{0.36\textwidth}
    \includegraphics[width=\linewidth]{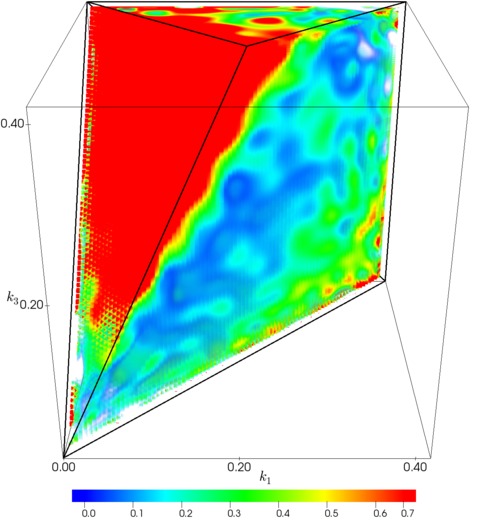}
    \caption{10 averaged $1024^3$ COLA runs}
  \end{subfigure}
  ~
  \begin{subfigure}[b]{0.36\textwidth}
    \includegraphics[width=\linewidth]{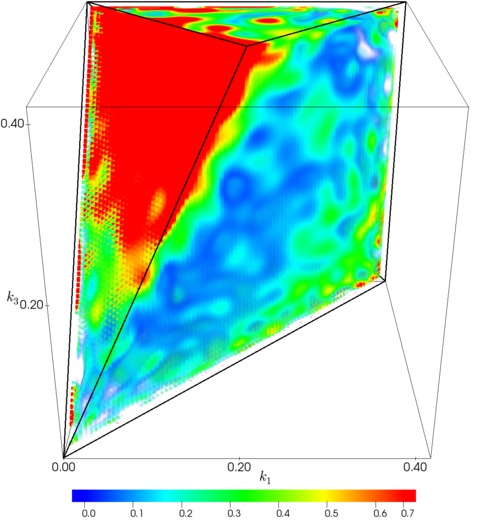}
    \caption{10 averaged $2048^3$ COLA runs}
  \end{subfigure}

  \begin{subfigure}[b]{0.36\textwidth}
    \includegraphics[width=\linewidth]{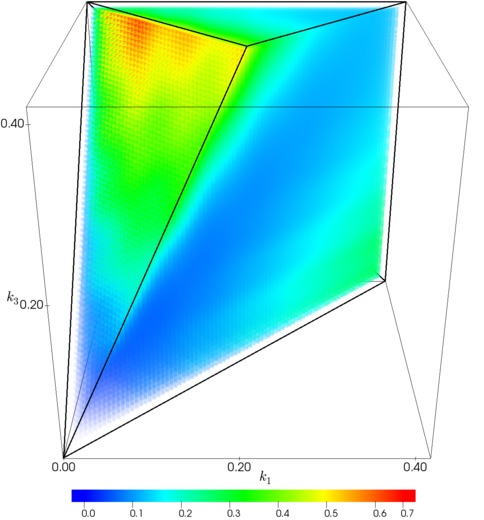}
    \caption{Tree-level bispectrum}
  \end{subfigure}
  \caption{
    The reconstructed bispectra from averaged 2LPT IC, and the
    desired signal, i.e. the tree-level bispectrum, plotted
    up to $k_{max}=0.41\,h\,\text{Mpc}^{-1}$ . The colour scale
    is chosen to show the full range of the tree-level bispectrum,
    leading to significant saturation for the simulation bispectra.
    With increeasing FFT grid size the IC bispectrum morphology
    approaches the theoretical one, but the amplitude remains
    grossly inflated.
  }
  \label{fig:z99}
\end{figure*}

The development of the \MODALLSS{} toolkit is to allow
straightforward comparisons between bispectra, either
from simulations, observational data, or theory. In that
cause we first test our method by estimating the bispectrum
of 2LPT initial conditions (IC) generated by \LPICOLA{},
using the fact that it should reproduce the tree-level
bispectrum.
We used a range of grid sizes to
generate the initial conditions, and to combat cosmic
variance at large scales we average over multiple
realisations. Similar to the test in \Cref{subsec:num}
we use \Cref{eqn:shape,fnl} to find that  
\begin{align}
  \mathcal{S}_{\beta,\alpha}&=\frac{\sum_n\beta^R_n\alpha^R_n}
                              {\sqrt{\sum_n(\beta^R_n)^2\sum_n(\alpha^R_n)^2}},
                              \nonumber \\
  f_{nl}^{\beta,\alpha}&=\frac{\sum_n\beta^R_n\alpha^R_n}{\sum_n(\alpha^R_n)^2}.
\end{align}
The correlators between the averaged runs and the tree-level
bispectrum are shown in \Cref{tab:z99}, and we also plot
the reconstructed simulation bispectra in \Cref{fig:z99}.

\begin{table*}
  \begin{tabular}{c|c|c|c|c|c|c|c|c}
    \multirow{2}{*}[-1ex]{$k_{max}\,(h\,\text{Mpc}^{-1})$}
    & \multicolumn{2}{c|}{10000 averaged $256^3$ runs} & \multicolumn{2}{c|}{10 averaged $512^3$ runs} & \multicolumn{2}{c|}{10 averaged $1024^3$ runs} & \multicolumn{2}{c}{10 averaged $2048^3$ runs} \\[1ex] \cline{2-9}
    \rule{0pt}{3ex} & $\mathcal{S}_{\beta,\alpha}$ & $f_{nl}^{\beta,\alpha}$ & $\mathcal{S}_{\beta,\alpha}$ & $f_{nl}^{\beta,\alpha}$ & $\mathcal{S}_{\beta,\alpha}$ & $f_{nl}^{\beta,\alpha}$ & $\mathcal{S}_{\beta,\alpha}$ & $f_{nl}^{\beta,\alpha}$ \\[1ex] \hhline{=|=|=|=|=|=|=|=|=}
    \rule{0pt}{3ex}0.4123 & 0.9300 & 11.12 & 0.9339 & 5.603 & 0.9469 & 3.072 & 0.9583 & 1.830 \\
    0.8296 & - & - & 0.9501 & 6.076 & 0.9613 & 3.228 & 0.9794 & 1.895 \\
    1.6690 & - & - & - & - & 0.9696 & 3.442 & 0.9830 & 1.950 \\
    3.3429 & - & - & - & - & - & - & 0.9870 & 2.064 \\[1ex] \hhline{-|-|-|-|-|-|-|-|-}
  \end{tabular}
  \caption{
    Comparisons between averaged 2LPT IC bispectra and the tree-level
    bispectrum, where the IC have been generated with different
    grid sizes. The poor shape correlation at low $k$ cannot be caused
    by cosmic variance alone due to the high number of runs used, and
    a clear trend of scale dependence can be seen in the $f_{nl}$ correlator.
  }
  \label{tab:z99}
\end{table*}

The poor shape correlation ($<95\%$) for low $k$ is a
strong indication that something is wrong with the IC,
but cosmic variance cannot be the only source of error
since a very large number of runs were used in the
$256^3$ case. We have also ruled out shot noise since
it is not the correct shape. Moreover the large amplitude
of the simulation bispectra leads to an inflated $f_{nl}$
in a way that is dependent on the size of the FFT grid
used. We propose this failure of the IC code to reproduce
the correct bispectrum is due to both (i) transients, as
discussed in \citep{transients,semiclassical}, and (ii)
grid effects. Similar problems were observed in
\citep{glass}, and subsequently alleviated by the use of
glass initial conditions. With more sophisticated
technology at hand now we shall investigate this further
in the near future.

\begin{figure*}
  \begin{subfigure}[b]{0.49\textwidth}
    \includegraphics[width=\linewidth]{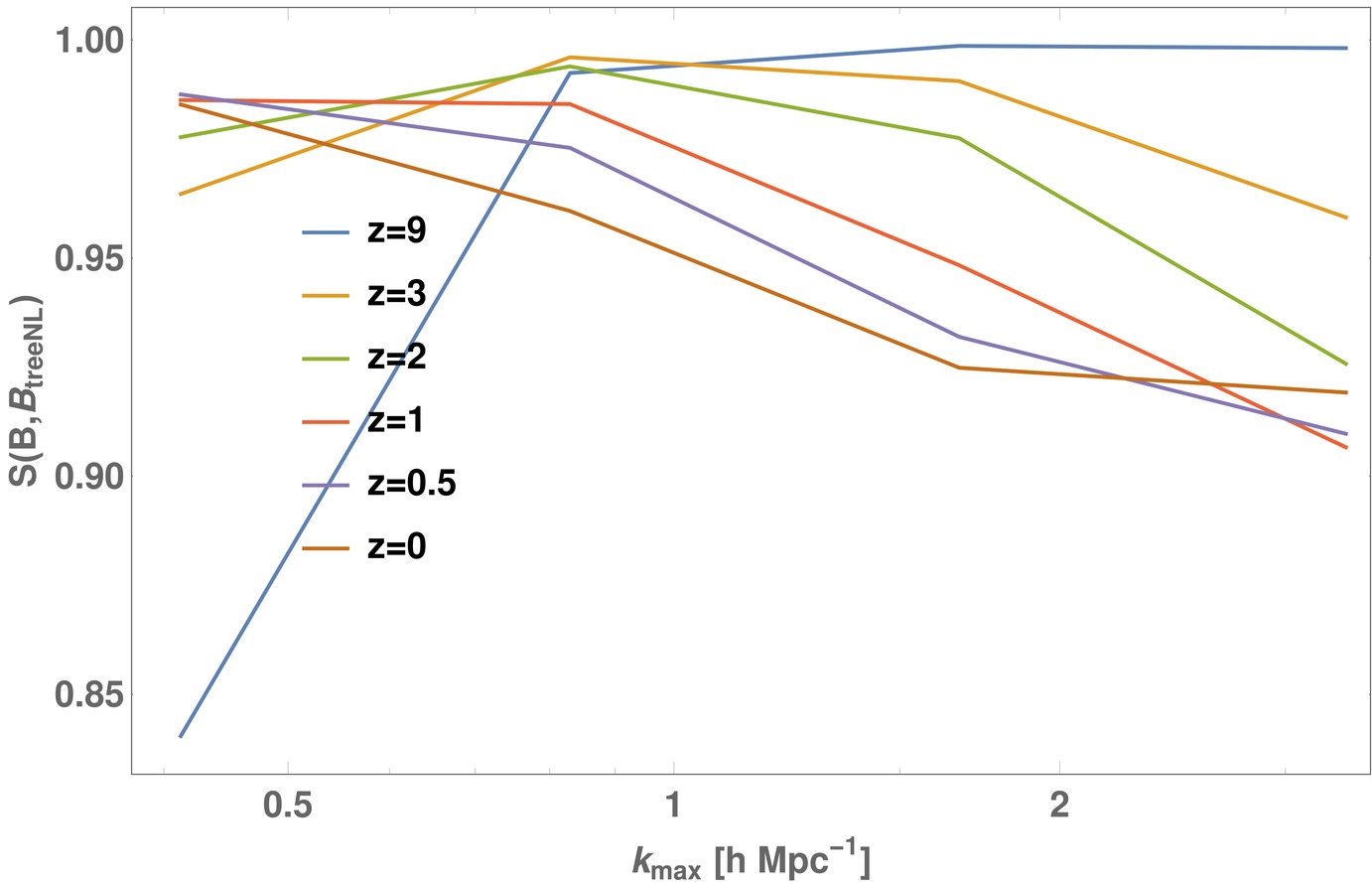}
    \caption{Shape correlator}
  \end{subfigure}
  ~
  \begin{subfigure}[b]{0.48\textwidth}
    \includegraphics[width=\linewidth]{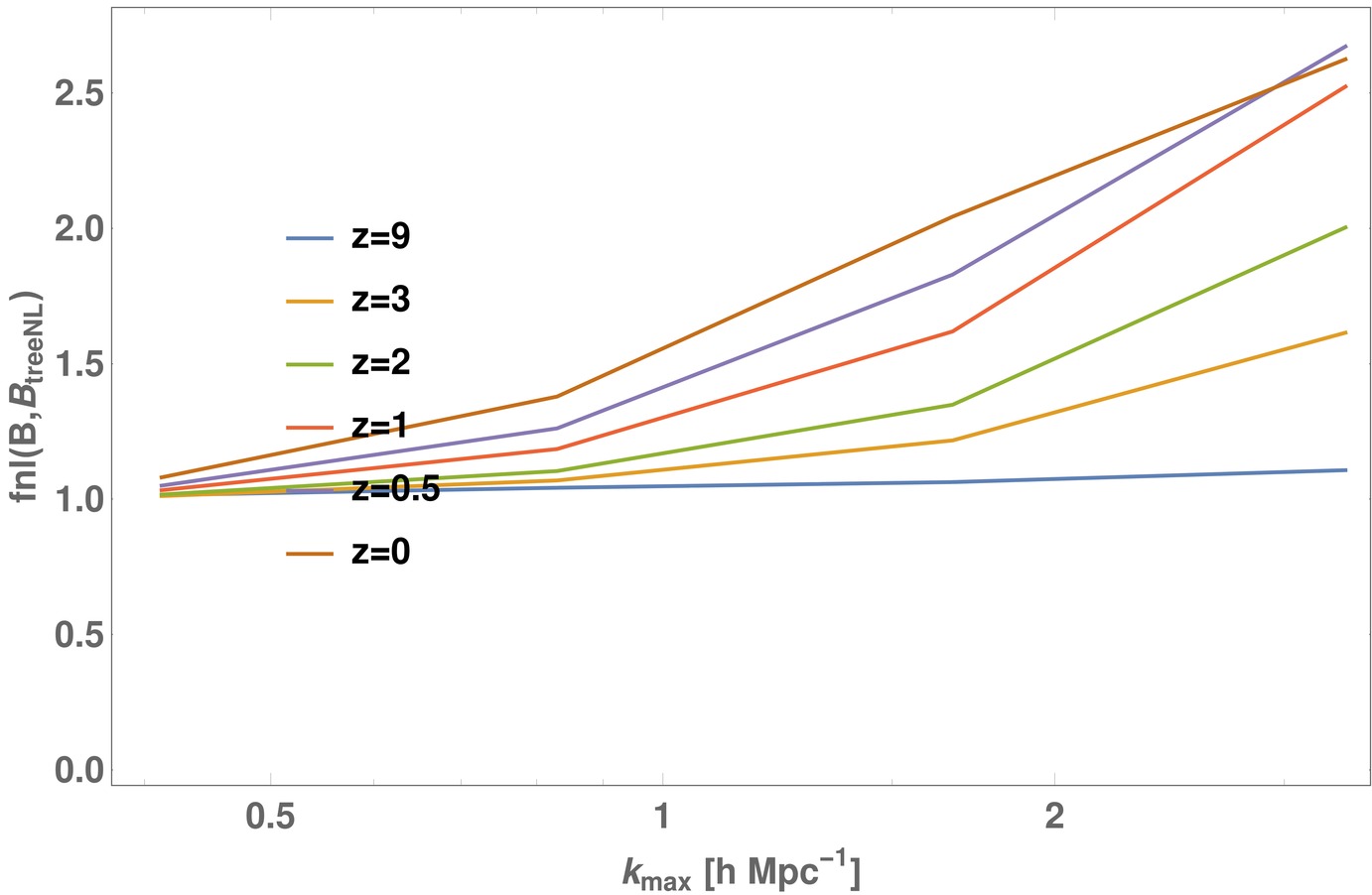}
    \caption{$f_{nl}$ correlator}
  \end{subfigure}
  \caption{
    Correlators between a $1280\,h^{-1}$ Mpc \GADGET{}
    simulation and the tree-level bispectrum at various
    redshifts. Transients is the likely suspect for the
    especially poor shape correlation at low $k$ at
    redshift $z=9$.  
  }
  \label{fig:bis_z}
\end{figure*}

Another obvious candidate for our tests is the redshift evolution
of a simulation. It is natural to expect a faithful adherence
to the tree-level bispectrum at earlier times, even at high $k$.
With the passage of time, and hence gravitational collapse, the
non-linear signal will eventually dominate at small scales,
leading to significant deviations from perturbation theory. This
is shown clearly in \Cref{fig:bis_z}, where we compare the
$1280\,h^{-1}$ Mpc \GADGET{} simulation to the tree-level bispectrum.
As the smallest FFT grid we use in bispectrum estimation is $256^3$
we unfortunately miss out on the observationally relevant scales
of $k\sim0.1\,h\,\text{Mpc}^{-1}$, but our efforts to recover the
tree-level bispectrum in larger simulations (i.e. 1280 and
$2560\,h^{-1}$ Mpc) have failed, probably due to the same issues
we encountered when we tried to extract the initial conditions
bispectra. Transients are the most likely explanation for the
poor shape correlation at low $k$, especially at redshift $z=9$,
as the correlation improves with time when these modes decay
away.

\section{Conclusions \label{sec:conclusions}}

In this paper we present the newly improved \MODALLSS{}
code for efficiently computing the bispectrum of any 3D input density
field. This code enables us to do high precision
analysis with the dark matter bispectrum from large N-body simulations or faster alternative codes, 
and to make detailed quantitative comparisons between theory and
simulations.  By exploiting highly optimised numerical libraries, we were able to incorporate 1000 separable modes in the bispectrum analysis (relative to 50 modes previously \citep{MODAL-LSS}), also including specially tailored modes to accurately recover the tree-level bispectrum.  This allows convergence to a much broader range of nonlinear gravitational and primordial bispectra and makes generic non-Gaussian searches feasible in huge future galaxy surveys.  

First,  we have addressed a few common areas where errors
in the \MODALLSS{} estimator can be significant, i.e. shot noise,
the covariance of the estimator, and aliasing effects
from using FFTs. Shot noise in the bispectrum is well-known
and required little discussion. The full covariance of the
\MODALLSS{} estimator was derived for the first time, but
the non-Gaussian contributions to the covariance appear to be analytically
intractable, even with the separable modal expansion, so we can
only estimate the Gaussian covariance, and we must tackle the problem numerically. 
While others have investigated of discrete FFT methods on
bispectrum estimation, we find that contrary to other
estimators the \MODALLSS{} estimator breaks down at the
same frequency as power spectra estimators, i.e. at the
Nyquist frequency $k_{Ny}$, rather than at $\frac{2}{3}k_{Ny}$.
We believe this is not a consequence of the \MODALLSS{}
method but rather a general result in bispectrum estimation
since the aliasing effects come from the discrete sampling
of the density field and not the use of FFTs itself.

With many large galaxy data-sets on the horizon, there is a pressing need
for fast mock catalogue codes. While these
fast codes are designed to only replicate the accuracy of
N-body codes at large scales without resolving finer structure,
we have found a simple and effective way to enhance their
performance. A comparison between the 2LPT, PM and COLA
algorithms against \GADGET{} shows 2LPT is deficient in
both the power spectrum and bispectrum, while the COLA
algorithm is successful in giving comparable performance to
PM with fewer time-steps. Noting that the drop in bispectrum
at large scales might be influenced by the power spectrum,
we attempted to rectify this by boosting the power spectrum
of the COLA simulation and saw a significant reduction in the power lost.

Finally we address the theoretical modelling of the dark
matter bispectrum by examining the full covariance of the
\MODALLSS{} estimator, showing that non-Gaussian contributions
begin to dominate at $k\sim0.1\,h\,\text{Mpc}^{-1}$ and plateaus
towards $\sim0.1\%$. This is a significant adjustment as the
non-Gaussian covariance is difficult to calculate even
numerically, leading to the use of only the Gaussian covariance
in most Fisher matrix forecasts. In principle, this will lead to gross underestimates
of the theoretical error and thus the ability to put constraints on
cosmological parameters. To show the power of the \MODALLSS{}
method in testing theoretical models against simulations
we have compared (i) 2LPT initial conditions against the
tree-level bispectrum, and (ii) a \GADGET{} simulation
against the tree-level bispectrum at various redshifts.
We have observed problematic transient modes and grid effects
that affect the initial conditions, where 
the tree-level bispectrum should be recovered after averaging over many
realisations. These effects propagate and persist to late times on the largest scales,
as shown in a \GADGET{} comparison, and must be addressed in the initial conditions. 

\section{Acknowledgements}
\label{sec:acknowledgements}

We are especially grateful to Tobias Baldauf for many enlightening conversations and for his
frequent useful advice.   We are also very grateful for discussions with Marc Manera
and Marcel Schimmittfull, who pioneered the first MODAL approach to the LSS bispectrum \citep{MODAL-LSS}.  Kacper Kornet and Juha Jaykka provided invaluable technical
support for \MODALLSS{} code optimisation and dealing with this large in-memory pipeline.  JRF and EPS acknowledge support from STFC Consolidated Grant ST/P000673/1.

This work was undertaken on the COSMOS Shared Memory system at
DAMTP, University of Cambridge operated on behalf of the STFC
DiRAC HPC Facility. This equipment is funded by BIS National
E-infrastructure capital grant ST/J005673/1 and STFC grants
ST/H008586/1, ST/K00333X/1.

This work used the COSMA Data Centric system at Durham University, 
operated by the Institute for Computational Cosmology on behalf of 
the STFC DiRAC HPC Facility (www.dirac.ac.uk). This equipment was 
funded by a BIS National E-infrastructure capital grant ST/K00042X/1, 
DiRAC Operations grant ST/K003267/1 and Durham University. DiRAC is 
part of the National E-Infrastructure.

\appendix

\section{Calculation of  $\gamma_{nm}$ with FFTs
  \label{appendix_C}}

\begin{figure*}
  \begin{subfigure}[b]{0.49\textwidth}
    \includegraphics[width=\textwidth]{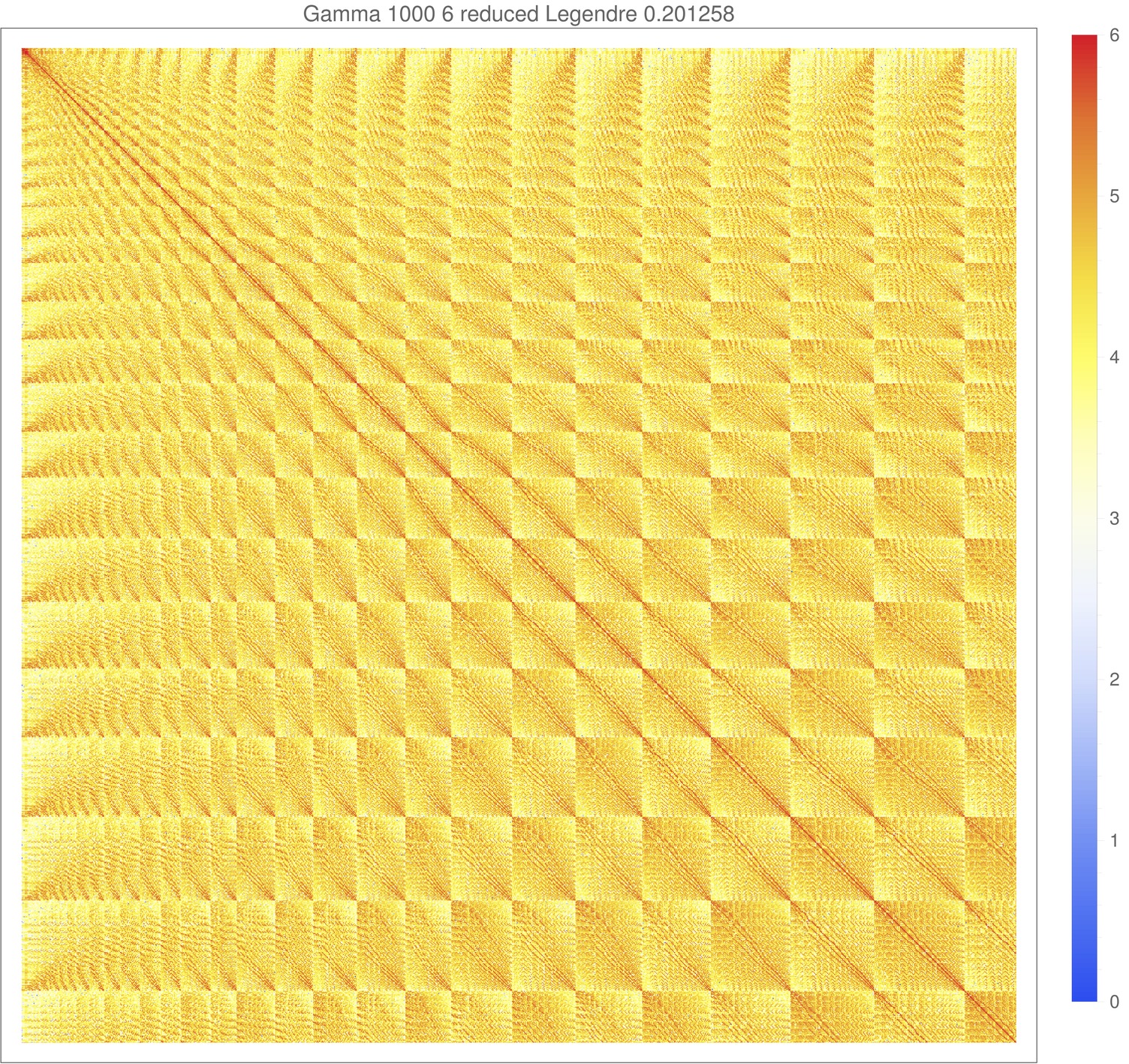} 
    \caption{$\gamma_{nm}$ calculated directly on the tetrapyd for 1000
      shifted Legendre polynomials using 42 grid points in each
      dimension. The abundance of off-diagonal features demonstrate
      the lack of orthogonality between modes on the tetrapyd,
      especially for high $n$.}
    \label{gamma_tetrapyd}
  \end{subfigure}
  ~
  \begin{subfigure}[b]{0.49\textwidth}
    \includegraphics[width=\textwidth]{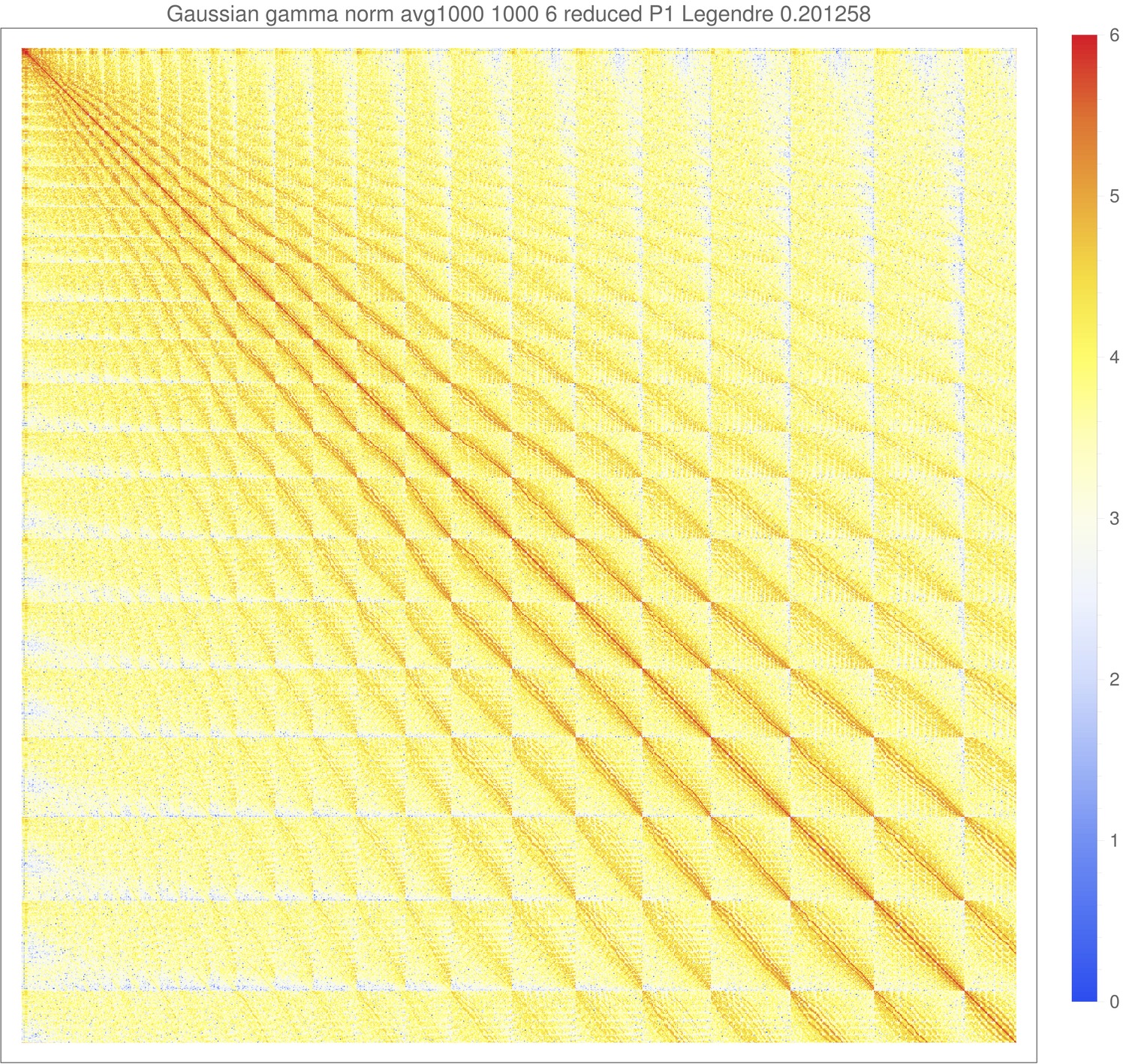}
    \caption{The same calculation but using FFTs over $128^3$ grids
      in real space, but keeping the same $k_{max}$. While this
      qualitatively agrees with the plot to the left, the off-diagonal
      elements differ wildly quantitatively, demonstrating much
      stronger orthogonality between the modes.}
    \label{gamma_fft}
  \end{subfigure}
  \caption{To better highlight the off-diagonal structure, in both of
    these figures we plot
    $\log_{10}\left|\frac{\gamma_{nm}}{\sqrt{\gamma_{nn}\gamma_{mm}}}\times10^6\right|$,
    such that the diagonal is always 6, and limit the plot range to $[0,6]$.
    This is important since we need the inverse of $\gamma_{nm}$ to rotate
    the \MODALLSS{} coefficients in the $Q$ basis to the $R$ basis, and even
    small numerical differences in the off-diagonal elements can create large
    deviations in the final result.
  }
  \label{fig:gamma}
\end{figure*}

As mentioned in the main text the integral
\begin{align}
  \gamma_{nm}\equiv\frac{V}{\pi}\int_{\mathcal{V}_B}dV_kQ_nQ_m
\end{align}
can be evaluated in two ways. The first is by direct
integration on the tetrapydal domain which gives the
most accurate answer. In \Cref{gamma_tetrapyd} we show
$\gamma_{nm}$ calculated in this way for 1000 modes using
shifted Legendre polynomials and 42 grid points in each
dimension.

Alternatively this can be done with the use of FFTs. It can
be shown that for any function $F(k_1,k_2,k_3)$ this expression
holds:
\begin{align}
  &\int\frac{d^3 k_1}{(2\pi)^3}\frac{d^3 k_2}{(2\pi)^3}
    \frac{d^3 k_3}{(2\pi)^3}(2\pi)^6\delta^2_D
    \left(\mathbf{k}_1+\mathbf{k}_2+\mathbf{k}_3\right)F \nonumber \\
  &\quad=\frac{V}{8\pi^4}\int_{\mathcal{V}_B}dk_1dk_2dk_3\,k_1k_2k_3F.
    \label{integrals}
\end{align}
Therefore we can write down an expression
for $\gamma_{nm}$ in terms of inverse Fourier Transforms:
\begin{align}
  \label{eqn:gamma_ffts}
  & \gamma_{nm} \nonumber \\
  & = (2\pi)^9\int_{\mathbf{k}_1,\mathbf{k}_2,\mathbf{k}_3}
    \delta_D(\mathbf{k}_1+\mathbf{k}_2+\mathbf{k}_3)
    \frac{Q_nQ_m}{k_1k_2k_3}
    \nonumber \\
  & = (2\pi)^6\int d^3 x\int\frac{\prod_i d^3k_i}{(2\pi)^9}
    e^{i(\mathbf{k}_1+\mathbf{k}_2+\mathbf{k}_3)\cdot\mathbf{x}}
    \frac{Q_nQ_m}{k_1k_2k_3}
    \nonumber \\
  & = \frac{(2\pi)^6}{6}\int \left[M_{r_1r_2}(\mathbf{x})
    M_{s_1s_2}(\mathbf{x})M_{t_1t_2}(\mathbf{x})+
    5\,\text{perms}\right]d^3 x,
\end{align}
where we have suppressed the arguments $(\frac{k_1}{k_{max}},
\frac{k_2}{k_{max}},\frac{k_3}{k_{max}})$ of $Q_n$ and $Q_m$
for brevity, and introduce the integrals
\begin{align}
  \label{eqn:M_r1r2}
  M_{r_1r_2}(\mathbf{x}) = \int\frac{d^3k}{(2\pi)^3}
  \frac{1}{k}q_{r_1}(k/k_{max})q_{r_2}(k/k_{max})
  e^{i\mathbf{k}\cdot\mathbf{x}}
\end{align}
resulting from the product $Q_nQ_m$. For $n\equiv\{r_1,s_1,t_1\}$
and $m\equiv\{r_2,s_2,t_2\}$ this product produces 36 terms, but
only 6 unique combinations, i.e.
\begin{itemize}
\item $(r_1r_2)(s_1s_2)(t_1t_2)$
\item $(r_1s_2)(s_1t_2)(t_1r_2)$
\item $(r_1t_2)(s_1r_2)(t_1s_2)$
\item $(r_1r_2)(s_1t_2)(t_1s_2)$
\item $(r_1s_2)(s_1r_2)(t_1t_2)$
\item $(r_1t_2)(s_1s_2)(t_1r_2)$,
\end{itemize}
hence the 6 permutations in the final line of
\Cref{eqn:gamma_ffts}. \Cref{gamma_fft} shows the result of such
a calculation with $128^3$ grids in real space, but keeping the
same $k_{max}$. The discrepancy in number of grid points arises
from aliasing considerations when putting particles on a grid,
as discussed in \Cref{sec:aliasing}. Although this is not relevant
here we only use up to $\frac{2}{3}k_{Ny}$ of FFT grids here for
consistency with our analysis of simulation data. Thus, both
methods effectively use the same number of grid points as far as
the tetrapyd is concerned.

Although \Cref{gamma_tetrapyd} and \Cref{gamma_fft} share
qualitatively similarities, demonstrating the same grid structure
and features along the main diagonal and its close neighbours,
the numerical values of the off-diagonal elements are much
smaller with the FFT calculation. Curiously this would suggest
the modes are more orthogonal to each other when used in
conjunction with FFTs. In rotating the \MODALLSS{} coefficients
from the $Q$ to $R$ basis we need to calculate $\lambda_{nm}$
(\Cref{eqn:RQconv}), given by $\gamma^{-1}=\lambda^T\lambda$.
Since the inverse of a matrix is highly susceptible even to
small changes in off-diagonal elements, big differences in the
final bispectrum estimation can result if one is not careful.
To illustrate this effect we made the following tests of the
FFT-based \MODALLSS{} code using randomly generated Gaussian
density fields. Gaussianity implies the lack of bispectrum
and higher order correlators, which has two consequences on
the \MODALLSS{} coefficients. First,
$\expval{\beta^Q_n}=\langle\beta^R_n\rangle=0$ due to the absence of
any bispectrum. Additionally, as shown in the \MODALLSS{}
covariance calculation (\Cref{betaQ2}), for a Gaussian density
field the $\beta^Q$ coefficients satisfy
$\expval{\beta^Q_m\beta^Q_n}=\gamma_{mn}$. To ensure the internal
consistency of the method we rotate this expression into the $R$
basis with the $\gamma_{nm}$ calculated with the two methods above
and check if we recover $\expval{\beta^R_m\beta^R_n}=\delta_{mn}$.
The conversion is achieved in the same manner as discussed in
\Cref{sec:ortho_basis} by first taking the Cholesky decomposition
of $\gamma$ to obtain $\lambda^{-1}$, then a further matrix
inversion gives $\lambda$. These are good sanity checks that our
numerical code is behaving as expected and that the algorithm
does indeed work. 

\begin{figure*}
  \begin{subfigure}[b]{0.49\textwidth}
    \includegraphics[width=\textwidth]{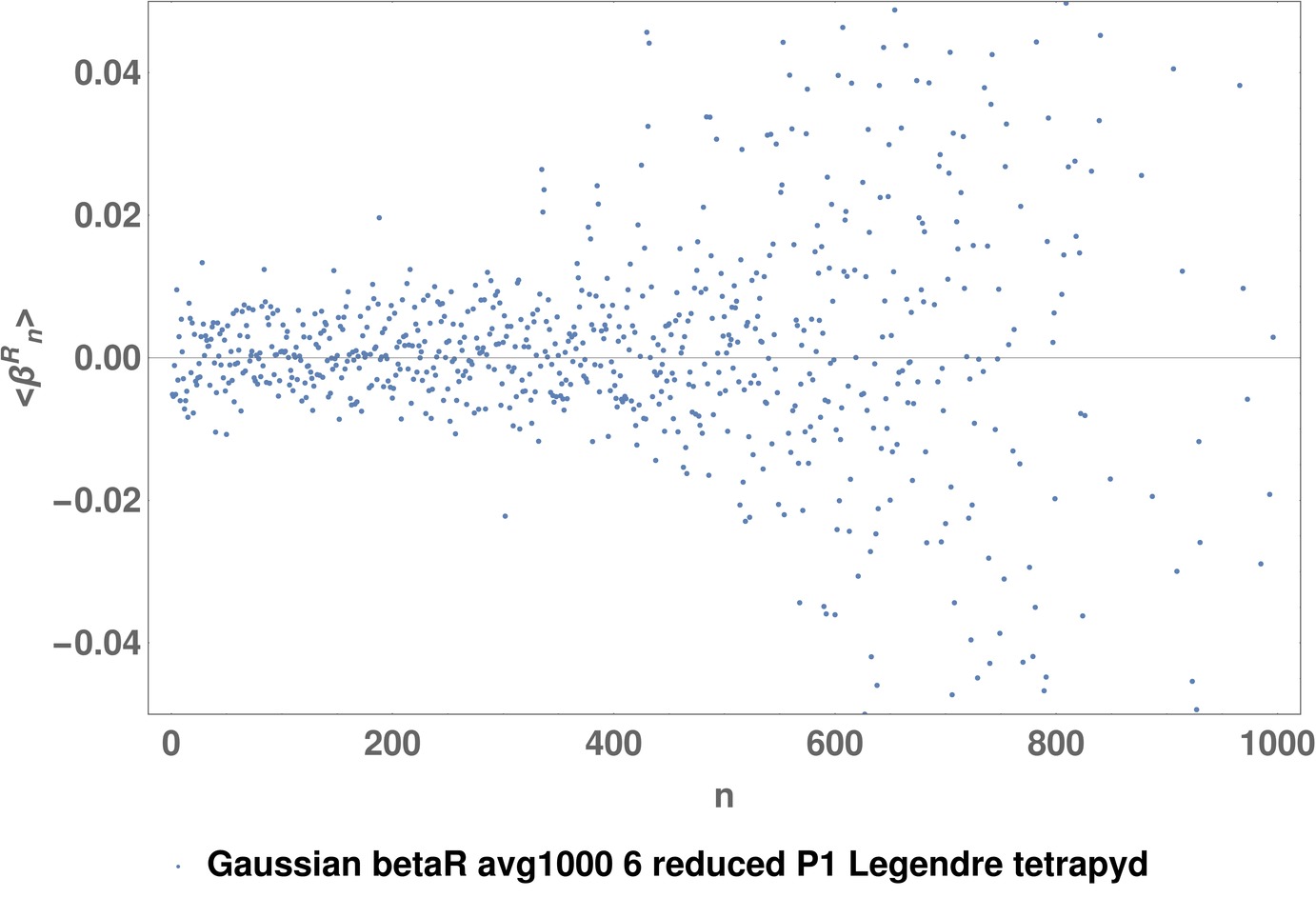} 
    \caption{$\gamma_{nm}$ calculated on the tetrapyd, giving
      $\langle\beta^R_n\rangle=0.016\pm0.202$. Although
      this is consistent with 0, it is clear the higher modes are
      strongly divergent from the mean which is an indication
      something is wrong.
    }
    \label{betaR_tetrapyd}
  \end{subfigure}
  ~
  \begin{subfigure}[b]{0.49\textwidth}
    \includegraphics[width=\textwidth]{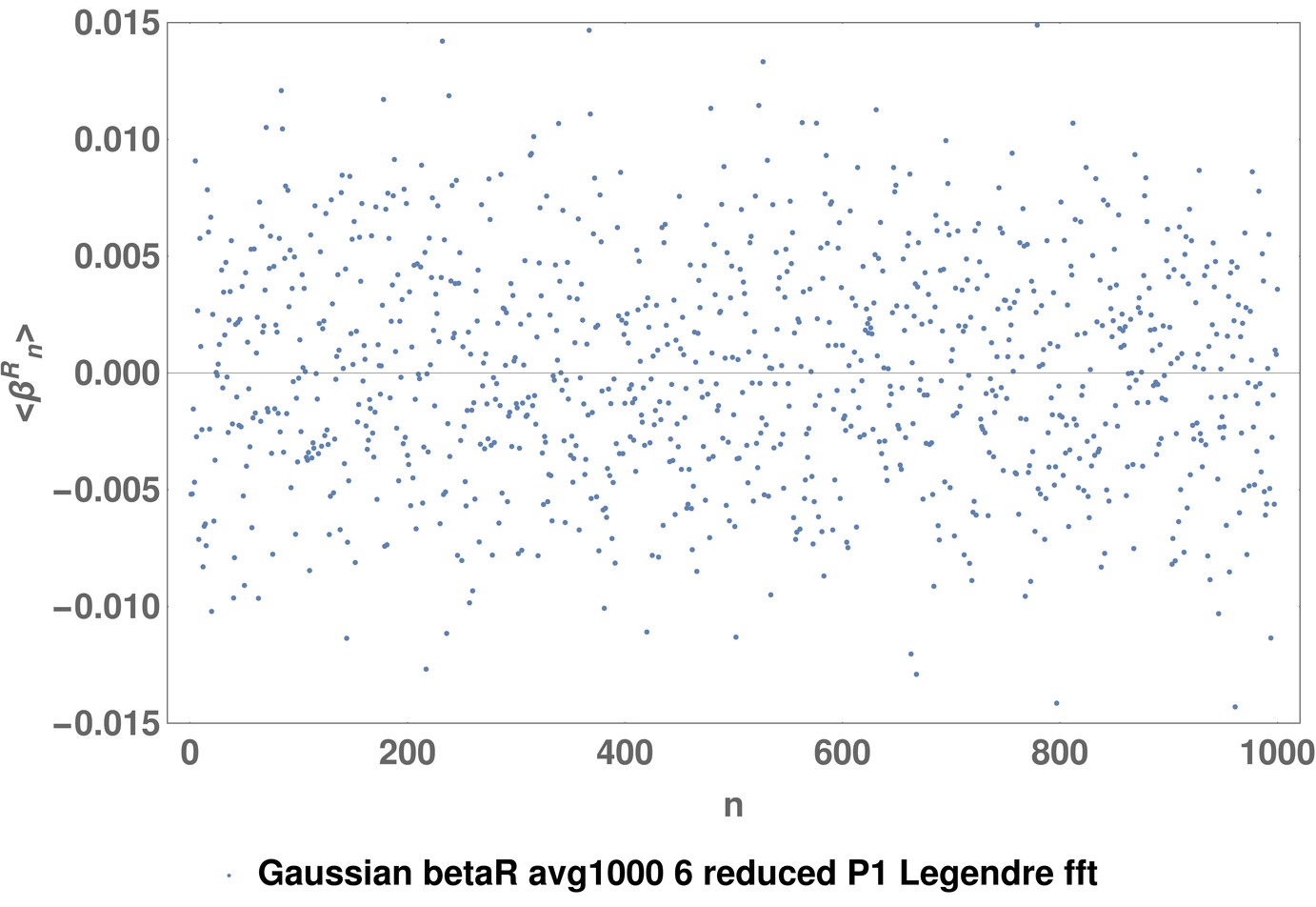}
    \caption{$\gamma_{nm}$ calculated with FFTs, giving
      $\langle\beta^R_n\rangle=0.0001\pm0.0048$. It is clear the
      $\beta^R_n$ thus obtained is much better behaved across the
      entire range of $n$, without any of the divergences seen to
      the left.
    }
    \label{betaR_fft}
  \end{subfigure}
  \caption{Testing the $\gamma_{mn}$ matrices by rotating $\beta^Q_n$
    into $\beta^R_n$ and checking $\langle\beta^R_n\rangle=0$.}
  \label{fig:betaR}
\end{figure*}

\begin{figure*}
  \begin{subfigure}[b]{0.49\textwidth}
    \includegraphics[width=\textwidth]{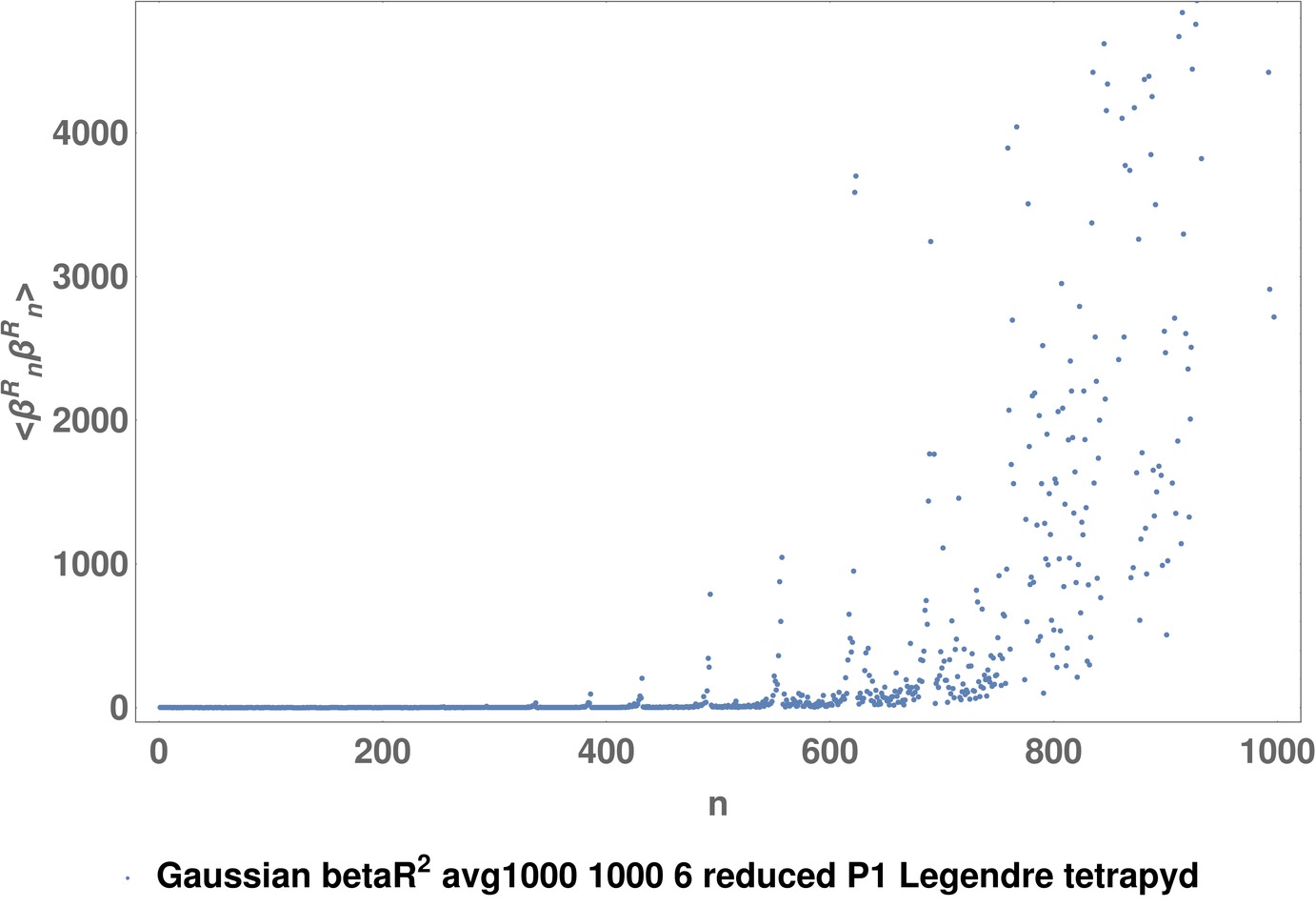} 
    \caption{$\gamma_{nm}$ calculated on the tetrapyd, giving
      $\expval{\beta^R_n\beta^R_n}=4000\pm22000$. There is
      no doubt that using this $\gamma_{nm}$ will lead to
      inconsistent bispectrum estimates.}
    \label{betaR2_tetrapyd}
  \end{subfigure}
  ~
  \begin{subfigure}[b]{0.49\textwidth}
    \includegraphics[width=\textwidth]{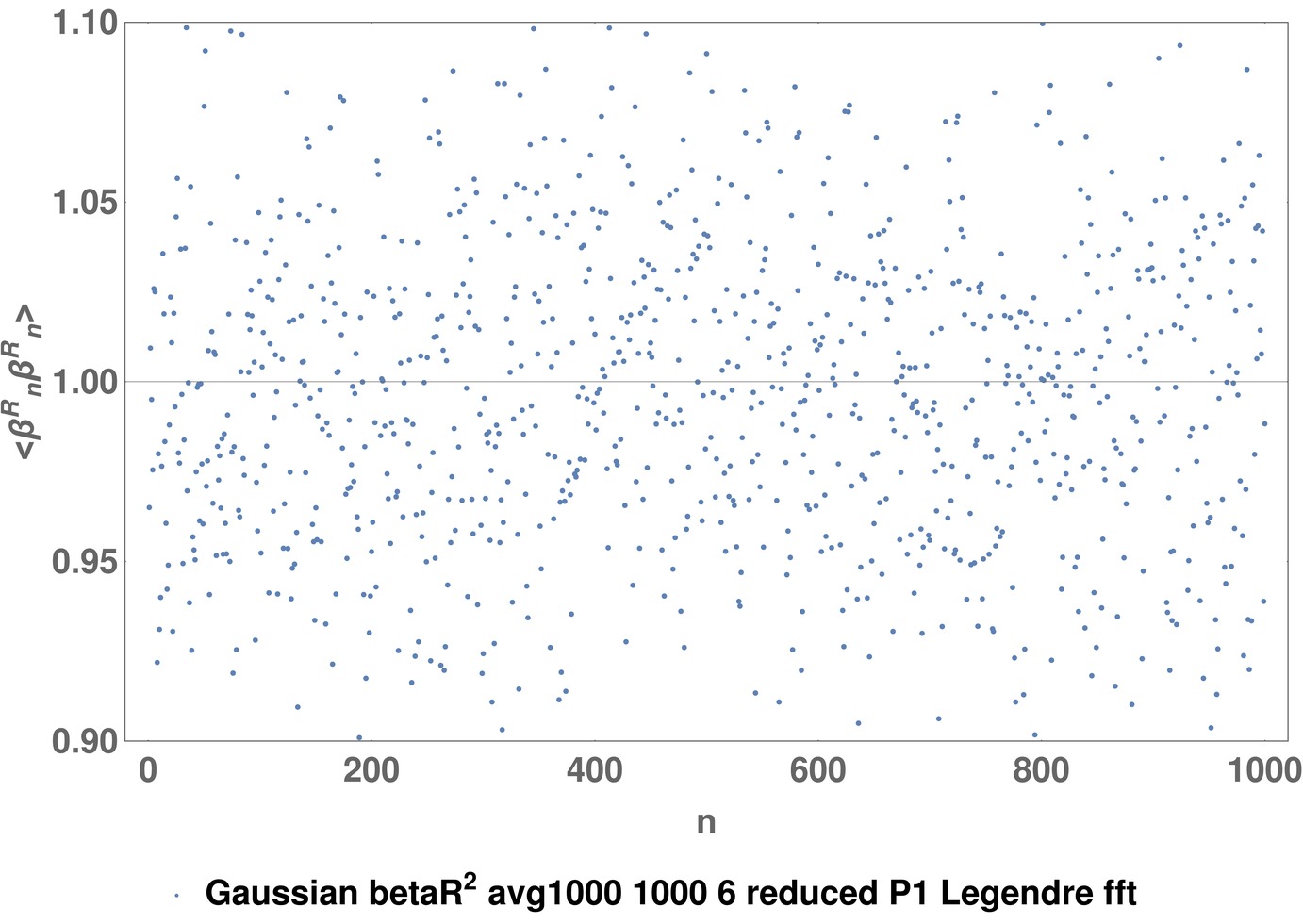}
    \caption{$\gamma_{nm}$ calculated with FFTs, giving
      $\expval{\beta^R_n\beta^R_n}=0.997\pm0.045$. This
      gives the correct mean and the correct order of
      magnitude in error since
      $\frac{1}{\sqrt{1000}}\sim3.3\%$.}
    \label{betaR2_fft}
  \end{subfigure}
  \caption{Testing the $\gamma_{mn}$ matrices by checking
    $\expval{\beta^R_n\beta^R_n}=1$.}
  \label{fig:betaR2}
\end{figure*}

The results of the $\langle\beta^R_n\rangle=0$ test is shown in
\Cref{fig:betaR} and the $\expval{\beta^R_n\beta^R_n}=1$
test in \Cref{fig:betaR2}. Here we used $128^3$ FFT grids and
42 tetrapyd points as above. The $\langle\beta^R_n\rangle=0$ test
is inconclusive as $\langle\beta^R_n\rangle$ calculated both
ways are consistent with 0, but when the $\gamma_{nm}$ calculated
with the tetrapyd is used a strong divergence from the mean is
observed at high $n$, which might be an indication that something
is amiss. On the other hand \Cref{fig:betaR2} clearly demonstrates
the problem with using the tetrapyd-based $\gamma_{nm}$, as even
stronger deviations are seen due to the inconsistent off-diagonal
terms. We conclude that if the incorrect $\gamma_{nm}$ is used
one would not bias the mean (i.e. the bispectrum estimation itself),
but would lead to hugely inflated covariances in the estimated
bispectrum.

\begin{figure*}
  \begin{subfigure}[b]{0.49\textwidth}
    \includegraphics[width=\textwidth]{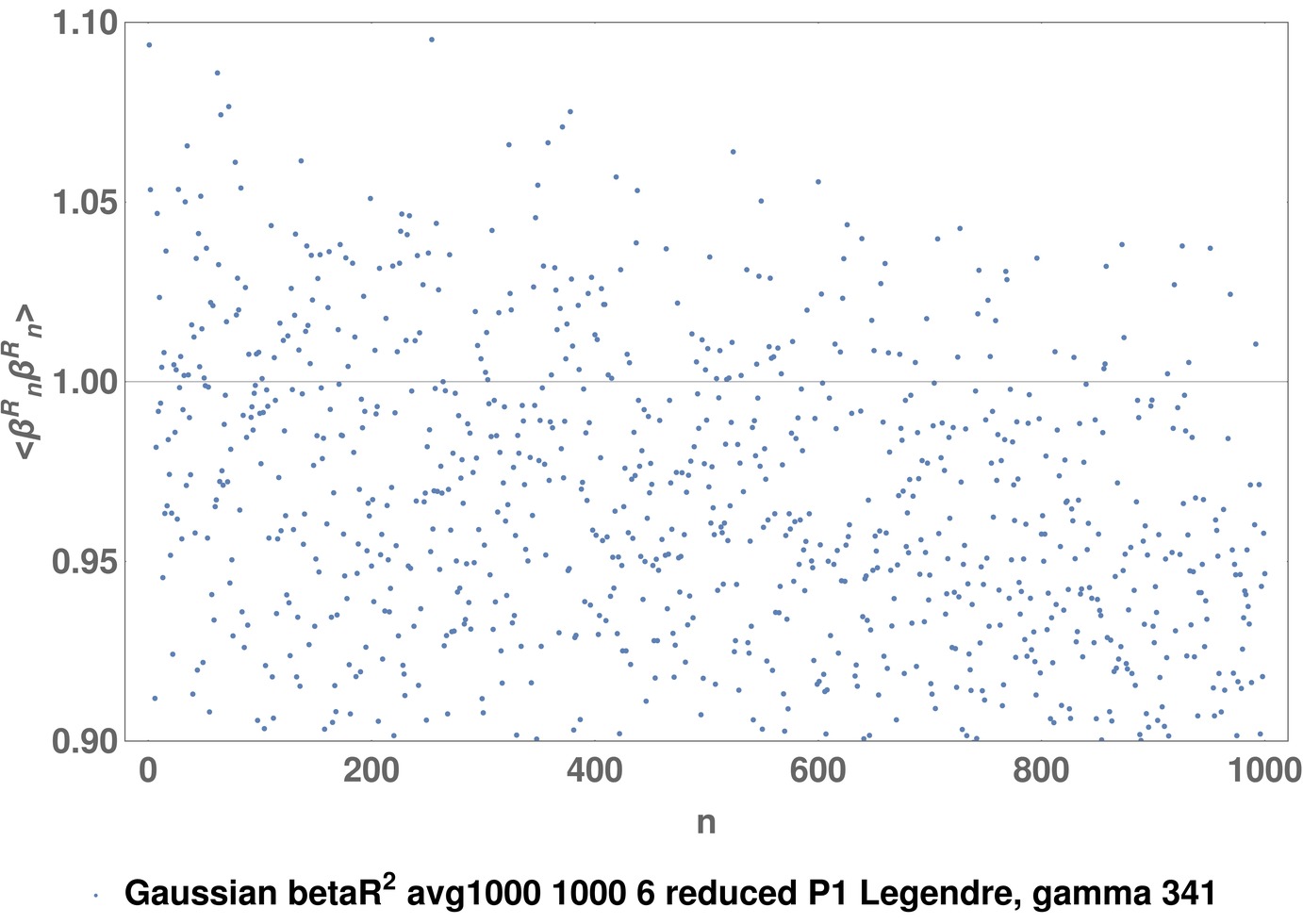} 
    \caption{341 grid points, giving $\expval{\beta^R_n\beta^R_n}=0.960\pm0.047$}
    \label{betaR2_341}
  \end{subfigure}
  ~
  \begin{subfigure}[b]{0.49\textwidth}
    \includegraphics[width=\textwidth]{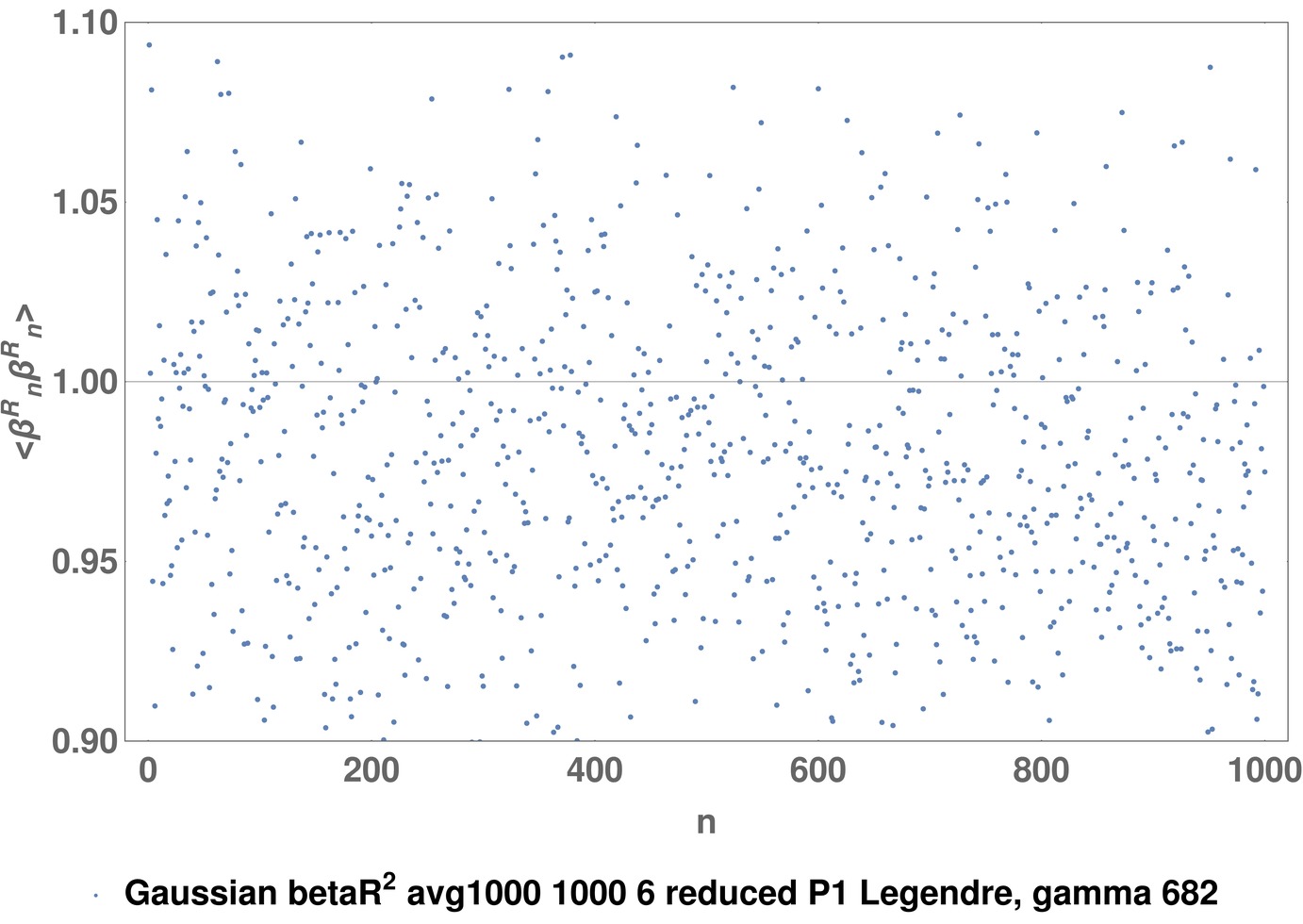}
    \caption{682 grid points, giving $\expval{\beta^R_n\beta^R_n}=0.977\pm0.045$}
    \label{betaR2_682}
  \end{subfigure}

  \begin{subfigure}[b]{0.49\textwidth}
    \includegraphics[width=\textwidth]{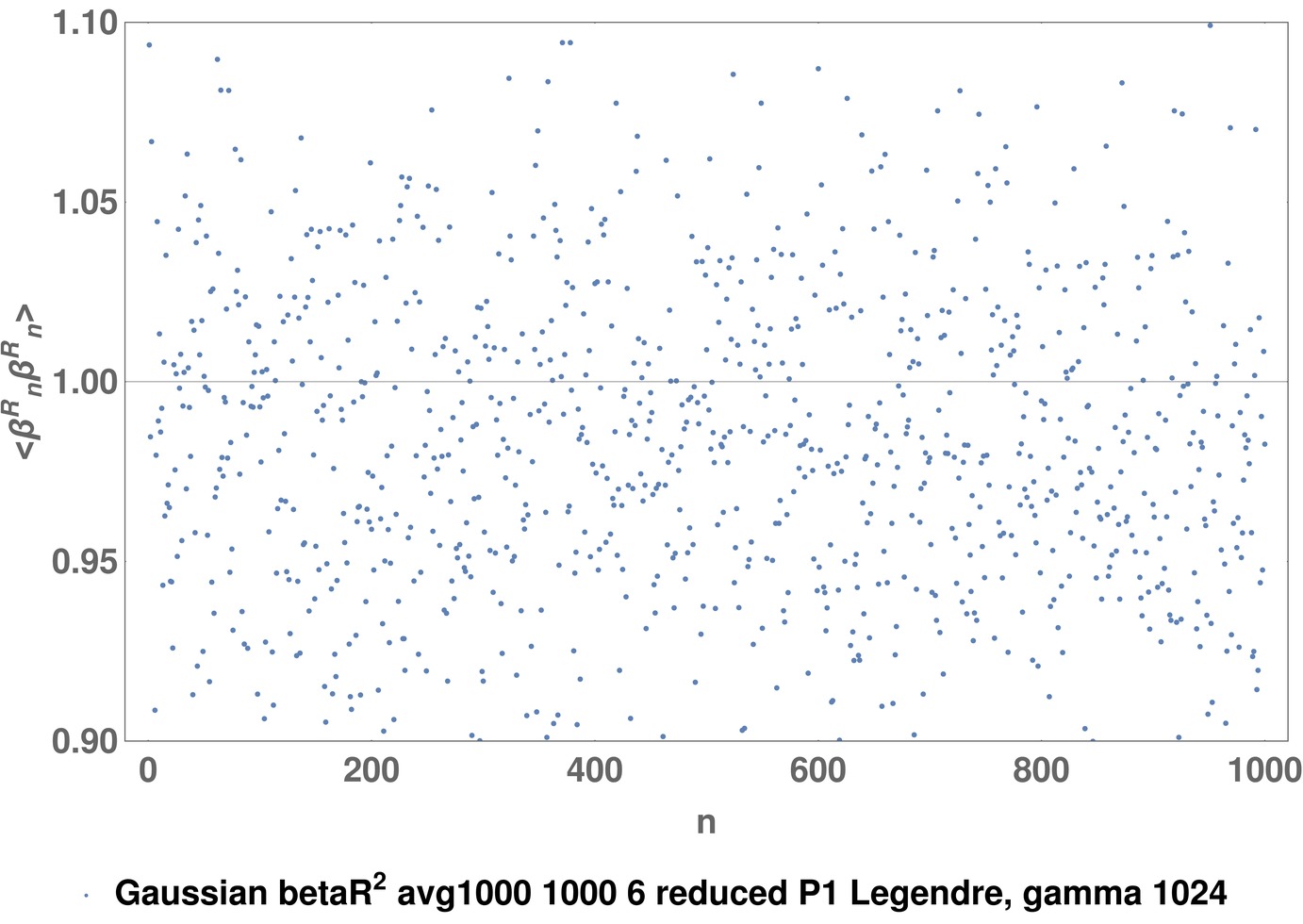} 
    \caption{1024 grid points, giving $\expval{\beta^R_n\beta^R_n}=0.981\pm0.045$}
    \label{betaR2_1024}
  \end{subfigure}
  ~
  \begin{subfigure}[b]{0.49\textwidth}
    \includegraphics[width=\textwidth]{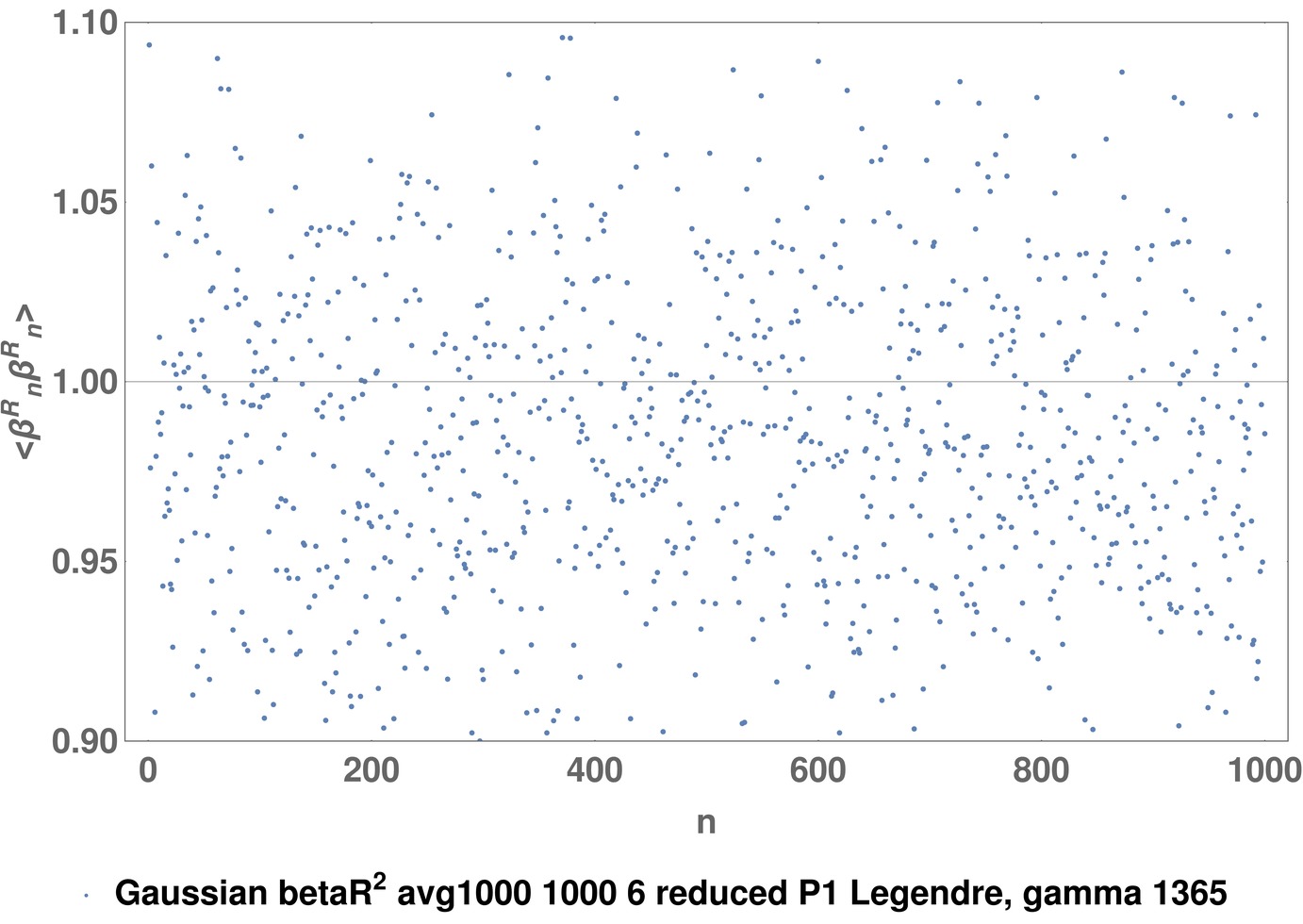}
    \caption{1365 grid points, giving $\expval{\beta^R_n\beta^R_n}=0.983\pm0.045$}
    \label{betaR2_1365}
  \end{subfigure}
  \caption{Checking $\expval{\beta^R_n\beta^R_n}=1$ with
    $\gamma_{mn}$ calculated on the tetrapyd with a range of
    grid points.}
  \label{fig:betaR2_gamma}
\end{figure*}

For grid sizes up to $512^3$ we can use the FFT method to calculate
$\gamma_{mn}$, but for $1024^3$ grids and above the computational
cost becomes impractically big. For this reason we have found a way
to use the tetrapyd-based $\gamma_{nm}$ to deliver consistent
results. This is illustrated in \Cref{fig:betaR2_gamma} where we
check $\expval{\beta^R_n\beta^R_n}=1$ with $1024^3$ FFT grids and 
$\gamma_{nm}$ computed on the tetrapyd using different number of
grid points. There is a clear improvement over the previous results
based on only 41 tetrapyd grid points, but although all 4 plots are
consistent with $\expval{\beta^R_n\beta^R_n}=1$ a downward trend at
high $n$ can be seen in the 341 and 682 case. However when 1024 or
more tetrapyd points are used this trend virtually disappears,
with only a marginal improvement in using 1365 points instead of 1024.
Therefore for large FFT grids we shall use the same number of
tetrapyd points as the FFT grid so as not to bias the bispectrum
covariance.

\begin{figure*}
  \begin{subfigure}[b]{0.49\textwidth}
    \includegraphics[width=\linewidth]{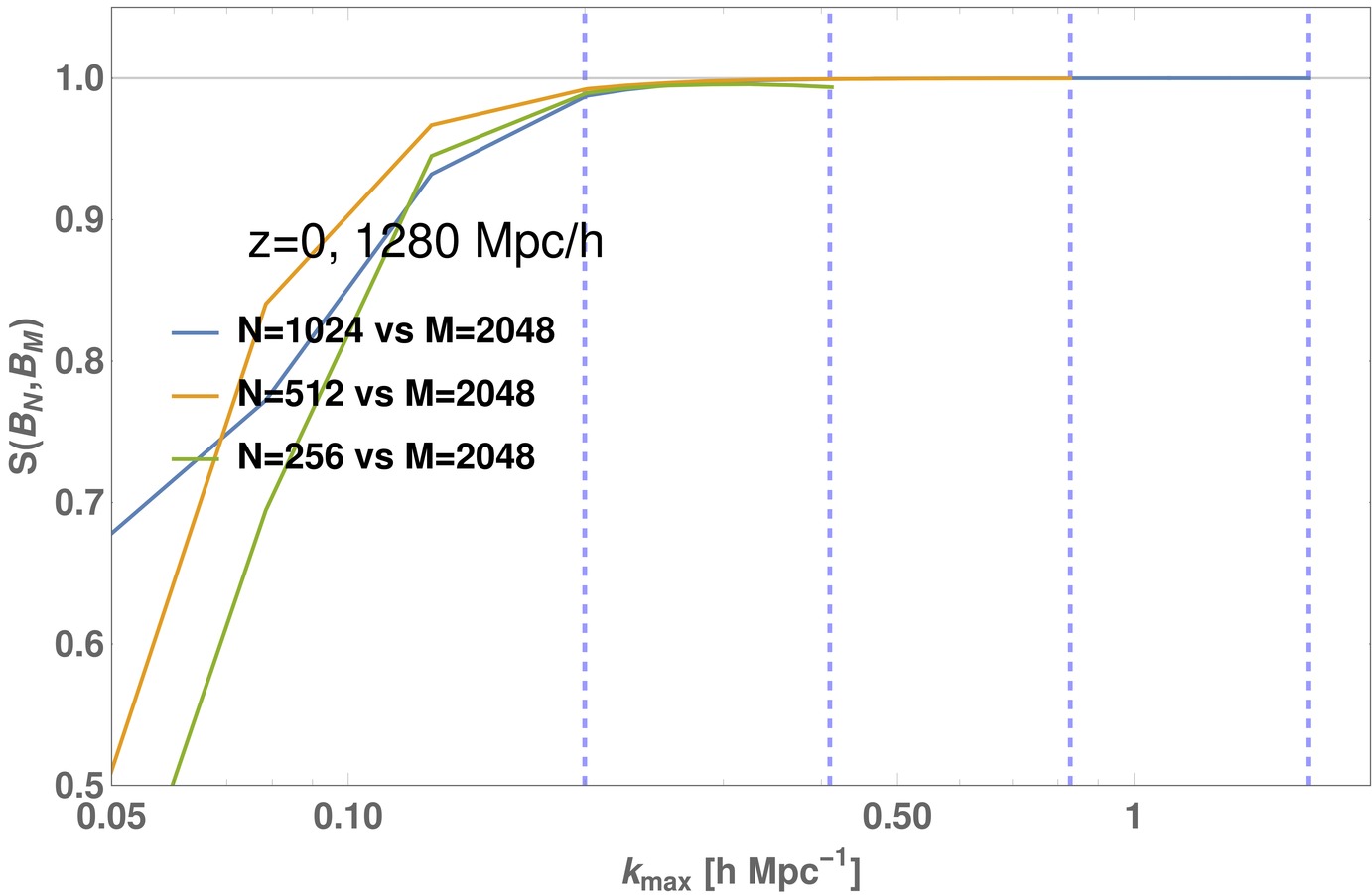}
    \caption{Shape correlator}
  \end{subfigure}
  ~
  \begin{subfigure}[b]{0.49\textwidth}
    \includegraphics[width=\linewidth]{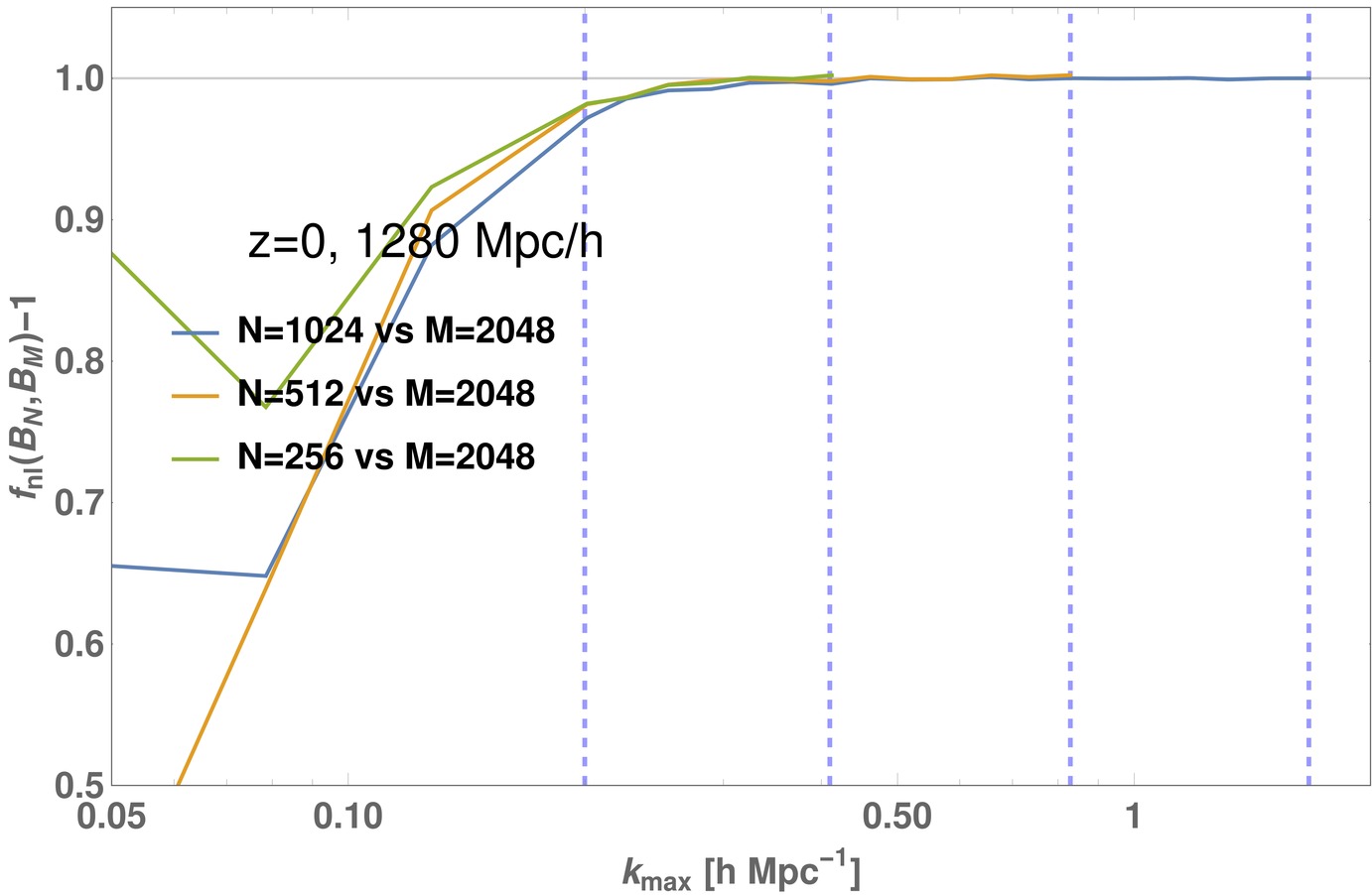}
    \caption{$f_{nl}$ correlator}
  \end{subfigure}
  \caption{
    Correlation coefficients between $\beta^R$ coefficients
    calculated with different $k_{max}$. This is achieved by
    reconstructing the estimated bispectrum to a lower $k_{max}$
    within the range of validity of the $\beta^R$ coefficients,
    and calculating the correlation coefficients directly
    using the resulting tetrapyds. The dashed blue lines
    represent the cutoff frequency corresponding to $128^3$,
    $256^3$, $512^3$ and $1024^3$ FFT grids, i.e. $41k_{F}$,
    $84k_{F}$, $169k_{F}$ and $340k_{F}$ respectively.
  }
  \label{fig:bis_recon}
\end{figure*}

Finally in this section we assess the effectiveness of this procedure
on a real signal, i.e. the $1280,\text{Mpc}$ \GADGET{} simulation at
redshift $z=0$ as presented in \Cref{work}. With the $\beta^R$
coefficients calculated up to a certain $k_{max}$ we can reconstruct
the bispectrum tetrapyd of the simulation to a lower one, and thus
compare the fidelity of bispectrum estimation when different FFT
grids and means of calculating $\gamma_{mn}$ are used. As shown in
\Cref{fig:bis_recon} the set of $\beta^R$ coefficients from a $2048^3$
grid is consistent with the others to 2\% level down to $41k_{F}$,
a very impressive result considering this accounts for
$\left(\frac{41}{681}\right)^3\sim0.02\%$ of the total tetrapyd. It is
therefore unnecessary to recalculate $\beta^R$ coefficients with
fewer FFT grid points, as long as we disregard the very tip of the
tetrapyd where the \MODALLSS{} method breaks down. We also restrict
ourselves to using $256^3$ grids or larger since it is clear that
reliable information cannot be obtained below $41k_{F}$. One therefore
has to carefully choose the box size of the simulation so that
the physically interesting $k$ scales are above this limit.

\bibliographystyle{apsrev4-1}
\bibliography{main}{}

\end{document}